\documentclass[12pt]{amsart}
\usepackage[leqno]{amsmath}
\usepackage[psamsfonts]{amssymb}
\usepackage{amsthm}
\usepackage[mathscr,psamsfonts]{eucal}
\usepackage{cmmib57}
\usepackage{amscd}
\usepackage{diagrams}
\usepackage{url}
\usepackage{color}

\IfFileExists{Ueur.fd}{\input{Ueur.fd}}{\input{ueur.fd}}
\IfFileExists{Ueur57.fd}{\input{Ueur57.fd}}{\input{ueur57.fd}}
\DeclareMathAlphabet{\eurm}{U}{eur}{m}{n}
\DeclareMathAlphabet{\eubf}{U}{eur}{b}{n}
\font\cyr=wncyr10 scaled 1200

\newcommand{\myssec}[1]{\subsection{#1}}

\newtheoremstyle
{MyThm}
{10pt}
{10pt}
{\itshape}
{\parindent}
{\bfseries}
{.}
{.5em}
{}
\swapnumbers
\theoremstyle{MyThm}
\newtheorem{Caution}{Caution}[section]
\newtheorem{Convention}[Caution]{Convention}
\newtheorem{Corollary}[Caution]{Corollary}
\newtheorem{Definition}[Caution]{Definition}
\newtheorem{Example}[Caution]{Example}
\newtheorem{Exercise}[Caution]{Exercise}
\newtheorem{Lemma}[Caution]{Lemma}
\newtheorem{Notation}[Caution]{Notation}
\newtheorem{Note}[Caution]{Note}
\newtheorem{Problem}[Caution]{Problem}
\newtheorem{Proposition}[Caution]{Proposition}
\newtheorem{Remark}[Caution]{Remark}
\newtheorem{Theorem}[Caution]{Theorem}
\newcommand{\bAs}{\begin{Assumption}\em}
\newcommand{\eAs}{\end{Assumption}}
\newcommand{\bCa}{\begin{Caution}\em}
\newcommand{\eCa}{\end{Caution}}
\newcommand{\bCr}{\begin{Corollary}\em}
\newcommand{\eCr}{\end{Corollary}}
\newcommand{\bCv}{\begin{Convention}\em}
\newcommand{\eCv}{\end{Convention}}
\newcommand{\bDf}{\begin{Definition}\em}
\newcommand{\eDf}{\end{Definition}}
\newcommand{\bDr}{\begin{Exercise}\em}
\newcommand{\eDr}{\end{Exercise}}
\newcommand{\bEx}{\begin{Example}\em}
\newcommand{\eEx}{\end{Example}}
\newcommand{\bLm}{\begin{Lemma}\em}
\newcommand{\eLm}{\end{Lemma}}
\newcommand{\bNo}{\begin{Notation}\em}
\newcommand{\eNo}{\end{Notation}}
\newcommand{\bNt}{\begin{Note}\em}
\newcommand{\eNt}{\end{Note}}

\newcommand{\bPb}{\begin{Problem}\em}
\newcommand{\ePb}{\end{Problem}}
\newcommand{\bPr}{\begin{Proposition}\em}
\newcommand{\ePr}{\end{Proposition}}
\newcommand{\bPs}{\begin{Postulate}\em}
\newcommand{\ePs}{\end{Postulate}}
\newcommand{\bRm}{\begin{Remark}\em}
\newcommand{\eRm}{\end{Remark}}

\newcommand{\bTh}{\begin{Theorem}}
\newcommand{\eTh}{\end{Theorem}}
\newcommand{\bEq}{\begin{equation}}
\newcommand{\eEq}{\end{equation}}
\newcommand{\beq}{\begin{equation*}}
\newcommand{\eeq}{\end{equation*}}

\newcommand{\bal}{\begin{align*}}
\newcommand{\bAl}{\begin{align}}
\newcommand{\bat}{\begin{alignat*}}
\newcommand{\bAt}{\begin{alignat}}
\newcommand{\bml}{\begin{multline*}}
\newcommand{\bMl}{\begin{multline}}
\newcommand{\bgt}{\begin{gather*}}
\newcommand{\bGt}{\begin{gather}}
\newcommand{\bCd}{\bEq\begin{CD}}
\newcommand{\eCd}{\end{CD}\eEq}
\newcommand{\bcd}{\beq\begin{CD}}
\newcommand{\ecd}{\end{CD}\eeq}
\newcommand{\bdg}{\beq\begin{diagram}}
\newcommand{\edg}{\end{diagram}\eeq}
\newcommand{\bDg}{\bEq\begin{diagram}}
\newcommand{\eDg}{\end{diagram}\eEq}
\newcommand{\bcn}{\begin{center}}
\newcommand{\ecn}{\end{center}}
\newcommand{\ben}{\begin{enumerate}}
\newcommand{\een}{\end{enumerate}}
\newcommand{\btb}{\begin{tabbing}}
\newcommand{\etb}{\end{tabbing}}
\newcommand{\bfz}{\begin{footnotesize}}
\newcommand{\efz}{\end{footnotesize}}
\newcommand{\bsz}{\begin{scriptsize}}
\newcommand{\esz}{\end{scriptsize}}
\newcommand{\Rn}{{\B R}}

\newcommand{\Al}{\forall}

\newcommand{\h}{\hbar}
\newcommand{\1}{\mathbf 1}

\newcommand{\der}{\partial}

\newcommand{\nab}{\nabla}

\newcommand{\Ela}{^\mathfrak{a}{}}
\newcommand{\Fla}{^{\flat}{}}
\newcommand{\Sha}{^{\sharp}{}}

\newcommand{\Prl}{^{\|}{}}
\newcommand{\prl}{_{\|}{}}
\newcommand{\Per}{^{\perp}{}}
\newcommand{\per}{_{\perp}{}}

\newcommand{\lang}{\langle}
\newcommand{\rang}{\rangle}

\newcommand{\mto}{\mapsto}

\newcommand{\sub}{\subset}
\newcommand{\com}{\circ}
\newcommand{\comm}{\!\circ\!}
\newcommand{\car}{\times}
\newcommand{\ten}{\otimes}

\newcommand{\wed}{\wedge}

\DeclareMathOperator{\con}{\lrcorner}

\newcommand{\eqv}{\,\equiv\,}
\newcommand{\seq}{\,\simeq\,}
\newcommand{\nid}{\not\equiv}
\DeclareMathOperator{\byd}{\,{\rm =}{\raisebox{.092ex}{\rm :}}\,}

\newcommand{\ucar}[1]{\underset{#1}{\times}}

\newcommand{\udrs}[1]{\underset{#1}{\oplus}}

\newcommand{\fr}[2]{\frac{#1}{#2}\,}
\newcommand{\tfr}[2]{\tfrac{#1}{#2}\,}
\newcommand{\co}[2]{_{#1}{}^{#2}}
\newcommand{\col}[3]{_{#1}{}^{#2}{}_{#3}}

\newcommand{\cu}[3]{_{#1#2}{}^{#3}}
\newcommand{\cul}[4]{_{#1#2}{}^{#3}{}_{#4}}

\newcommand{\Ga}[2]{_{#1}{}^{#2}_0}

\newcommand{\ga}[1]{_0{}^{#1}_0}

\newcommand{\ENDE}{{\,\text{\footnotesize\qedsymbol}}}

\newcommand{\sep}[1]{{\quad\text{\rm{#1}}\quad}}
\newcommand{\ssep}[1]{{\qquad\text{\rm{#1}}\qquad}}

\newcommand{\sst}{\;\;|\;\;}

\newcommand{\ar}[1]{\url{http://arXiv.org/abs/#1}}

\newcommand{\Gcu}[3]{_{#1#2}{}^{#3}_0}

\newcommand{\im}{{{}{\rm im \, }}}

\DeclareMathOperator{\Alt}{{{Alt}}}

\DeclareMathOperator{\Grass}{{{Grass}}}

\DeclareMathOperator{\Span}{{{span}}}

\DeclareMathOperator{\fib}{{{fib}}}

\DeclareMathOperator{\id}{{{id}}}
\DeclareMathOperator{\map}{{{map}}}

\DeclareMathOperator{\proj}{{{proj}}}

\newcommand{\f}[1]{{\boldsymbol{#1}}}

\newcommand{\ul}[1]{{\underline{#1}}}
\newcommand{\ba}[1]{{{\bar{#1}}}}
\newcommand{\bau}[1]{{\text{\b{$#1$}}}{}}

\newcommand{\ha}[1]{{\hat{#1}}}

\newcommand{\ti}[1]{{\tilde{#1}}}
\newcommand{\wti}[1]{{\widetilde{#1}}}
\newcommand{\dt}[1]{{\dot{#1}}}

\newcommand{\br}[1]{\breve{#1}{}}

\newcommand{\bma}{\left(\begin{matrix}}
\newcommand{\ema}{\end{matrix}\right)}

\newcommand{\M}[1]{{\mathscr{#1}}}

\newcommand{\B}[1]{{\mathbb{#1}}}
\newcommand{\baB}[1]{{\bar{{\mathbb{#1}}}}}

\newcommand{\K}[1]{\text{\cyr{#1}}}

\newcommand{\alp}{\alpha}
\newcommand{\bet}{\beta}
\newcommand{\gam}{\gamma}
\newcommand{\del}{\delta}
\newcommand{\eps}{\epsilon}

\newcommand{\tht}{\theta}

\newcommand{\kap}{\kappa}

\newcommand{\lam}{\lambda}

\newcommand{\sig}{\sigma}

\newcommand{\ups}{\upsilon}
\newcommand{\vphi}{\varphi}

\newcommand{\ome}{\omega}
\newcommand{\Gam}{\Gamma}

\newcommand{\Lam}{\Lambda}

\newcommand{\Sig}{\Sigma}

\newcommand{\Ups}{\Upsilon}

\newcommand{\Ome}{\Omega}

\allowdisplaybreaks[4]

\title[Geometric structures of the 
phase space]{Geometric structures of the 
classical 
\\
general relativistic phase space}

\author[J. Jany\v{s}ka, M. Modugno]
	{Josef Jany\v{s}ka and Marco Modugno}

\address{
{\ }
\newline
Department of Mathematics and Statistics, Masaryk University
\newline
Jan\'a\v{c}kovo n\'am 2a, 602 00 Brno, Czech Republic
\newline
E-mail: {\tt janyska@math.muni.cz}
\newline
{\ }
\newline
Department of Applied Mathematics, Florence University
\newline
Via S. Marta 3, 50139 Florence, Italy
\newline
E-mail: {\tt marco.modugno@unifi.it}
}



\thanks{
 This research has been supported
by the Ministry of Education of the Czech Republic under the project
MSM0021622409,
by the Grant agency of the Czech Republic under the project GA
201/05/0523,
by MIUR of Italy under the project PRIN 2005 ``Simmetrie e
Supersimmetrie Classiche e Quantistiche",
by GNFM of INdAM
and by Florence University.
}

\begin{document}
\maketitle
\begin{abstract}
 This paper is concerned with basic geometric properties of the
phase space of a classical general relativistic particle, regarded as
the 1st jet space of motions, i.e. as the 1st jet space of timelike
1--dimensional submanifolds of spacetime.
 This setting allows us to skip constraints.

 Our main goal is to determine the geometric conditions by which the
Lorentz metric and a connection of the phase space yield contact
and Jacobi structures.
 In particular, we specialise these conditions to the cases when the
connection of the phase space is generated by the metric and an
additional tensor.
 Indeed, the case generated by the metric and the electromagnetic
field is included, as well.
\end{abstract}

\maketitle
\tableofcontents

\section{Introduction}
\label{Introduction}
 Usually the classical phase space of general relativity is defined to
be the subspace of the tangent bundle of spacetime consisting of
future oriented timelike vectors.
 In this phase space we can achieve the standard symplectic formalism
for general relativistic analytical mechanics.
 However, only time normalised vectors have physical meaning, hence
one is forced to introduce the corresponding constraint with all
associated odd consequences.

 On the other hand, there is another possible approach to the phase
space, by defining it as the 1st jet space of motions regarded as
timelike 1--dimensional submanifolds of spacetime.
 This viewpoint deals with the essential projective nature of the
general relativistic phase space and allows us to skip constraints.

 The present paper is aimed at analysing a few basic properties of
the phase space according to the 2nd approach. 
 Clearly, the most natural goal is to study the contact and
Jacobi structures induced by the Lorentz metric.
 On the other hand, we could show that the electromagnetic field
yields additional terms to the purely metric geometric objects of the
phase space.
 Indeed, the analytical mechanics for a general relativistic particle
effected by the electromagnetic fields is to be formulated with these
extended geometric objects.

 For this reason, in the present paper, we determine the conditions by
which the geometric objects induced on the phase space by a general
phase connection yield contact and Jacobi structures.
 Of course, we consider the purely metric framework as particular
case.

 Actually, the literature on jets of submanifolds is less popular
and more sophisticated than that on jets of fibred manifolds.
 So, here we sketch a few essential notions on jets of submanifolds in
order to make the paper selfconsistent.

 We shall be concerned with several geometric structures on the
tangent space and on the phase space of spacetime.
 Some of these structures are well known and some are less standard.
 Thus, we recall some definitions and introduce new objects, as well.
 Indeed, the new objects are suitable to capture the non easy objects
arising on the phase space.

\smallskip

 Throughout the paper, all manifolds and maps will be smooth.
 If
$\f M$
and
$\f N$
are manifolds, then the sheaf of local maps
$\f M \to \f N$
is denoted by
$\map(\f M, \, \f N) \,;$
if
$p : \f F \to \f B$
is a fibred manifold, then the sheaf of local sections 
$\f B \to \f F$
is denoted by
$\sec(\f B, \f F) \,;$ 
if 
$p' : \f F' \to \f B'$ 
is another fibred manifold, then the sheaf of local fibered maps 
$\f F \to \f F'$ 
is denoted by 
$\fib(\f F,\f F') \,.$

\smallskip

 Let
$\f M$
be an $m$--dimensional manifold.

 If
$\Ome$
is a 2--form
and
$\Lam$
a 2--vector, then we define the ``musical" morphisms 
\bat{4}
\Ome\Fla &: \sec(\f M,T\f M) &&\to \sec(\f M,T^*\f M) &&: 
X &&\mto i_X \, \Ome \,,
\\
\Lam\Sha &: \sec(\f M,T^*\f M) &&\to \sec(\f M,T\f M) &&: 
\alp &&\mto i_\alp \, \Lam \,.
\end{alignat*}

\smallskip

 Let
$\f M$
be a $2n$--dimensional manifold.

 We recall that a {\em symplectic structure} 
\cite[pag. 90]{LibMar87} is defined by a 2--form
$\Ome$
such that
\beq
d\Ome = 0 \,,
\qquad
\Ome^n \nid 0 \,,
\eeq
and that a {\em Poisson structure} 
\cite[pag. 107]{LibMar87}
is defined by a 2--vector
$\Lam$
such that
\beq
[\Lam, \Lam] = 0 \,,
\eeq
where 
$[\,,]$
denotes the Schouten bracket.
 Moreover, if
$\Lam^n \nid 0 \,,$
then we say that the Poisson structure is {\em regular}.

 The musical morphism 
$\Ome\Fla$
of a symplectic structure turns out to be an isomorphism.
 If a Poisson structure is regular, then the musical morphism
$\Lam\Sha$
turns out to be an isomorphism.
 We say that a symplectic structure and a Poisson structure are {\em
mutually dual} if
\beq
(\Ome\Fla)^{-1} = \Lam\Sha \,.
\eeq

 Indeed, if 
$\Ome$
is a symplectic 2--form and
$\Lam$
the 2--vector given by
$\Lam\Sha = (\Ome\Fla)^{-1} \,,$ 
then 
$\Lam$ 
defines a regular Poisson structure.

\smallskip

 Let
$\f M$
be a 
$(2n + 1)$--dimensional manifold.

 We define a \emph{covariant pair} to be a pair
$(\ome, \Ome)$
consisting of a 1--form
$\ome$
and
a 2--form
$\Ome$
of constant rank
$2r \,,$
with
$0 \le r \le n \,,$
such that 
$\ome \wed \Ome^r \nid 0 \,,$ 
and a \emph{contravariant pair} to be a pair
$(E, \Lam)$
consisting of a vector field
$E$
and a 2--vector
$\Lam$
of constant rank
$2s \,,$
with
$0 \le s \le n \,,$
such that 
$E \wed \Lam^s \nid 0 \,.$
 Thus, by definition, we have
$\Ome^r \nid 0 \,,\; \Ome^{r+1} \eqv 0$
and
$\Lam^s \nid 0 \,,\; \Lam^{s+1} \eqv 0 \,.$

 We say that the pairs
$(\ome, \Ome)$
and
$(E, \Lam)$
are \emph{regular} if, respectively,
\beq
\ome \wed \Ome^n \nid 0
\ssep{and}
E \wed \Lam^n \nid 0 \,.
\eeq

 We recall that a {\em cosymplectic structure} 
\cite{deLTuy96} is
defined by a covariant pair
$(\ome,\Ome)$
such that
\beq
d\ome = 0 \,,
\qquad
d\Ome = 0 \,,
\qquad
\ome \wed \Ome^n \nid 0 \,,
\eeq
and a {\em contact structure} \cite{JanMod07} is defined by a
covariant pair
$(\ome, \Ome)$
such that
\beq
\Ome = d\ome \,,
\qquad
\ome \wed \Ome^n \nid 0 \,.
\eeq

 Equivalently, we can define a {\em contact structure}
\cite[pag. 285]{LibMar87} as a 1--form
$\ome$
such that
\beq
\ome \wed (d\ome)^n \nid 0 \,.
\eeq

 More generally, we define an {\em almost--cosymplectic--contact
structure} \cite{JanMod07} by a covariant pair
$(\ome,\Ome)$
such that
\beq
d\Ome = 0 \,,
\qquad
\ome \wed \Ome^n \nid 0 \,.
\eeq

 Thus, an almost--cosymplectic--contact structure becomes a
cosymplectic structure when 
$d\ome = 0$ 
and a contact structure when
$\Ome = d\ome \,.$

 We recall that a {\em Jacobi structure} \cite[pag. 337]{LibMar87} is
defined by a contravariant pair
$(E, \Lam)$
such that
\beq
[E, \Lam] = 0 \,,
\qquad
[\Lam, \Lam] = - 2 E \wed \Lam \,.
\eeq
 In the particular case when 
$E \eqv 0 \,,$
we obtain
$[\Lam, \Lam] = 0$
and the corresponding structure, given by 
$(E, \Lam) \eqv (0, \Lam) \,,$  
is a  {\em Poisson structure\/}.

 Further, a {\em coPoisson structure\/} \cite{JanMod07} is defined by
a contravariant pair
$(E, \Lam)$
such that
\bEq\label{coPoisson structure}
[E, \Lam] = 0 \,,
\qquad
[\Lam, \Lam] = 0 \,.
\eEq

 More generally, we define \cite{JanMod07} an {\em
almost--coPoisson--Jacobi structure} by a contravariant pair
$(E, \Lam) \,,$ 
along with a 1--form
$\ome \,,$
called {\em fundamental 1--form}, such that
\beq
[E,\Lam] = 
- E \wed \Lam\Sha(L_E \ome) \,,
\qquad
[\Lam,\Lam] = 
2 \, E \wed \big((\Lam\Sha \ten \Lam\Sha) (d\ome)\big)
\,.
\eeq

 The covariant pair
$(\ome,\Ome)$
and the contravariant pair
$(E,\Lam)$
are said to be \emph{mutually dual} if they are regular, the maps
$\Ome\Fla_{|\im\Lam\Sha} : \im\Lam\Sha \to
\im\Ome\Fla$
and
$\Lam\Sha_{|\im\Ome\Fla} : \im\Ome\Fla \to
\im\Lam\Sha$
are isomorphisms and
\beq
(\Ome\Fla_{|\im\Lam\Sha})^{-1} = 
\Lam\Sha_{|\im\Ome\Fla} \,,
\quad
(\Lam\Sha_{|\im\Ome\Fla})^{-1} = 
\Ome\Fla_{|\im\Lam\Sha} \,,
\quad 
i_E\Ome =0 \,,
\quad 
i_\ome\Lam = 0 \,,
\quad 
i_E\ome =1 \,.
\eeq

\bTh\label{Th1.1}
{\rm \cite{JanMod07}}
 Let 
$(\ome,\Ome)$ 
and 
$(E,\Lam)$ 
be mutually dual covariant and contravariant pairs.
 Then,

\smallskip

{\rm (1)}
$(\ome,\Ome)$ 
is an almost-cosymplectic-contact structure if and only if 
$(E,\Lam)$ 
is an almost-coPoisson-Jacobi structure along with the fundamental
1-form 
$\ome \,;$

\smallskip

{\rm (2)}
$(\ome,\Ome)$ 
is a cosymplectic structure 
if and only if 
$(E,\Lam)$ 
is a coPoisson structure;

\smallskip

{\rm (3)}
$(\ome,\Ome)$ 
is a contact structure if and only if 
$(E,\Lam)$ 
is a Jacobi structure.
\hfill\ENDE
\eTh

 Actually, the geometric structures arising in this paper, in the
framework of the Einstein's phase space, involve mainly the concepts
of contact and regular Jacobi structures (eventually, of
almost--cosymplectic--contact and regular almost--coPoisson--Jacobi
structures).
 On the other hand, the analogous geometric structures arising in the
framework of the Galilei's phase space 
\cite{JanMod06}
involve mainly the concepts of cosymplectic and coPoisson structures.

\smallskip

 In order to make our theory explicitly independent from scales, we
introduce the ``spaces of scales'' \cite{JanModVit07}.
 Roughly speaking, a space of scales
$\B S$
has the algebraic structure of
$\Rn^+$
but has no distinguished `basis'.
 We can define the tensor product of spaces of scales and the tensor
product of spaces of scales and vector spaces.
 We can define rational tensor powers
$\B S^{m/n}$
of a space of scales
$\B S \,.$
 Moreover, we can make a natural identification
$\B S^* \seq \B S^{-1} \,.$

 The basic objects of our theory (the metric field, the phase
2--form, the phase 2--vector, etc.) will be valued into {\em scaled}
vector bundles, that is into vector bundles multiplied tensorially
with spaces of scales.
 In this way, each tensor field carries explicit information on its
``scale dimension".

 Actually, we assume the following basic spaces of scales:
the space of {\em time intervals}
$\B T\,,$
the space of {\em lengths}
$\B L $ and the space of {\em masses}
$\B M \,.$
 We assume  the {\em speed of light}
$c \in \B T^{-1} \ten \B L$ and the {\em Planck constant} 
$\h \in \B T^{-1} \ten \B L^2 \ten \B M$
as  ``universal scales".
 Moreover, we will consider a {\em particle} of {\em mass}
$m \in \B M$
and {\em charge}
$q \in \B T^{-1} \ten \B L^{3/2} \ten \B M^{1/2} \,.$

\section{Geometry of spacetime}
\label{Geometry of spacetime}
\setcounter{equation}{0}
 We start by recalling a few basic properties of the Einstein
spacetime and of its tangent bundle, \cite{Jan05,JanMod95}.
\subsection{Spacetime}
\label{Spacetime}
 We assume \emph{spacetime} to be an oriented 4--dimensional manifold
$\f E$
equipped with a scaled Lorentzian metric
$g : \f E \to \B L^2 \ten (T^*\f E \ten T^*\f E) \,,$
with signature
$(-+++) \,;$
we suppose spacetime to be time oriented.
 The dual metric will be denoted by
$\ba g : \f E \to \B L^{-2} \ten (T\f E \ten T\f E) \,.$
 Actually, our results are essentially valid for any dimension 
$n \ge 3$ 
and a pseudo-Riemannian metric of signature 
$(1,n-1) \,.$

 A \emph{spacetime chart} is defined to be an ordered chart
$(x^0, x^i) \in \map(\f E, \, \Rn \car \Rn^3)$
of
$\f E \,,$
which fits the orientation of spacetime and such that the vector field
$\der_0$
is timelike and time oriented and the vector fields
$\der_1, \der_2, \der_3$
are spacelike.
 In the following, we shall always refer to spacetime charts.
Greek indices
$\lam, \mu, \dots$
will span spacetime coordinates, while Latin indices
$i, j, \dots$
will span spacelike coordinates.
 We have the coordinate expressions
\bat{3}
g
&=
g_{\lam\mu} \, d^\lam \ten d^\mu \,,
&&\ssep{with}
g_{\lam\mu}
&&\in \map(\f E, \, \B L^2 \ten \Rn) \,,
\\
\ba g
&=
g^{\lam\mu} \, \der_\lam \ten \der_\mu\,,
&&\ssep{with}
g^{\lam\mu}
&&\in \map(\f E, \, \B L^{-2} \ten \Rn) \,.
\end{alignat*}

 The fact that the metric
$g$
is a scaled tensor implies that the induced objects of the phase
space turn out to be scaled.
 In order to obtain unscaled objects of the phase space we could
replace
$g$
with the rescaled Lorentz metric
$G \byd \tfr m\h g : \f E \to \B T \ten (T^*\f E \ten T^*\f E) \,.$
 We leave to the reader the possible task to perform explicitly this
easy rescaling throughout the paper.
\subsection{Spacetime connections}
\label{Spacetime connections}
 Let us sketch the notion of general connection of the phase space
and related objects.
 This generality is intended as a preparation for the notion of
general connection of the phase space (which is not a vector bundle,
hence it does not carry the usual elementary definition of
connection) and later in order to allow general theorems concerning
geometric structures of the phase space.

 We define a \emph{spacetime connection} to be a connection of the
bundle
$T\f E \to \f E \,.$

 A spacetime connection can be expressed, equivalently, by a tangent
valued form
$K : T\f E \to T^*\f E \ten TT\f E \,,$
which is projectable over
$\f 1 : \f E \to T^*\f E \ten T\f E \,,$
or by the complementary vertical valued form
$\nu[K] : T\f E \to T^*T\f E \ten VT\f E \,.$
 Their coordinate expressions are of the type
\beq
K =
d^\lam \ten (\der_\lam + K\co\lam\nu \, \dt\der_\nu) \,,
\quad
\nu[K] =
(\dt d^\nu - K\co\lam\nu \, d^\lam) \ten \dt\der_\nu \,,
\sep{with}
K\co\lam\nu \in \map(T\f E, \, \Rn) \,,
\eeq
where
$(\der_\lam,\dt \der_\lam)$
and
$(d^\lam,\dt d^\lam)$
are the induced bases of local sections of
$TT\f E\to T\f E$
and
$T^*T\f E\to T\f E \,,$
respectively.

 Let us consider a spacetime connection
$K \,.$

\smallskip

 The connection
$K$
is said to be \emph{linear} if it is a linear fibred morphism over
$\f 1 : \f E \to T^*\f E \ten T\f E \,.$
 Thus, the connection
$K$
is linear if and only if its coordinate  expression is of the type
\[
K\co\lam\nu = K\col\lam\nu\mu \, \dt x^\mu \,,
\ssep{with}
K\col\lam\nu\mu \in \map(\f E, \, \Rn) \,.
\]

 The \emph{torsion} of
$K$
is defined to be the vertical valued 2--form
\beq
T \eqv T[K] \byd 2 \, [v, K] : T\f E \to \Lam^2T^*\f E \ten VT\f E
\,,
\eeq
where
$[\,,]$
is the Fr\"olicher-Nijenhuis bracket and
$v : T\f E \to T^*\f E \ten VT\f E$
is the natural vertical valued 1--form with coordinate expression
$v = d^\lam \ten \dt\der_\lam \,.$
 We have the coordinate expression
\beq
T = T\cu\lam\mu\nu \, d^\lam \wed d^\mu \ten \dt\der_\nu =
- 2 \, \dt\der_\mu K\co\lam\nu \, d^\lam \wed d^\mu \ten \dt\der_\nu
\,.
\eeq

 In the linear case, 
$T$
can be regarded as the section
$T : \f E \to \Lam^2T^*\f E \ten T\f E$
given, for each vector fields
$X$
and
$Y \,,$ 
by
$T(X,Y) = \nab_X Y - \nab_Y X - [X,Y]$
and its coordinate expression turns out to be given by the usual
formula
\beq
T = - (K\col \lam\nu\mu - K\col \mu\nu\lam) \, 
d^\lam \wed d^\mu \ten \der_\nu \,.
\eeq

 Thus, the connection
$K$
is linear and torsion free if and only if its coordinate expression
is given by
$K\co\lam\nu = 
K\col\lam\nu\mu \, \dt x^\mu \,,$
with
$K\col\lam\nu\mu = K\col\mu\nu\lam \in \map(\f E, \, \Rn) \,.$

 The \emph{curvature} of
$K$
is defined to be the vertical valued 2--form
\beq
R \eqv R[K] \byd - [K, K] : T\f E \to \Lam^2T^*\f E \ten VT\f E \,,
\eeq
where
$[\,,]$
is the Fr\"olicher-Nijenhuis bracket.
 We have the coordinate expression
\beq
R = 
R\cu\lam\mu\nu \, d^\lam \wed d^\mu \ten \dt\der_\nu =
- 2 \,
(\der_\lam K\co\mu\nu + K\co\lam\rho \, \dt\der_\rho K\co\mu\nu) \,
d^\lam \wed d^\mu \ten \dt\der_\nu \,.
\eeq

 In the linear case, 
$R$
can be regarded as the section
$R : \f E \to \Lam^2T^*\f E \ten T\f E \ten T^*\f E$
given, for each vector fields
$X, Y$
and
$Z \,,$ 
by
$R(X,Y,Z) = \nab_X \nab_Y Z - \nab_Y \nab_X Z - \nab_{[X,Y]} Z$
and its coordinate expression turns out to be given by the usual
formula
\beq
R =
R\cul\lam\mu\nu\sig \, d^\lam \wed d^\mu
        \ten \der_\nu \ten d^\sig =- 2 \,
(\der_\lam K\col\mu\nu\sig + K\col\lam\rho\sig \, K\col\mu\nu\rho) \,
d^\lam \wed d^\mu \ten \der_\nu \ten d^\sig \,.
\eeq

 We denote by 
$K[g]$ 
the \emph{Levi--Civita connection}, i.e. the torsion free
linear spacetime connection such that 
$\nab g=0 \,.$
 We have the usual coordinate expression
\beq
K\col \mu\lam\nu =
- \tfr12 g^{\lam\rho}\,(\der_\mu g_{\rho\nu}
+ \der_\nu g_{\rho\mu} - \der_\rho g_{\mu\nu}) \,.
\eeq
\subsection{Differential operators}
\label{Differential operators}
 Let us recall some differential operators associated with tangent
valued forms and apply them to a spacetime connection, its curvature
and the metric.

 A spacetime connection 
$K$ 
and its curvature
$R \,,$
regarded as tangent valued forms of
$T\f E \,,$ 
yield \cite{KolMicSlo93} \emph{Lie derivatives} of forms of
$T\f E \,.$ 
 Namely, if
$\phi \in \sec(T\f E, \Lam^rT^*T\f E) \,,$
then we define
$L_K \, \phi \in \sec(T\f E, \Lam^{r+1}T^*T\f E)$
and
$L_R \, \phi \in \sec(T\f E, \Lam^{r+2}T^*T\f E)$
as
\bEq\label{definitions of Lie derivatives}
L_K \, \phi \byd 
\big(i_K \, d - d \, i_K\big) \, \phi 
\sep{and}
L_R \,\phi \byd 
(i_R \, d + d \, i_R) \, \phi \,.
\eEq

 In the particular case when
$\phi$
is horizontal, i.e.
$\phi \in \fib(T\f E, \Lam^rT^*\f E) \,,$
we have the coordinate expressions
\bal
L_K \, \phi 
&= 
(\der_{\lam_1} \, \phi_{\lam_2 \dots \lam_{r+1}} +
K\co{\lam_1}\mu \, \dt\der_{\mu} \, \phi_{\lam_2 \dots \lam_{r+1}})
d^{\lam_1} \wed \dots \wed d^{\lam_{r+1}} \,,
\\
L_R \, \phi 
&=
R\cu{\lam_1}{\lam_2}\mu \, 
\dt\der_{\mu} \, \phi_{\lam_3 \dots \lam_{r+2}} \, 
d^{\lam_1} \wed \dots \wed d^{\lam_{r+2}} \,.
\end{align*}

 In the further particular case when 
$K$
(hence also
$R$)
and
$\phi$
are linear, the above expressions become
\bal
L_K \, \ome 
&=
(\der_{\lam_1} \, \phi_{\lam_2 \dots \lam_{r+1} \, \nu} +
K\col{\lam_1}\mu\nu \, \phi_{\lam_2 \dots \lam_{r+1} \; \mu})
\, \dt x^\nu \, d^{\lam_1} \wed \dots \wed d^{\lam_{r+1}} \,,
\\
L_R \, \phi 
&=
R\cul{\lam_1}{\lam_2}\mu\nu \, \phi_{\lam_3 \dots \lam_{r+2} \; \mu}
\, \dt x^\nu \, d^{\lam_1} \wed \dots \wed d^{\lam_{r+2}} \,.
\end{align*}

 On the other hand, a linear spacetime connection 
$K$
yields the \emph{covariant exterior differential} \cite{KolMicSlo93}
of cotangent valued forms of 
$\f E \,.$
 Namely, if
$\phi \in \sec(\f E, \Lam^r T^*\f E \ten T^*\f E) \,,$ 
then we define
$d_K\phi \in \sec(\f E, \Lam^{r+1} T^*\f E \ten T^*\f E) \,,$ 
through the equality
\bml
d_K\phi(X_1, \cdots, X_{r+1})(Y) = \sum_{i=1}^{r+1}(-1)^{i+1}
\nab_{X_i}(\phi (X_1, \cdots, \ha X_i, \cdots, X_{r+1}))(Y)
\\
+ \sum_{i<j}(-1)^{i+j}\phi
([X_i,X_j], X_1, \dots, \ha X_i, \dots, \ha X_j, \dots, X_{r+1})(Y),
\end{multline*}
for each vector fields 
$X_1, \cdots, X_{r+1}, Y$ 
of 
$\f E \,,$
the vector fields
$\ha X_i$ 
being omitted.

 We have the coordinate expression
\beq
d_K \phi = (r+1) \,
(\der_{\lam_1} \, \phi_{\lam_2 \dots \lam_{r+1} \; \nu} +
K\col{\lam_1}\mu\nu \, \phi_{\lam_2 \dots \lam_{r+1} \; \mu}) \,
d^{\lam_1} \wed \dots \wed d^{\lam_{r+1}} \ten d^\nu \,.
\eeq

 Thus, we can compare the Lie derivative and the covariant exterior
differential, by considering any cotangent valued $r$--form of
$\f E$
as a linear horizontal $r$--form of 
$T\f E \,.$
 Indeed, the following result holds.

\bLm\label{Lm1.2}\cite{Jan05}
 Let 
$K$ 
be a spacetime connection and
$\phi$ 
a linear horizontal $r$--form of 
$T\f E \,.$
 Then, the Lie derivative 
$L_K \, \phi$ 
is a linear horizontal
$(r+1)$--form of 
$T\f E$ 
if and only if 
$K$ 
is linear.
 Moreover, in such a case, we have
\beq
L_K \, \phi \cong \fr1{r+1} \, d_K \phi \,.
\eeq
\vglue-1.6\baselineskip{\ }\hfill\ENDE
\eLm

\bNt\label{Note: dKg LKg}
 We can apply 
$d_K$ 
and 
$L_K$ 
to the scaled metric 
$g$
as follows.

 We can regard 
$g$ 
as an
$(\B L^2 \ten T^*\f E)$--valued
1--form of 
$\f E \,.$
 Then, if
$K$
is linear, we obtain the covariant exterior differential 
$d_K g \,,$ 
which is an 
$(\B L^2 \ten T^*\f E)$--valued
2--form given by
\bAl\label{expression of d K g}
(d_K g)(X,Y)(Z)
&=
\big(\nab_{X}(Y\Fla) - \nab_{Y}(X\Fla) - ([X,Y]\Fla)\big)(Z)
\\ \nonumber
&=
\big(\nab_X g)(Y,Z) - (\nab_Y g)(X,Z) + g(T(X,Y) \,, Z\big) \,,
\end{align}
for each vector fields 
$X,Y,Z \,,$
where
$g\Fla : T\f E \to \B L^2 \ten T^*\f E$
denotes the musical map.
 We have the coordinate expression
\bal
d_K g
&=
2 \, (\nab_\lam g_{\mu\rho} - g_{\sig\rho}\, K\col\lam\sig\mu) \,
(d^\lam \wed d^\mu)\ten d^\rho
\\
&= 
2 \, (\der_\lam g_{\mu\rho} + g_{\sig\mu} \, K\col\lam\sig\rho) \,
 (d^\lam \wed d^\mu)\ten d^\rho\,.
\end{align*}

 On the other hand, we can regard the musical map
$g\Fla$
as a scaled linear horizontal 1--form
(the \emph{metric Liouville 1--form})
of 
$T\f E \,,$ 
with coordinate expression
$g\Fla = g_{\lam\mu} \, \dt x^\lam \, d^\mu \,.$
 Then, we obtain the Lie derivative
$L_Kg\Fla \,,$
which is a scaled horizontal 2--form of 
$T\f E \,,$
with coordinate expression
\beq
L_K \, g\Fla = 
(\der_\lam g_{\rho\mu} \, \dt x^\rho +
g_{\rho\mu}\, K\co\lam\rho)\, d^\lam\wed d^\mu \,,
\eeq
hence, if 
$K$ 
is linear,
\beq
L_K \, g\Fla = 
(\der_\lam g_{\rho\mu} + 
g_{\sig\mu} \, K\col\lam\sig\rho) \, \dt x^\rho \, d^\lam \wed d^\mu
\,.
\eeq

Thus, if
$K$
is linear, we have
\beq
L_K \, g\Fla \cong \fr12 d_Kg \,.
\eeq
\vglue-1.4\baselineskip{\ }\hfill\ENDE
\eNt
\subsection{Spacetime 2--forms and 2--vectors}
\label{Spacetime 2--forms and 2--vectors}
 Let us analyse the spacetime 2--form and 2--vector generated by the
metric and a spacetime connection.

 Let us consider a \emph{spacetime connection}
$K$ 
and the natural vertical valued 1--form
$\ups : T\f E \to T^*\f E \ten VT\f E \,,$
with coordinate expression
$\ups = d^\lam  \ten \dt\der_\lam \,.$

 We define the \emph{spacetime 2--form} and the \emph{spacetime
2--vector} of
$T\f E$
associated with
$g$
and
$K$
to be, respectively, the sections \cite{Jan95,Jan01,Jan05}
\bat{2}
\Ups 
&= 
\Ups[g, K] \byd
g \con (\nu[K] \wed \ups) 
&&:
T\f E \to \B L^2 \ten \Lam^2 T^*T\f E \,,
\\
\Xi 
&= 
\Xi[g, K] \byd
\ba g \con (K \wed \ups) 
&&: 
T\f E \to \B L^{-2} \ten \Lam^2 TT\f E \,,
\end{alignat*}
with coordinate expressions
\beq
\Ups =
g_{\lam\mu} \, (\dt d^\lam - K\co\nu\lam \, d^\nu) \wed d^\mu
\ssep{and}
\Xi =
g^{\lam\mu} \, 
(\der_\lam + K\co\lam\nu \, \dt\der_\nu) \wed \dt\der_\mu \,,
\eeq
and, if 
$K$ 
is linear,
\beq
\Ups =
g_{\lam\mu} \, 
(\dt d^\lam - K\col\nu\lam\rho \, \dt x^\rho \, d^\nu) \wed
d^\mu
\ssep{and}
\Xi =
g^{\lam\mu} \, 
(\der_\lam + K\col\lam\nu\rho \, \dt x^\rho \, \dt\der_\nu)
\wed \dt\der_\mu \,.
\eeq

 The scaled 2--form and the scaled 2--vector
\bat{3}
\eta 
&\byd \Ups \wed \Ups \wed \Ups \wed \Ups 
&&:
T\f E \to \B L^8 &&\ten \Lam^8T^*T\f E
\\
\ba\eta 
&\byd \Xi \wed \Xi \wed \Xi \wed \Xi 
&&:
T\f E \to \B L^{-8} &&\ten \Lam^8TT\f E
\end{alignat*}
turn out to be a scaled volume form and a scaled volume vector, with
coordinate expressions
\bat{2}
\eta
&=
4! \, |g| \, 
&&\dt d^0 \wed \dt d^1 \wed \dt d^2 \wed \dt d^3
\wed d^0 \wed d^1 \wed d^2 \wed d^3
\\
\ba\eta
&=
4! \, |\ba g| \, 
&&\dt\der_0 \wed \dt\der_1 \wed \dt\der_2 \wed \dt\der_3
\wed \der_0 \wed \der_1 \wed \der_2 \wed \der_3 \,.
\end{alignat*}

 We have \cite{Jan05}
$ i_\Xi \, \Ups = - 4$ 
and 
$\Xi\Sha = (\Ups\Fla)^{-1} \,;$ 
thus,
$\Ups$ 
and 
$\Xi$ 
are mutually dual.
 Here, we naturally identify 
$\Ups\Fla : TT\f E \to \B L^2 \ten T^*T\f E$ 
with
$\Ups\Fla : \B L^{-2} \ten TT\f E \to T^*T\f E \,.$

\smallskip

 In view of forthcoming considerations, let us consider a \emph{linear
spacetime connection}
$K \,,$
the scaled tangent space
$\ti T\f E \byd \B T^* \ten T\f E$
and the natural vertical valued 1--form
$\ti\ups : \ti T\f E \to
T^*\ti T\f E \ten V\ti T\f E \,,$
with coordinate expression
$\ti\ups = d^\lam_0  \ten \dt\der^0_\lam \,.$

 We observe that 
$K$
induces naturally a linear connection 
$\ti K$ 
on the tensor product bundle
$\B T^* \ten T\f E \,,$ 
via the tensor product of the natural flat linear connection of the
trivial bundle
$\f E \car \B T^* \to \f E$ 
and the linear connection
$K \,.$
 Indeed, the connection
$\ti K$
can be regarded as a tangent valued form 
$\ti K : \ti T\f E \to T^*\f E \ten T\ti T\f E \,,$
or as the complementary vertical valued form
$\nu[\ti K] : T\ti T\f E \to V\ti T\f E \,,$
with coordinate expressions
\beq
\ti K = 
d^\lam \ten 
(\der_\lam + K\col\lam\mu\nu \, \dt x^\nu_0 \, \dt\der^0_\mu )
\ssep{and}
\nu[\ti K] = (\dt d^\mu_0 - K\col\lam\mu\nu \, \dt x^\nu_0 \, d^\lam)
\ten \dt\der^0_\mu \,.
\eeq

 We can refrase the previous construction by replacing, respectively,
$T\f E \,,$
$K$
and
$\ups$
with
$\ti T\f E \,,$
$\ti K$
and
$\ti\ups \,.$

 We define the \emph{scaled spacetime 2--form}  and the \emph{scaled
spacetime 2--vector} of
$\ti T\f E \,,$
associated with
$g$
and
$\ti K \,,$
to be the scaled sections
\bat{2}
\wti\Ups 
&= 
\wti\Ups[g, \ti K] \byd
g \con \big(\nu[\ti K] \wed \ti\ups\big)
&&:
\ti T\f E \to
(\B T^* \ten \B L^2) \ten \Lam^2 T^*\ti T\f E \,,
\\
\wti\Xi 
&= 
\wti\Xi[g, \ti K] \byd
\ba g \con (\ti K \wed \ti\ups) 
&&: 
\ti T\f E \to 
(\B T \ten \B L^{-2}) \ten \Lam^2 T\ti T\f E \,,
\end{alignat*}
with coordinate expressions
\beq
\wti\Ups =
g_{\lam\mu} \, u^0 \ten (\dt d^\lam_0 -
K\col\nu\lam\rho \, \dt x^\rho_0 \, d^\nu) \wed d^\mu
\ssep{and}
\wti\Xi =
g^{\lam\mu} \, u_0 \ten (\der_\lam +
K\col\lam\nu\rho \, \dt x^\rho_0 \, \dt\der^0_\nu) 
	\wed \dt\der^0_\mu \,.
\eeq

 We have \cite{Jan05}
$ i_{\wti\Xi} \, \wti\Ups = - 4$ 
and 
$\wti\Xi\Sha = (\wti\Ups\Fla)^{-1} \,;$ 
thus,
$\wti\Ups$ 
and 
$\wti\Xi$ 
are mutually dual.
 Here, we naturally identify 
$\wti\Ups\Fla : T\ti T\f E \to \B T^*\ten \B L^2 \ten T^*\ti T\f E$ 
with
$\wti\Ups\Fla : 
\B T\ten \B L^{-2} \ten T\ti T\f E \to T^*\ti T\f E \,.$

\section{Geometric structures of the tangent bundle}
\label{Geometric structures of the tangent bundle}
\setcounter{equation}{0}
 Next, we recall a few results \cite{Jan05} on symplectic and
Poisson structures induced on the tangent bundle of spacetime by the
metric 
$g$
and a spacetime connection 
$K \,,$ 
and add some new results as well.
\subsection{General spacetime connection case}
\label{General spacetime connection case}
 Let us consider a \emph{spacetime connection} 
$K \,,$ 
its curvature
$R = R[K] \,,$
the spacetime 2--form
$\Ups = \Ups[g,K]$ 
and the spacetime 2--vector
$\Xi = \Xi[g,K] \,.$
 Let
$I = \dt x^\lam \, \dt\der_\lam$ 
be the Liouville vector field of
$T\f E \,.$

\bTh \label{Th2.1}\emph{\cite{Jan05}}
 The following conditions are equivalent:

\smallskip

{\rm (1)} 
$L_I \, L_K \, g\Fla = 0$ 
and  
$L_R \, g\Fla = 0 \,.$
\qquad\qquad
{\rm (2)} 
$d\Ups = 0 \,.$
\qquad\qquad
{\rm (3)} 
$[\Xi \,, \Xi] = 0 \,.$
\hfill\ENDE
\eTh

\bCr\label{Cr2.2}
 $\Ups$ 
is a scaled symplectic 2--form and
$\Xi$
a scaled Poisson 2--vector if and only if
$L_I \, L_K \, g\Fla = 0$ 
and  
$L_R \, g\Fla = 0 \,.$
\hfill\ENDE
\eCr 

\bCr
 The condition 
$L_K \, g\Fla = 0$ 
implies
$L_I \, L_K \, g\Fla = 0$ 
and
$L_R g\Fla = 0 \,,$ 
hence it implies that
$\Ups$
is a scaled symplectic 2--form and 
$\Xi$
a scaled Poisson 2--vector.
\hfill\ENDE
\eCr

\bLm
 The 2-form 
$\Ups + L_K \, g\Fla$ 
is exact and more precisely we have
\beq
\Ups + L_K \, g\Fla = dg\Fla \,.
\eeq

 Hence, we have
\beq
d\Ups = - dL_K \, g\Fla \,.
\eeq
\eLm

\begin{proof}
 We have
\bal
\Ups + L_K \, g\Fla
&=
	g_{\lam\mu} \, (\dt d^\lam - K\co\nu\lam \, d^\nu) \wed d^\mu
+ (\der_\lam g_{\rho\mu}\, \dt x^\rho + g_{\rho\mu}\, K\co\lam\rho) \,
		d^\lam \wed d^\mu
\\
&=
\dt \der_\lam(g_{\rho\mu}\, \dt x^\rho) \, \dt d^\lam \wed d^\mu
+ \der_\lam(g_{\rho\mu}\, \dt x^\rho)\, d^\lam \wed d^\mu 
\\
&=
	d(g_{\rho\mu} \, \dt x^\rho \, d^\mu) 
= 
d g\Fla \,. 
\end{align*}
\vglue-1.5\baselineskip
\end{proof}

\bTh
 The following conditions are equivalent:

\smallskip

{\rm (1)} 
$L_K \, g\Fla = 0 \,;$
\qquad\qquad
{\rm (2)} 
$\Ups = dg\Fla \,.$
\hfill\ENDE
\eTh
\subsection{Linear spacetime connection case}
\label{Linear spacetime connection case}
 Let us consider a \emph{linear spacetime connection}
$K$
and the induced scaled spacetime 2--form 
$\Ups = \Ups[g,K]$
and scaled spacetime 2--vector
$\Xi = \Xi[g,K] \,.$

\bTh\label{Th2.7}\emph{\cite{Jan05}}
 The following equalities are equivalent:

\smallskip

{\rm (1)} 
$L_K \, g\Fla = 0 \,;$
\qquad\quad
{\rm (2)} 
$d_K g = 0 \,;$
\qquad\quad
{\rm (3)} 
$\Ups = dg\Fla \,;$
\qquad\quad
{\rm (4)} 
$[\Xi \,, \Xi] = 0 \,.$
\hfill\ENDE
\eTh

\bCr \label{Cr2.4}
 $\Ups$ 
is a scaled symplectic 2--form and
$\Xi$
a scaled Poisson 2--vector if and only if
$d_K g = 0 = L_K \, g\Fla \,.$
\hfill\ENDE
\eCr

\bNt
 If
$K$ 
is torsion free, then
$\nab g$ 
is symmetric if and only if
$d_K g = 0 = L_K \, g\Fla \,.$

 In fact, in virtue of Note \ref{Note: dKg LKg}, we have
$d_K g = 0 = L_K g\Fla$
if and only if
$(\nab_X g)(Y,Z) = (\nab_Y g)(X,Z) \,.$
\hfill\ENDE
\eNt

\bTh \label{Th2.10}\emph{\cite{Jan05}}
 If
$K$ 
is torsion free, then the following conditions are equivalent:

\smallskip

{\rm (1)}
$\nab g$
is symmetric;
\qquad\qquad
{\rm (2)} 
$d\Ups = 0 \,$;
\qquad\qquad
{\rm (3)} 
$[\Xi \,, \Xi] = 0 \,.$
\hfill\ENDE
\eTh

\bCr\label{Cr2.7}
 Let
$K$
be torsion free.
 Then,
$\Ups$ 
is a scaled symplectic 2--form and
$\Xi$
a scaled Poisson 2--vector if and only if
$\nab g$ 
is symmetric.
\hfill\ENDE
\eCr

\bCr
 If
$K = K[g]$
is the Levi--Civita connection, then we obtain the metric scaled
symplectic 2--form and scaled Poisson 2--vector
\beq
\Ups = \Ups[g] \byd \Ups[g,K[g]]
\ssep{and}
\Xi = \Xi[g] \byd \Xi[g,K[g]] \,,
\eeq
respectively.
 In this case, we have
$\Ups = dg\Fla \,.$
\hfill\ENDE
\eCr
\subsection{Non-metric spacetime connection case}
\label{Non-metric spacetime connection case}
 We have seen that the metric 
$g$ 
yields naturally the metric scaled symplectic 2--form
$\Ups = \Ups[g]$ 
and the metric scaled Poisson 2--vector
$\Xi = \Xi[g] \,.$
 Now, we discuss the case of ``non-metric" connections 
$K$
yielding a scaled spacetime 2--form
$\Ups \byd \Ups[g,K] \,,$ 
which fulfills the equality 
$\Ups = dg\Fla \,.$

\bNt The equality
\beq
K = K[g] + \Phi
\eeq
yields a bijection between spacetime connections
$K$
and sections
$\Phi: T\f E\to T^*\f E\ten VT\f E \,,$
with coordinate expression
$\Phi = \Phi\co\lam\mu \, d^\lam \ten \dt\der_\mu \,,$
where
$	\Phi\co\lam\mu \in \map(T\f E,\Rn) \,.$

 On the other hand, by identifying
$VT\f E$ 
with 
$T\f E \car_\f E T\f E \,,$
the equalities
\beq
\bau\Phi \byd \proj_2 \com \Phi \,,
\quad
\Phi = \ups (\bau\Phi)
\ssep{and}
\phi \byd g\Fla (\bau\Phi) \,,
\quad
\bau\Phi = g\Sha{}^2 (\phi)
\eeq
yield bijections between the sections
$\Phi: T\f E\to T^*\f E\ten VT\f E$
and the fibred morphisms
$\bau\Phi : T\f E \to T^*\f E \ten T\f E$
and
$\phi : T\f E \to \B L^2 \ten (T^*\f E \ten T^*\f E) \,.$
\hfill\ENDE
\eNt

 Let us consider a \emph{spacetime connection}
$K = K[g] + \Phi = K[g] + \ups(\bau\Phi)$
and the induced scaled spacetime 2--form 
$\Ups = \Ups[g,K] \,.$

\bTh
 The following conditions are equivalent:

\smallskip

{\rm (1)}
$\Ups = dg\Fla \,;$
\qquad\qquad
{\rm (2)}
$L_\Phi \, g\Fla = 0 \,;$
\qquad\qquad
{\rm (3)}
$\phi$
is symmetric.
\eTh

\begin{proof}
 In virtue of Theorem \ref{Th2.7}, 
$\Ups = dg\Fla$
if and only if 
$L_K \, g\Fla = 0 \,,$ 
i.e. if and only if 
$L_\Phi \, g\Fla = 0 \,,$
i.e., in coordinates, if and only if
$g_{\lam\rho} \, \Phi\co\mu\rho \, d^\lam \wed d^\mu = 0 \,.$
 Hence,
$L_\Phi \, g\Fla = 0$
if and only if
$\phi$
is symmetric. 
\end{proof}

\bEx
 The connection 
$K = K[g] + \ups \,$ 
is a ``non--metric" non--linear spacetime connection yielding the
scaled symplectic 2--form
$\Ups = dg\Fla \,.$
\hfill\ENDE
\eEx

\bNt
 It is easy to see that 
$K$ 
is linear if and only if 
$\Phi$ 
is linear, i.e. if and only if
$\phi$ 
can be identified with a section
$\phi : \f E \to \B L^2 \ten T^*\f E \ten T^*\f E \ten T^*\f E$
which is symmetric in the 1st two factors. 

 Moreover, if
$\phi$ 
is symmetric in all factors, then
$K$ 
turns out to be torsion free.
\hfill\ENDE
\eNt

\bNt
 We recall \cite{Wey21} that a linear torsion free connection
$K$ 
is projectively equivalent to the metric connection 
$K[g]$
if and only if 
$K$ 
and 
$K[g]$ 
have the same unparametrized  geodesics, i.e., in coordinates, if
and only if 
\beq
K\col\lam\nu\mu = K[g]\col\lam\nu\mu + \del^\nu_\lam\,\psi_\mu
	+ \del^\nu_\mu \, \psi_\lam \,,
\ssep{where}
\psi_\lam \in \map(\f E,\Rn) \,.
\eeq

 It is easy to see that any torsion free linear spacetime connection
$K$ 
projectively equivalent to 
$K[g]$ 
is such that 
$\Ups = dg\Fla$ 
and it is associated with a symmetric 
(0,3)--tensor field of the type
$\phi = g \odot \psi \,,$ 
for a certain spacetime 1-form 
$\psi \,.$ 

 But not all torsion free linear spacetime connections 
$K$ 
such that 
$\Ups = dg\Fla$ 
are projectively equivalent to 
$K[g] \,.$
 Hence, the condition 
$d_K g = 0$ 
is more general than the condition for connections projectively
equivalent to 
$K[g] \,.$
\hfill\ENDE
\eNt

\bNt
 Let 
$K$ 
and 
$K'$ 
be two spacetime connections and consider the difference tensor
$\Phi \byd K - K' : T\f E \to T^*\f E \ten VT\f E \,.$
 Then, the following conditions are equivalent:

1) $\Ups[g,K] = \Ups[g,K'] \,,$
 
2) $\Xi[g,K] = \Xi[g,K'] \,,$
 
3) $\phi \byd g\Fla(\bau \Phi)$
is symmetric.

 Thus, the relation 
$\Ups[g,K] = \Ups[g,K'] \,,$
or equivalently, the relation
$\Xi[g,K] = \Xi[g,K'] \,,$
defines an equivalence relation on the space of spacetime
connections.
\hfill\ENDE
\eNt

 We are mainly interested in the spacetime connections which are
equivalent to
$K[g] \,,$
because they yield an exact spacetime 2--form
$\Ups \,.$

\section{Geometry of the Einstein phase space}
\label{Geometry of the Einstein phase space}
\setcounter{equation}{0}
 Next, we study the geometric properties of the phase space of a
classical particle in the Lorentzian framework, by adding several new
results with respect to
\cite{JanMod95}.
\subsection{Jets of submanifolds}
\label{Jets of submanifolds}
 In view of the definition of the phase space, let us consider a
manifold
$\f M$
of dimension
$n$
and recall a few basic facts concerning jets of submanifolds.

 Let
$k \geq 0$
be an integer.
 A {\em $k$--jet\/} of 1--dimensional submanifolds of
$\f M$
at
$x \in \f M$
is defined to be an equivalence class of 1--dimensional submanifolds
touching each other at
$x$
with a contact of order
$k \,.$
 The $k$--jet of a 1--dimensional submanifold
$s : \f N \sub \f M$
at
$x \in \f N$
is denoted by
$j_ks(x) \,.$
 The set of all $k$--jets of all 1--dimensional submanifolds at
$x \in \f M$
is denoted by
$J_{k \, x}(\f M,1) \,.$
 The set
$J_k(\f M,1) \byd \bigsqcup_{x \in \f M} J_{k \, x}(\f M,1)$
is said to be the {\em $k$--jet space\/} of 1--dimensional
submanifolds of
$\f M \,.$
 In particular, for
$k = 0 \,,$
we have the natural identification
$J_0(\f M,1) = \f M \,,$
given by
$j_0s(x) = x \,,$
for each 1--dimensional submanifold
$s : \f N \sub \f M \,.$
 For each integers
$k \geq h \geq 0 \,,$
we have the natural projection
$\pi^k_h : J_k(\f M,1) \to J_h(\f M,1) :
j_ks(x) \mto j_hs(x) \,.$

 For each 1--dimensional submanifold
$s : \f N \sub \f M$
and each integer
$k\ge 0 \,,$
we have the map
$j_ks : \f N \to J_k(\f M,1) : x \mto j_ks(x) \,.$

 A chart of
$\f M$
is said to be {\em divided\/} if the set of its coordinate functions
is divided into two subsets of 1 and
$n-1$
elements.
 Our typical notation for a divided chart will be
$(x^0,x^i) \,,$
with
$1 \le i \le n-1 \,.$
A divided chart and a 1--dimensional submanifold
$s : \f N \sub \f M$
are said to be {\em related\/} if the map
$\br x^0 \byd x^0|_\f N \in \map(\f N, \, \Rn)$
is a chart of
$\f N \,.$
 In such a case, the submanifold
$\f N$
is locally characterised by
$s^i \com (\br x^0)^{-1} \byd
(x^i \com s) \com (\br x^0)^{-1} \in \map(\Rn, \Rn) \,.$
 In particular, if the divided chart is adapted to the submanifold,
then the chart and the submanifold are related.

 Let us consider a divided chart
$(x^0, x^i)$
of
$\f M \,.$

 Then, for each submanifold
$s : \f N \sub \f M$
which is related to this chart, the chart yields naturally the local
fibred chart
$(x^0, x^i; \, x^i_\ul \alp)_{1 \leq |\ul\alp| \leq k} \in
\map(J_k(\f M,1), \; \Rn^n \car \Rn^{k(n-1)})$
of
$J_k(\f M,1) \,,$
where
$\ul\alp \byd (h)$
is a multi--index, of ``range" 1 and ``length"
$|\ul\alp| = h \,,$
and the functions
$x^i_\ul\alp$
are defined by
$x^i_\ul\alp \com j_1\f N \byd \der_{0 \dots 0} \, s^i \,,$
with
$1 \leq |\ul\alp| \leq k \,.$

 We can prove the following facts:

1) the above charts
$(x^0, x^i; \, x^i_{\ul\alp})$
yield a smooth structure of
$J_k(\f M, 1) \,;$

2) for each 1--dimensional submanifold
$s : \f N \sub \f M$
and for each integer
$k \geq 0 \,,$
the subset
$j_ks(\f N) \sub J_k(\f M, 1)$
turns out to be a smooth 1--dimensional submanifold;

3) for each integers
$k \geq h \geq 1 \,,$
the maps
$\pi^k_h : J_k(\f M,1) \to J_h(\f M,1)$
turn out to be smooth bundles.

 We shall always refer to such diveded charts
$(x^0, x^i)$
of
$\f M$
and to the induced fibred charts
$(x^0, x^i; \, x^i_\ul\alp)$
of
$J_k(\f M,1) \,.$

 Let
$m_1 \in J_1(\f M,1) \,,$
with
$m_0 = \pi^1_0 (m_1) \in \f M \,.$
Then, the tangent spaces at
$m_0$
of all 1--dimensional submanifolds
$\f N \,,$
such that
$j_1s(m_0) = m_1 \,,$
coincide.
Accordingly, we denote by
$T[m_1] \sub T_{m_0} \f M$
the tangent space at
$m_0$
of the above 1--dimensional submanifolds
$\f N$
generating
$m_1 \,.$
 We have the natural fibred isomorphism
$J_1(\f M,1) \to \Grass(\f M,1) :
m_1 \mto T[m_1]  \sub T_{m_0} \f M$
over
$\f M$
of the 1st jet bundle with the Grassmannian bundle of dimension 1.
 If
$s : \f N \sub \f M$
is a submanifold, then we obtain
$T[j_1s] = \Span\lang \der_0 + \der_0s^i \der_i \rang \,,$
with reference to a related chart.
\subsection{Phase space}
\label{Phase space}
 Let us introduce the phase space of a classical particle and its
basic contact structure induced by the Lorentz metric.

 A \emph{motion} is defined to be a 1--dimensional timelike
submanifold
$s : \f T \sub \f E \,.$
Let us consider a motion
$s : \f T \sub \f E \,.$

 For every arbitrary choice of a ``\emph{proper time origin}"
$t_0 \in \f T \,,$
we obtain the ``\emph{proper time scaled function}" given by the
equality
$\sig : \f T \to \baB T :
t \mto \fr1c \int_{[t_0, t]} \|\fr{ds}{d\br x^0}\| \, d\br x^0 \,.$

 This map yields, at least locally, a bijection
$\f T \to \baB T \,,$
hence a (local) affine structure of
$\f T$
associated with the vector space
$\baB T \,.$
 Indeed, this (local) affine structure does not depend on the choice
of the proper time origin and of the spacetime chart.

 Let us choose a time origin
$t_0 \in \f T$
and consider the associated proper time scaled function
$\sig : \f T \to \baB T$
and the induced linear isomorphism
$T\f T \to \f T \car \baB T \,.$
 Moreover, let us consider a spacetime chart
$(x^\lam)$
and the induced chart
$(\br x^0) \in \map(\f T, \Rn) \,.$
 Let us set
$\der_0 s^\lam \byd \fr{d s^\lam}{d\br x^0} \,.$

 The \emph{1st differential} of the motion
$s$
is the map
$ds \byd \fr{ds}{d\sig} : \f T \to \B T^* \ten T\f E \,.$

 We have
$g(ds, \, ds) = - c^2$
and the coordinate expression
\beq
ds =
\fr{d s^\lam}{d\sig} \, (\der_\lam \comm s) =
\fr{c_0 \,
u^0 \ten \big((\der_0 \comm s) + \der_0 s^i \, (\der_i \comm s)\big)}
{\sqrt{ |(g_{00} \comm s) +
2 \, (g_{0j} \comm s) \, \der_0 s^j +
(g_{ij} \comm s) \, \der_0 s^i \, \der_0 s^j|}} \,.
\eeq

 We define the \emph{phase space} of a classical particle to be the
subspace 
$\M J_1\f E \sub J_1(\f E,1)$
consisting of all 1--jets of motions.

 For each 1--dimensional submanifold
$s : \f T \sub \f E$
and for each
$x \in \f T \,,$
we have
$j_1s(x) \in \M J_1\f E$
if and only if
$T[j_1s(x)] = T_x\f T$
is timelike.

 Any spacetime chart
$(x^0, x^i)$
is related to each motion
$s : \f T \to \f E \,.$
 Hence, the fibred chart
$(x^0, x^i, x^i_0)$
is defined on tubelike open subsets of
$\M J_1\f E \,.$

 We shall always refer to the above fibred charts.

 The \emph{velocity} of a motion
$s : \f T \sub \f E$
is defined to be its 1-jet
$j_1s : \f T \to \M J_1(\f E,1) \,.$

 We define the \emph{contact map} to be the unique fibred morphism
$\K d : \M J_1\f E \to \B T^* \ten T\f E$
over
$\f E \,,$
such that
$\K d \com j_1s = ds \,,$
for each motion
$s : \f T \to T\f E \,.$
 We have the coordinate expression
$\K d =
c \, \alp^0 \,  (\der_0 + x^i_0 \, \der_i) \,,$
where we have set
$\alp^0 \byd
1 /\sqrt{|g_{00} + 2 \, g_{0j} \, x^j_0 + g_{ij} \, x^i_0 \, x^j_0|}
\,.$

 We have
$g \, (\K d, \K d) = - c^2 \,.$

 The fibred morphism
$\K d : \M J_1\f E \to \B T^*\ten T\f E$
is injective.
 Indeed, it makes
$\M J_1\f E \sub \B T^*\ten T\f E$
the fibred submanifold over
$\f E$
characterised by the constraint
$g_{\lam\mu} \, \dt x^\lam_0 \, \dt x^\mu_0 = - (c_0)^2 \,.$

\smallskip

 We define the \emph{time form} to be the 1--jet based scaled
1--form
$\tau
\byd - \fr1{c^2} g\Fla(\K d) :
\M J_1\f E \to \B T\ten T^*\f E \,,$
with coordinate expression
$\tau = \tau_\lam \, d^\lam \,,$
where
$\tau_\lam = - \fr{\alp^0}{c}
\, (g_{0\lam}+ g_{i\lam}\, x^i_0) \,.$

 We have
$\tau (\K d) = 1$
and 
$\ba g(\tau,\tau) = - \fr1{c^2} \,.$

\smallskip

 We define the \emph{complementary contact map} to be the linear
fibred morphism
$\tht \byd 1 - \K d \ten \tau :
\M J_1\f E \ucar{\f E} T\f E \to T\f E \,,$
over
$\M J_1\f E \,,$
given by
$\tht (v) = v - \tau(v) \, \K d \,.$
 We have the coordinate expressions
$\tht =
d^\lam \ten \der_\lam +
(\alp^0)^2 \, (g_{0\lam} + g_{i\lam} \, x^i_0) \,
d^\lam \ten (\der_0 + x^j_0 \, \der_j) \,.$
\subsection{Orthogonal splittings}
\label{Orthogonal splittings}
 We have a natural orthogonal splitting of the tangent and
cotangent spaces of spacetime pullbacked over the phase space.

 We define

- the \emph{$\K d$--horizontal} tangent space of spacetime,

- the \emph{$\tau$--vertical} tangent space of spacetime,

- the \emph{$\tau$--horizontal} cotangent space of spacetime,

- the \emph{$\K d$--vertical} cotangent space of spacetime

\noindent
to be, respectively, the vector subbundles over
$\M J_1\f E$
\bat{2}
H_\K d\f E
&\byd
\{(e_1, X) \in \M J_1\f E \ucar{\f E} T\f E \sst X \in T[e_1]\}
&&\sub 
\M J_1\f E \ucar{\f E} T\f E \,,
\\
V_\tau\f E
&\byd
\{(e_1, X) \in \M J_1\f E \ucar{\f E} T\f E \sst X \in T[e_1]\Per\}
&&\sub 
\M J_1\f E \ucar{\f E} T\f E \,,
\\
H^*_\tau\f E
&\byd
\{(e_1, \ome) \in \M J_1\f E \ucar{\f E} T^*\f E  \sst
\lang\ome, T[e_1]\Per\rang = 0\}
&&\sub 
\M J_1\f E \ucar{\f E} T^*\f E \,,
\\
V^*_\K d\f E
&\byd
\{(e_1, \ome) \in \M J_1\f E \ucar{\f E} T^*\f E  \sst
\lang\ome, T[e_1]\rang = 0\}
&&
\sub \M J_1\f E \ucar{\f E} T^*\f E \,,
\end{alignat*}
where
$T[e_1]$
is the tangent space associated with the 1st jet
$e_1$
and
$T[e_1]\Per$
is its orthogonal.

 Indeed, 
$H_\K d\f E$
and
$H^*_\tau\f E$
are generated by
$\K d$
and
$\tau \,,$
respectively; moreover,
$V_\tau\f E$ 
and
$V^*_\K d\f E$
are generated by
$\tht$
and
$\tht^* \,,$
respectively.

\bPr
 We have the natural orthogonal linear fibred splittings over
$\M J_1\f E$
\beq
\M J_1\f E \ucar{\f E} T\f E =
H_\K d\f E \udrs{\M J_1\f E} V_\tau\f E
\ssep{and}
\M J_1\f E \ucar{\f E} T^*\f E =
H^*_\tau\f E \udrs{\M J_1\f E} V^*_\K d\f E 
\eeq
and the corresponding projections
\bat{5}
&\pi\Prl = \tau \ten \K d 
&&:
\M J_1\f E \ucar{\f E} T\f E \to H_\K d\f E
&&\ssep{and}
&&\pi\prl = \K d \ten \tau 
&&:
\M J_1\f E \ucar{\f E} T^*\f E \to H^*_\tau\f E \,,
\\
&\pi\Per = \tht 
&&:
\M J_1\f E \ucar{\f E} T\f E \to V_\tau\f E
&&\ssep{and}
&&\pi\per = \tht^* 
&&:
\M J_1\f E \ucar{\f E} T^*\f E \to V^*_\K d\f E \,.
\end{alignat*}
\vglue-1.6\baselineskip{\ }\hfill\ENDE
\ePr

 We have a natural identification
$(H_\K d\f E)^* = H^*_\tau\f E$
and
$(V_\tau\f E)^* = V^*_\K d\f E \,.$

\bLm
 The restrictions of
$g$
and
$\ba g$ 
to the components of the above splitting are
\bat{3}
&g\prl \byd g \com (\pi\Prl, \pi\Prl) 
= - c^2 \, \tau \ten \tau
&&\ssep{and}
&&\ba g\Prl \byd \ba g \com (\pi\prl, \pi\prl)
= - \fr1{c^2} \, \K d \ten \K d \,,
\\
&g\per \byd g \com (\pi\Per, \pi\Per)
= g + c^2 \, \tau \ten \tau
&&\ssep{and}
&&\ba g\Per \byd \ba g \com (\pi\per, \pi\per)
= \ba g + \fr1{c^2} \, \K d \ten \K d \,.
\end{alignat*}
\vglue-1.6\baselineskip{\ }\hfill\ENDE
\eLm

 We have the mutually dual local bases
$(b_0, \, b_i)$
and
$(\bet^0, \, \bet^i)$
adapted to the above splittings, where
\bat{2}
b_0
& \byd \der_0 + x^i_0 \, \der_i
&&\in
\fib(\M J_1\f E, \, H_\K d\f E) \,,
\\
b_i
&\byd 
\der_i - c \, \alp^0 \, \tau_i \, (\der_0 + x^j_0 \, \der_j)
&&\in
\fib(\M J_1\f E, \, V_\tau\f E) \,,
\\[1mm]
\bet^0
&\byd 
d^0 + c \, \alp^0 \, \tau_i \, (d^i - x^i_0 \, d^0 )
&&\in 
\fib(\M J_1\f E, \, H^*_\tau\f E) \,,
\\ \nonumber
\bet^i
&\byd
d^i - x^i_0 \, d^0
&&\in 
\fib(\M J_1\f E, \, V^*_\K d\f E) \,.
\end{alignat*}

 If we put
\beq
\br\del^\lam_0 = \del^\lam_0 + \del^\lam_i \, x^i_0 \,,
\qquad
\br\del^i_\mu = \del^i_\mu - \del^0_\mu \, x^i_0 \,,
\eeq
then we can write shortly
\bal
b_0 
&= 
\br\del^\lam_0 \, \der_\lam \,,
\quad
b_i = 
(\del^\lam_i - c \, \alp^0 \, \tau_i \, \br\del^\lam_0) \, \der_\lam
\,,
\quad
\bet^0 = 
(\del^0_\mu + c \, \alp^0 \, \tau_i \, \br\del^i_\mu) \, d^\mu \,,
\quad
\bet^i = \br\del^i_\mu \, d^\mu \,.
\end{align*}

\smallskip

 We have the inverse relations
\beq
\der_\lam =
c \, \alp^0 \, \tau_\lam \, b_0 + \br\del^i_\lam \, b_i \,,
\qquad
d^\mu =
\br\del^\mu_0 \, (\bet^0 - c \, \alp^0 \, \tau_j \, \bet^j) +
\del^\mu_j \, \bet^j \,.
\eeq

\bLm
 We have the equalities
\bat{3}
\br g_{0\lam}
&\byd
g \, (b_0, \der_\lam) = 
g_{\rho\lam} \, \br\del^\rho_0\,,
&&
\br g^{0\lam}
\byd
\ba g \, (\bet^0, d^\lam)
= - (\alp^0)^2 \, \br\del^\lam_0 \,,
\\[1mm] 
\br g_{i\lam}
&\byd 
g(b_i, \, \der_\lam) = 
g_{i\lam} + (\alp^0)^2 \, \br g_{0i} \, \br g_{0\lam} \,,
&&
\br g^{i\lam}
\byd 
\ba g(\bet^i, \, d^\lam) = 
\br\del^i_\rho\, g^{\rho\lam}\,,
\\[1mm]
\ha g_{00}
&\byd
g \, (b_0, b_0) = - 1/(\alp^0)^2 =
g_{\rho\sig}\,\br\del^\rho_0\,\br\del^\sig_0\,,\qquad
&&
\ha g^{00}
\byd
\ba g \, (\bet^0, \bet^0) = - (\alp^0)^2 \,,
\\[1mm]
\ha g_{ij}
&\byd g(b_i, \, b_j) = 
g_{ij} + (\alp^0)^2 \, \br g_{0i} \, \br g_{0j} \,,
&&
\ha g^{ij}
\byd
\ba g(\bet^i, \, \bet^j) =
\br\del^j_\sig \, \br g^{i\sig}
= \br\del^i_\rho \, \br\del^j_\sig\, g^{\rho\sig}
\,,
\\
\ha g_{0j}
&\byd 
g(b_0, \, b_j) = 0\,,
&&
\ha g^{0j} 
\byd
\ba g(\bet^0,\bet^j) = 0 \,.
\end{alignat*}
\vglue-1.6\baselineskip{\ }\hfill\ENDE
\eLm

\bLm
 We have the coordinate expressions
\bat{3}
\pi\Prl
&= 
- (\alp^0)^2 \, \br g_{0\lam} \, \br\del^\mu_0 \,
d^\lam \ten \der_\mu \,,
\qquad
&&\pi\prl
&&= - (\alp^0)^2 \, \br g_{0\lam} \, \br\del^\mu_0 \,
\der_\mu \ten d^\lam \,,
\\
\pi\Per
&=
\br g^{i\mu} \, \br g_{i\lam} \, d^\lam \ten \der_\mu \,,
\qquad
&&\pi\per
&&=
\br g_{i\lam} \, \br g^{i\mu} \, \der_\mu \ten d^\lam \,,
\\
g\prl
&= 
- (\alp^0)^2 \, \br g_{0\lam} \, \br g_{0\mu} \, d^\lam \ten d^\mu
\,,
\qquad
&&\ba g\Prl
&&= - (\alp^0)^2 \, \br\del^\lam_0 \, \br\del^\mu_0 \, 
\der_\lam \ten \der_\mu \,,
\\
g\per
&= 
(g_{\lam\mu} + (\alp^0)^2 \, \br g_{0\lam} \, \br g_{0\mu}) \,
 d^\lam \ten d^\mu
\,,
\qquad
&&\ba g\Per
&&=
(g^{\lam\mu} + (\alp^0)^2 \, \br\del^\lam_0 \, \br\del^\mu_0) \,
\der_\lam \ten \der_\mu \,.
\end{alignat*}
\vglue-1.6\baselineskip{\ }\hfill\ENDE
\eLm

 Later, we shall be frequently involved with the following useful
technical identities.

\bLm\label{Lemma: useful identities}
 We have the following identities
\bgt
\br g_{0\lam} \, d^\lam = \ha g_{00} \, \bet^0 \,,
\qquad
\br g_{i\lam} \, d^\lam = \ha g_{ij} \, \bet^j \,,
\qquad
\br g^{0\lam} \, \der_\lam = \ha g^{00} \, b_0 \,,
\qquad
\br g^{i\lam} \, \der_\lam = \ha g^{ij} \, b_j \,,
\\
\br g_{i\lam} \, \br g^{i\mu}  =
\del^\mu_\lam + (\alp^0)^2 \, \br g_{0\lam} \, \br\del^\mu_0 \,,
\qquad
\br g_{0i} \, \br g^{i\lam} =  \br\del^\lam_0 -
\ha g_{00} \, g^{0\lam} \,,
\\
\br g_{i\mu} \, g^{0\mu} = (\alp^0)^2 \, \br g_{0i} \,,
\qquad
\br g_{0\nu} \, \br g^{i\nu} = 0 \,,
\qquad
\br g_{i\nu} \, \br g^{0\nu} = 0 \,,
\\
\br g_{i\nu} \, \br\del^\nu_0 = 0 \,,
\qquad
\br g_{i\mu} \, \br\del^i_\lam = 
g_{\lam\mu} + (\alp^0)^2 \, \br g_{0\lam} \, \br g_{0\mu} \,,
\qquad
\br g^{i\lam} \, \br\del^j_\lam = \ha g^{ij} \,,
\\
\ha g^{\lam\nu} \, \ha g_{\mu\nu} =
\ha g^{\nu\lam} \, \ha g_{\nu\mu} = \del^\lam_\mu \,,
\qquad
\ha g^{ih} \, \ha g_{jh} =
\ha g^{hi} \, \ha g_{hj} = \del^i_j \,,
\\ \nonumber
\ha g^{ij} \, g_{j\sig} = \br\del^i_\sig 	- \br g^{i0} \, \br g_{0\sig}
\,,
\qquad
\ha g^{ij}\, \del^\rho_{j} = 
\br g^{i\rho} 	- \br g^{i0}\, \br\del^\rho_0 \,,
\qquad
\br g_{0\nu} \, \br \del^\nu_0 = \ha g_{00} \,,
\qquad
\ha g^{ij}\, \br g_{0j} = - \ha g_{00} \, \br g^{i0} \,,
\\[3mm]
\der^0_i \alp^0 = (\alp^0)^3 \, \br g_{0i} \,,
\qquad
\der_\lam \alp^0 = \tfr12 (\alp^0)^3 \, \der_\lam \ha g_{00} \,.
\end{gather*}
\vglue-1.6\baselineskip{\ }\hfill\ENDE
\eLm
\myssec{Splitting of the spacetime 2--form}
\label{Splitting of the spacetime 2--form}
 Let us consider a linear spacetime connection 
$K$ 
and the induced scaled spacetime 2--forms
$\Ups = \Ups[g,K]$ 
and
$\wti\Ups = \wti\Ups[g,\ti K]$
(see Section \ref{Spacetime 2--forms and 2--vectors}).
 Then, we can split
$\Ups$
and 
$\wti\Ups \,,$ 
over the phase space, into the parallel and orthogonal components
(according to the splitting of
$T\f E$
and
$T^*\f E$
achieved in the above Section \ref{Orthogonal splittings}) as follows.

\bPr
 The pullback of
$\Ups \,,$
with respect to
$\M J_1\f E \ucar{\f E} T\f E \to T\f E \,,$
splits as
\beq
\Ups \eqv \Ups\prl + \Ups\per \,,
\eeq
where the parallel and orthogonal components are defined by
\bal
\Ups\prl
&\byd
g \con \big(\nu[K] \wed \pi\prl\big)
: \M J_1\f E \ucar{\f E} T\f E \to
\B L^2 \ten \Lam^2 T^*T\f E \,,
\\
\Ups\per
&\byd
g \con \big(\nu[K] \wed \pi\per\big)
: \M J_1\f E \ucar{\f E} T\f E \to
\B L^2 \ten \Lam^2 T^*T\f E \,,
\end{align*}
and have the coordinate expressions
\bal
\Ups\prl
&= 
- c^2 \, \tau_\lam \, \tau_\mu \,
(\dt d^\lam - K\col\nu\lam\rho \, \dt x^\rho \, d^\nu)
\wed d^\mu
\\
&= 
- (\alp^0)^2 \, \br g_{0\lam} \, \br g_{0\mu} \,
(\dt d^\lam - K\col\nu\lam\rho \, \dt x^\rho \, d^\nu)
\wed d^\mu \,,
\\[3mm]
\Ups\per
&= 
(g_{\lam\mu} + c^2 \, \tau_\lam \, \tau_\mu) \,
(\dt d^\lam - K\col\nu\lam\rho \, \dt x^\rho \, d^\nu)
\wed d^\mu
\\
&= 
\big(g_{\lam\mu} +
(\alp^0)^2 \, \br g_{0\lam} \, \br g_{0\mu}\big) \,
(\dt d^\lam - K\col\nu\lam\rho \, \dt x^\rho \, d^\nu)
\wed d^\mu \,.
\end{align*}
\ePr

\begin{proof}
 It follows from the coordinate expressions of
$\pi\prl$ 
and 
$\pi\per \,.$ 
\end{proof}

\bPr
 The pullback of
$\wti\Ups \,,$
with respect to
$\M J_1\f E \ucar{\f E} \ti T\f E \to \ti T\f E \,,$
splits as
\beq
\wti\Ups \eqv \wti\Ups\prl + \wti\Ups\per \,,
\eeq
where the parallel and orthogonal components are defined by
\bal
\wti\Ups\prl
&\byd
g \con \big(\nu[\ti K] \wed \ti\pi\prl\big)
: \M J_1\f E \ucar{\f E} (\B T^*\ten T\f E) \to
(\B T^* \ten \B L^2) \ten \Lam^2 T^*\ti T\f E \,,
\\
\wti\Ups\per
&\byd
g \con\big(\nu[\ti K] \wed \ti\pi\per\big)
: \M J_1\f E \ucar{\f E} \ti T\f E \to
(\B T^* \ten \B L^2) \ten \Lam^2 T^* \ti T\f E \,,
\end{align*}
with 
$\ti\pi\prl \byd \id\ten\pi\prl$ 
and 
$\ti\pi\per \byd \id\ten\pi\per \,,$
and have the coordinate expressions
\bal
\wti\Ups\prl
&= 
- c^2 \, \tau_\lam \, \tau_\mu \, u^0 \ten
(\dt d^\lam_0 - K\col\nu\lam\rho \, \dt x^\rho_0 \, d^\nu)
\wed d^\mu
\\
&= 
- (\alp^0)^2 \, \br g_{0\lam} \, \br g_{0\mu} \, u^0 \ten
(\dt d^\lam_0 - K\col\nu\lam\rho \, \dt x^\rho_0 \, d^\nu)
\wed d^\mu \,,
\\[3mm]
\wti\Ups\per
&= 
(g_{\lam\mu} + c^2 \, \tau_\lam \, \tau_\mu) \, u^0 \ten
(\dt d^\lam_0 - K\col\nu\lam\rho \, \dt x^\rho_0 \, d^\nu)
\wed d^\mu
\\
&= 
(g_{\lam\mu} + (\alp^0)^2 \, \br g_{0\lam} \, \br g_{0\mu})\, u^0 \ten
(\dt d^\lam_0 - K\col\nu\lam\rho \, \dt x^\rho_0 \, d^\nu)
\wed d^\mu \,.
\end{align*}
\vglue-1.6\baselineskip{\ }\hfill\ENDE
\ePr
\subsection{Vertical bundle of the phase space}
\label{Vertical bundle of the phase space}
 The metric
$g$
yields an isomorphism of the vertical space of the phase space with
the $\tau$--vertical subspace of spacetime. 
 This isomorphism can be regarded as analogous to the isomorphism
which holds in the case of a fibred manifold.

 Let
$V\M J_1\f E \sub T\M J_1\f E$
be the vertical tangent subbundle over
$\f E \,.$
 The vertical prolongation of the contact map yields the mutually
inverse linear fibred isomorphisms
\beq
\nu_\tau : \M J_1\f E \to \B T \ten V^*_\tau\f E \ten V\M J_1\f E
\ssep{and}
\nu^{-1}_\tau :
\M J_1\f E \to V^*\M J_1\f E \ten \B T^* \ten V_\tau\f E \,,
\eeq
with coordinate expressions
$\nu_\tau =  \fr1{c_0 \, \alp^0} \, u_0 \ten \bet^i \ten \der^0_i$
and
$\nu^{-1}_\tau = c_0 \, \alp^0 \, u^0 \ten d^i_0 \ten b_i \,.$

 Thus, for each
$Y \in \sec(\M J_1\f E, V\M J_1\f E)$ 
and 
$X \in \sec(\f E,T\f E) \,,$ 
we obtain
\beq
\nu^{-1}_\tau(Y) \in \fib(\M J_1\f E, \, \B T^* \ten V_\tau\f E)
\ssep{and}
\nu_\tau(X) \in \sec(\M J_1\f E, \, \B T \ten V\M J_1\f E) \,,
\eeq  
with coordinate expressions
\beq
\nu^{-1}_\tau(Y) = c \, \alp^0 \, Y^i_0 \, b_i
\sep{and}
\nu_\tau(X) = \fr 1{c\,\alp^0} \, \ti X^i \, \der^0_i \,,
\ssep{where} 
\ti X^i = X^i - x^i_0 \, X^0 \,.
\eeq
\subsection{Phase connection}
\label{Phase connections}
 Let us introduce the general notion of connection of the phase
space and discuss the relation with spacetime connections.

 We define a \emph{phase connection} to be a connection of the bundle
$\M J_1\f E \to \f E \,.$

 A phase connection can be represented, equivalently, by a tangent
valued form\linebreak
$\Gam : \M J_1\f E \to T^*\f E \ten T\M J_1\f E \,,$
which is projectable over
$\1 : \f E \to T^*\f E \ten T\f E \,,$
or by the complementary vertical valued form
$\nu[\Gam] :
\M J_1\f E \to T^*\M J_1\f E \ten V\M J_1\f E \,,$
or by the vector valued form
$\nu_\tau[\Gam] \byd \nu^{-1}_\tau \com \nu[\Gam] : \M J_1\f E \to
T^*\M J_1\f E \ten (\B T^* \ten V_\tau\f E) \,.$
 Their coordinate expressions are
\bgt
\Gam =
d^\lam \ten(\der_\lam + \Gam\Ga\lam i \, \der_i^0) \,,
\qquad
\nu[\Gam] =
(d^i_0 - \Gam\Ga\lam i \, d^\lam) \ten \der^0_i \,,
\\
\nu_\tau[\Gam] =
c \, \alp^0 \, (d^i_0 - \Gam\Ga\lam i \, d^\lam) \ten b_i \,,
\ssep{with}
\Gam\Ga\lam i \in \map(\M J_1\f E, \, \Rn) \,.
\end{gather*}

 We define the \emph{curvature} of a phase connection
$\Gam$
to be the vertical valued 2--form
\beq
R[\Gam] \byd - [\Gam, \, \Gam] :
\M J_1\f E \to
\Lam^2 T^*\f E \ten V\M J_1\f E \,,
\eeq
where
$[\,,]$
is the Fr\"olicher--Nijenhuis bracket.
 We have the coordinate expression
\beq
R[\Gam] = - 2 \,
(\der_\lam \Gam\Ga\mu i + \Gam\Ga\lam j \, \der^0_j \Gam\Ga\mu i) \,
d^\lam \wed d^\mu \ten \der^0_i \,.
\eeq

\bTh\label{Th3.1}
{\rm \cite{JanMod96}}
 For each linear spacetime connection
$K \,,$
there is a unique phase connection
$\Gam \,,$
such that the following diagram commutes
\newdiagramgrid{1}
{1.5, 1.5, 1.5, 1.5}
{.6, .8}
\bdg[grid=1]
T\M J_1\f E
&\rTo^{\nu[\Gam]}
&V\M J_1\f E
&\rTo^{\nu^{-1}_\tau}
&\B T^* \ten V_\tau\f E
\\
\dTo^{(\tau[\M J_1\f E], \, T\K d)}
&&&& \uTo_{\pi\Per}
\\
\M J_1\f E \ucar{\f E} T(\B T^* \ten T\f E)
&& \rTo_{(\id[\M J_1\f E] \car \nu[\ti K])}
&& \M J_1\f E \ucar{\f E} (\B T^* \ten T\f E)
\edg

 Indeed, we have the coordinate expression
\bal
\Gam\Ga \lam i
&=
\br \del^i_\rho K\col\lam\rho\sig \, \br\del^\sig_0 \,.
\end{align*}

 Thus, the above correspondence yields a natural map
$\chi : K \mto \Gam$
between the set of linear spacetime connections and the set of
phase connections.
\hfill\ENDE
\eTh

\bNt\label{Nt3.2}
 We have the following identity
\cite{JanMod96}
\bal
R[\chi(K)]\cu\lam\mu i_0 
&= 
\br\del^i_\rho \, R[K]\cul\lam\mu\rho\sig \, \br\del^\sig_0 \,.
\end{align*}
\vglue-1.6\baselineskip{\ }\hfill\ENDE
\eNt
\subsection{Dynamical phase connection}
\label{Dynamical phase connection}
 Let us introduce the general notion of dynamical connection of the
phase space and discuss the relation with phase connections and
spacetime connections.

 Let
$\M J_2\f E$
be the space of 2-jets of motions.
 We can see that this space can be naturally regarded as the
affine subundle
$\M J_2\f E \sub \B T^* \ten T\M J_1\f E \,,$
which projects on
$\K d : \M J_1\f E \to \B T^* \ten T\f E \,.$

 A \emph{dynamical phase connection} is defined to be a section
$\gam : \M J_1\f E \to \M J_2\f E \,,$
or, equivalently, a section
$\gam : \M J_1\f E \to \B T^* \ten T\M{J}_1\f E \,,$
which projects on
$\K d \,.$
 The coordinate expression of a dynamical connection is of the type
\beq
\gam = c \, \alp^0 \,
(\der_0 + x^i_0 \, \der_i + \gam\ga i \, \der^0_i) \,,
\ssep{with}
\gam\ga i \in \map(\M J_1\f E, \, \Rn) \,.
\eeq

 If
$\gam$
is a dynamical phase connection, then we have
$\gam \con \tau = 1 \,.$

 If
$\Gam$
is a phase connection, then the section
$\gam \byd \gam[\Gam] \byd 
\K d\con\Gam:\M J_1\f E \to \B T^*\ten T\M J_1\f E$
turns out to be a dynamical phase connection, whose coordinate
expression is given by
\beq
\gam\ga i = \Gam\Ga 0i + \Gam\Ga ji \, x^j_0
= \Gam\Ga\rho i\,\br\del^\rho_0\,.
\eeq

 Hence, a linear spacetime connection
$K$
yields the dynamical phase connection\linebreak
$\gam \byd \gam[K] \byd \K d \con \chi(K) \,.$
 Its coordinate expression is
\bgt
\gam\ga i =
\br\del^i_\rho K\col \sig\rho\tau \, \br\del^\sig_0\, \br\del^\tau_0
 \,.
\end{gather*}
\subsection{Phase 2--form and 2--vector}
\label{Phase 2--form and 2--vector}
 Let us introduce the general notions a phase 2--forms and phase
2--vectors associated with a phase connection and discuss the relation
with spacetime connections.

 If
$\Gam$
is a phase connection, then we define the scaled \emph{phase 2--form}
and the scaled \emph{phase 2--vector} associated with
$g$
and
$\Gam$
to be, respectively, the sections
\bat{3}
\Ome 
&\byd 
\Ome[g,\Gam] 
&&\byd
g \con \big(\nu_\tau[\Gam] \wed \tht\big) 
&&:
\M J_1\f E \to (\B T^* \ten \B L^2) \ten \Lam^2T^*\M J_1\f E \,,
\\
\Lam 
&\byd 
\Lam[g, \Gam] 
&&\byd
\ba g \con (\Gam \wed \nu_\tau) 
&&:
\M J_1\f E \to (\B T \ten \B L^{-2}) \ten \Lam^2T\M J_1\f E \,.
\end{alignat*}
 Their coordinate expressions are
\beq
\Ome =
c \, \alp^0 \, \br g_{i\mu} \,
(d^i_0 - \Gam\Ga \lam i \, d^\lam) \wed d^\mu
\ssep{and}
\Lam =
\fr1{c \, \alp^0} \, \br g^{j\lam} \,
 (\der_\lam + \Gam\Ga \lam i \, \der^0_i) \wed \der^0_j \,.
\eeq

 There is a unique dynamical phase
connection
$\ba\gam \,,$
such that
$\ba\gam \con \Ome[g,\Gam] = 0 \,.$
Namely,
$\ba\gam = \gam[\Gam] : \M J_1\f E \to \B T^* \ten T\M J_1\f E \,.$

\bLm\label{Lm3.3}
 If
$\Gam$
is a phase connection, then we have
$i_\Lam \, \Ome = - 3 \,.$
\eLm

\begin{proof}
 We have
$i_\Lam \, \Ome =
- \br g^{i\mu} \, \br g_{i\mu} =
- 3 \,.$ 
\end{proof}

 Thus, a linear spacetime connection
$K$
yields the scaled phase 2--form and the scaled phase 2--vector
\beq
\Ome \byd \Ome[g,K] \byd \Ome\big[g, \chi(K)\big]
\ssep{and}
\Lam \byd \Lam[g,K] \byd \Lam\big[g, \chi(K)\big] \,.
\eeq
 Their coordinate expressions are
\bal
\Ome
&= 
c \, \alp^0 \, \br g_{i\mu} \,
\big(d^i_0 -
\br\del^i_\rho\, K\col \lam\rho\sig \, \br\del^\sig_0 \,
d^\lam\big) \wed d^\mu \,,
\\[3mm]
\Lam 
&=
\fr1{c \, \alp^0} \, \br g^{j\lam} \,
(\der_\lam + \br\del^i_\rho K\col \lam\rho\sig \,
\br\del^\sig_0 \, \der^0_i) \wed \der^0_j \,.
\end{align*}
\myssec{Spacetime and phase 2--forms and 2--vectors}
\label{Spacetime and phase 2--forms and 2--vectors}
 Let
$K$
be a linear spacetime connection.
 Then, we can compare the scaled phase 2--form 
$\Ome \byd \Ome[g, K]$
and the scaled phase 2--vector
$\Lam \byd \Lam[g, K]$
with the scaled spacetime 2--form 
$\wti\Ups \byd \wti\Ups[g, \ti K]$
and the scaled spacetime 2--vector
$\wti\Xi \byd \wti\Xi[g, \ti K] \,,$
respectively, in the following way.

\bPr\label{Pr4.11}
 The contact map
$\K d$
yields the following scaled 2--forms of
$\M J_1\f E$
\bal
\K d^* \, \wti\Ups
&:
\M J_1\f E \to (\B T^* \ten \B L^2) \ten \Lam^2T^*\M J_1\f E\,,
\\
\K d^* \, \wti\Ups\prl
&:
\M J_1\f E \to (\B T^* \ten \B L^2) \ten \Lam^2T^*\M J_1\f E\,,
\\
\K d^* \, \wti\Ups\per
&:
\M J_1\f E \to (\B T^* \ten \B L^2) \ten \Lam^2T^*\M J_1\f E \,,
\end{align*}
which fulfill the equality
\beq
\K d^* \, \wti\Ups =
\K d^* \, \wti\Ups\prl + \K d^* \, \wti\Ups\per \,.
\eeq

 We have the coordinate expressions
\bal
\K d^* \, \wti\Ups
&=
c \, \alp^0 \, \Big(\br g_{i\mu} \, d^i_0 +
\big(
\tfr12 \, (\alp^0)^2 \, \br g_{0\mu} \, \der_\nu \ha g_{00}
-
g_{\lam\mu} \, \br\del^\rho_0\, K\col\nu\lam\rho \big) \,
d^\nu\Big) \wed d^\mu \,,
\\
\K d^* \, \wti\Ups\prl
&=
c \, (\alp^0)^3 \, \br g_{0\mu} \, \big(
\tfr12 \, \der_\nu \ha g_{00} + \br g_{0\lam} 
\, \br\del^\rho_0\, K\col\nu\lam\rho 
\big) \, d^\nu \wed d^\mu \,,
\\
\K d^* \, \wti\Ups\per
&=
c \, \alp^0 \, \br g_{i\mu} \,
\Big(
d^i_0 -
\br\del^i_\lam \,\br\del^\rho_0\, K\col\nu\lam\rho \,
d^\nu\Big) \wed d^\mu \,.
\end{align*}
\ePr

\begin{proof}
 The equality
\beq
(x^\lam, \; \dt x^0_0, \; \dt x^i_0) \com \K d =
(x^\lam, \; c_0 \, \alp^0, \; c_0 \, \alp^0 \, x^i_0)
\eeq
yields
\beq
\K d^* \dt x^0_0 = c_0 \, \alp^0 \,,
\quad
\K d^* \dt x^i_0 = c_0 \, \alp^0 \, x^i_0
\eeq
and
\bal
\K d^*
\dt d^0_0
&=
c_0 \, \der_\lam\alp^0 \, d^\lam + c_0 \, \der^0_j\alp^0 \, d^j_0
\\
&=
\tfr12 \, c_0 \, (\alp^0)^3 \, \der_\lam \ha g_{00} \, d^\lam +
c_0 \, (\alp^0)^3 \, \br g_{0j} \, d^j_0
\\[2mm]
\K d^* \dt d^i_0
&=
c_0 \, \der_\lam\alp^0 \, x^i_0 \, d^\lam +
c_0 \, \der^0_j\alp^0 \, x^i_0 \, d^j_0 +
c_0 \, \alp^0 \, d^i_0
\\
&=
\tfr12 \, c_0 \, (\alp^0)^3 \,
\der_\lam(\ha g_{00}) \, x^i_0 \, d^\lam +
c_0 \, (\alp^0)^3 \, \br g_{0j} \, x^i_0 \, d^j_0 +
c_0 \, \alp^0 \, d^i_0 \,.
\end{align*}

 Then, we obtain
\bal
\K d^* \wti\Ups
&=
\K d^* \big(g_{\lam\mu} \, u^0 \ten (\dt d^\lam_0 -
K\col\nu\lam\rho \, \dt x^\rho_0 \, d^\nu) \wed d^\mu\big)
\\[3mm]
&=
g_{0\mu} \, u^0 \ten
\big(
\tfr12 \, c_0 \, (\alp^0)^3 \, \der_\lam \ha g_{00} \, d^\lam +
c_0 \, (\alp^0)^3 \, \br g_{0i} \, d^i_0
\big) \wed d^\mu
\\
&\quad
+ g_{j\mu} \, u^0 \ten
\big(
\tfr12 \, c_0 \, (\alp^0)^3 \,
\der_\lam \ha g_{00} \, x^j_0 \, d^\lam +
c_0 \, (\alp^0)^3 \, \br g_{0i} \, x^j_0 \, d^i_0 +
c_0 \, \alp^0 \, d^j_0
\big) \wed d^\mu
\\
&\quad
- c_0 \, \alp^0 \,
g_{\lam\mu} \, (K\col \nu\lam 0 + K\col \nu\lam j \, x^j_0) \,
u^0 \ten d^\nu \wed d^\mu
\\[3mm]
&=
c^0 \, \alp^0 \, \br g_{i\mu} \, u^0 \ten d^i_0 \wed d^\mu
+
\tfr12 c_0 \, (\alp^0)^3 \, \br g_{0\mu} \,
\der_\lam \ha g_{00} \, u^0 \ten d^\lam  \wed d^\mu
\\
&\quad
- c_0 \, \alp^0 \,
g_{\lam\mu} \, (K\col \nu\lam 0 + K\col \nu\lam j \, x^j_0) \,
u^0 \ten d^\nu \wed d^\mu \,.
\end{align*}

 Moreover, by recalling the equality
$(\alp^0)^2 \, (\br g_{00} + \br g_{0i} \, x^i_0) = - 1 \,,$
we obtain
\bal
\K d^* \wti\Ups\prl
&=
- \K d^* \big(
(\alp^0)^2 \, \br g_{0\lam} \, \br g_{0\mu} \, u^0 \ten
(\dt d^\lam_0 - K\col\nu\lam\rho \, \dt x^\rho_0 \, d^\nu) \wed d^\mu
\big)
\\[3mm]
&=
- (\alp^0)^2 \, \br g_{00} \, \br g_{0\mu} \,
\big(
\tfr12 \, c_0 \, (\alp^0)^3 \, \der_\lam \ha g_{00} \, d^\lam +
c_0 \, (\alp^0)^3 \, \br g_{0j} \, d^j_0
\big) \wed d^\mu
\\
&\quad
- (\alp^0)^2 \, \br g_{0i} \, \br g_{0\mu} \, u^0 \ten
\big(
\tfr12 \, c_0 \, (\alp^0)^3 \,
\der_\lam \ha g_{00} \, x^i_0 \, d^\lam +
c_0 \, (\alp^0)^3 \, \br g_{0j} \, x^i_0 \, d^j_0 +
c_0 \, \alp^0 \, d^i_0
\big) \wed d^\mu
\\
&\quad
+ c_0 \, (\alp^0)^3 \, \br g_{0\lam} \, \br g_{0\mu} \,
(K\col\nu\lam0 + K\col\nu\lam j \, x^j_0) \,
u^0 \ten d^\nu \wed d^\mu
\\[3mm]
&=
\tfr12 \, c_0 \, (\alp^0)^3 \, \br g_{0\mu} \, \der_\lam \ha g_{00}
\, u^0 \ten d^\lam \wed d^\mu +
c^0 \, (\alp^0)^3 \, \br g_{0\lam} \, \br g_{0\mu} \,
K\col\nu\lam\rho \, \br\del^\rho_0 \,
u^0 \ten d^\nu \wed d^\mu \,.
\end{align*}

 Finally, we obtain
\bal
\K d^* \wti\Ups\per
&=
\K d^* \wti\Ups -
\K d^* \wti\Ups\prl
\\
&=
c_0 \, \alp^0 \, \br g_{i\mu} \, u^0 \ten
(d^i_0 -
\br\del^i_\rho \, K\col\nu\rho\sig \, \br\del^\sig_0 \, d^\nu)
\wed d^\mu \,. 
\end{align*}
\vglue-1.5\baselineskip
\end{proof}

\bTh\label{Th3.12}
 We have
\beq
\Ome = \K d^* \, \wti\Ups\per \,.
\eeq
\eTh

\begin{proof}
 The Theorem follows by a direct comparison of the coordinate
expressions
\bal
 \Ome
&=
c \, \alp^0 \, \br g_{i\mu} \,
(d^i_0 - \Gam\Ga \lam i \, d^\lam) \wed d^\mu \,,
\\
\K d^* \, \wti\Ups\per
&=
c \, \alp^0 \, \br g_{i\mu} \,
\Big(
d^i_0 -
\br\del^i_\lam \, K\col\nu\lam\rho \, \br\del^\rho_0
\, d^\nu\Big) \wed d^\mu \,,
\end{align*}
where we put 
$\Gam\Ga \nu i =
\br\del^i_\lam \, K\col\nu\lam\rho \, \br\del^\rho_0 \,.$ 
\end{proof}

\bCr
 $\Lam$
is the unique scaled phase 2--vector such that the following diagram
commutes
\bdg
\M J_1\f E 
&\rTo^{\Lam} 
&& (\B T\ten \B L^{-2}) \ten \Lam^2T\M J_1\f E
\\
\dTo^{\K d}
&&&\dTo_{\id\ten \Lam^2T\K d}
\\
\B T^*\ten T\f E 
&\rTo^{\wti\Xi}
&& (\B T\ten \B L^{-2}) \ten \Lam^2T(\B T^*\ten T\f E)
\edg
\eCr

\begin{proof}
 We have 
\beq
\wti\Xi \com \K d = g^{\lam\mu} \, u_0 \ten 
	(\der_\lam + c_0\, \alp^0 \, K\col \lam\nu\rho
	\, \br\del^\rho_0 \, \dt\der^0_\nu)\wed\dt\der^0_\mu
\eeq
and
\bal
(\id\ten\Lam^2 T\K d)\com \Lam 
&= 
	g^{\lam\mu} \,u_0\ten \der_\lam \wed \dt\der^0_\mu 
\\
&\quad
+
c_0 \, \alp^0 \, \bigg((\alp^0)^2 \, 
	(\tfr12 \br g^{j\lam} \, \der_\lam \hat g_{00} \,
	\br\del^\nu_0 \, \del^\mu_j
+ \br\del^\lam_0 \, \Gam\Ga\lam i \, 	\del^\nu_i \br\del^\mu_0 + 
	\br g^{j\lam} \, \Gam\Ga\lam i \, \br g_{0i} \,
	\del^\mu_j \, \br\del^\nu_0) 
\\
&\quad
+
g^{0\lam} \, \Gam\Ga\lam i 	\,	 \del^\nu_i \, \br\del^\mu_0 
+ \br g^{j\lam} \, \Gam\Ga\lam i 	\, \del^\mu_j \,\del^\nu_i\bigg) \,
	\dt\der^0_\nu\wed \dt\der^0_\mu \,. 
\end{align*}

 By comparing the coefficients standing by 
$u_0 \ten \dt\der_0 \wed \dt\der_j$ 
and
$u_0 \ten \dt\der_i \wed \dt\der_j \,,$ 
we get
\bal
(g^{j\lam} \, K\col\lam{0}\rho - g^{0\lam} \, K\col\lam j \rho)
		\, \br\del^\rho_0 
&= 
(\alp^0)^2 \, \big( 
	\tfr12 g^{j\lam} \, \der_\lam \hat g_{00} 
- \br\del^\lam_0 \, \Gam\Ga\lam j
+ \br g^{j\lam} \, \Gam\Ga\lam p \, \br g_{0p}\big)
- g^{0\lam} \, \Gam\Ga\lam j
\end{align*}
and
\bal
(g^{j\lam} \, K\col \lam j \rho - g^{i\lam} \, K\col\lam j \rho)
	\, \br\del^\rho_0 
&= 
x^i_0 \, [(\alp^0)^2 \, \big( 
	\tfr12 g^{j\lam} \, \der_\lam \hat g_{00} 
  - \br\del^\lam_0 \, \Gam\Ga\lam j
  + \br g^{j\lam} \, \Gam\Ga\lam p \, \br g_{0p}\big)
  - g^{0\lam} \, \Gam\Ga\lam j]
\\
&\quad
-
x^j_0 \, [(\alp^0)^2 \, \big( 
	\tfr12 g^{i\lam} \, \der_\lam \hat g_{00} 
  - \br\del^\lam_0 \, \Gam\Ga\lam i
  + \br g^{i\lam} \, \Gam\Ga\lam p \, \br g_{0p}\big)
  - g^{0\lam} \, \Gam\Ga\lam i]
\\
&\quad
+
\br g^{j\lam} \, \Gam\Ga\lam i - \br g^{i\lam} \, \Gam\Ga\lam j \,,
\end{align*}
respectively. 

 Then, inserting the first equality into the second one, we get
\bal
\br g^{j\lam} \, \Gam\Ga\lam i - \br g^{i\lam} \, \Gam\Ga\lam j
&=
	(g^{j\lam} \, K\col\lam i \rho 
- g^{i\lam} \, K\col\lam j \rho) \, \br\del^\rho_0
\\
&\quad
-
  x^i_0 \, (g^{j\lam} \, K\col \lam 0 \rho 
- g^{0\lam} \, K\col\lam j \rho)) \, \br\del^\rho_0
+ x^j_0 \, (g^{i\lam} \, K\col\lam 0 \rho 
- g^{0\lam} \, K\col\lam i \rho) ) \, \br\del^\rho_0
\end{align*}
which is satisfied if and only if
$
\Gam\Ga\lam i = 
\br\del^i_\nu\, K\col\lam\nu\rho \,\br\del^\rho_0\,,
$
i.e., if and only if
$\Gam = \chi(K) \,.$ 
\end{proof}

\section{Contact and Jacobi structures: general case}
\label{Contact and Jacobi structures: general case}
\setcounter{equation}{0}
 In this section, we consider the metric
$g \,,$
a general phase connection
$\Gam$
and the induced objects
$\Ome \byd \Ome[g, \Gam] \,,$
$\Lam \byd \Lam[g, \Gam] \,,$
$\gam \byd \gam[\Gam] \,.$
 Then, we analyse the conditions by which
$g$
and
$\Gam$
induce scaled almost--cosymplectic--contact, or contact, structures  
and scaled almost--coPoisson--Jacobi, or Jacobi, structures (see
Introduction) on the phase space.
\subsection{Regularity and duality}
\label{Regularity and duality}
 Let us analyse the non--degeneracy of the scaled covariant pair
$(- c^2 \, \tau, \Ome)$ 
and of the scaled contravariant pair
$(-\tfr1{c^2}\,\gam,\Lam)$
and the condition for their duality \cite{JanMod07}.

\bLm
 The section
$- c^2 \, \tau \wed \Ome \wed \Ome \wed \Ome :
\M J_1\f E \to (\B T^{*4} \ten \B L^8) \ten \Lam^7 T^*\M J_1\f E$
is a scaled volume form, with coordinate expression
\beq
- c^2 \, \tau \wed \Ome \wed \Ome \wed \Ome = 
 3! \, c^4 \,
(\alp^0)^4 \, |g| \, d^1_0 \wed d^2_0 \wed d^3_0 \wed d^0 \wed d^1
\wed d^2 \wed d^3 \,.
\eeq

 Hence, the pair
$(- c^2 \, \tau, \Ome)$
is a regular covariant pair.
\eLm

\begin{proof}
 The equalities
$\tau = \tau_\lam \, d^\lam =
- \fr{\alp^0} c \, \br g_{0\lam} \, d^\lam$
and
$\Ome =
c \, \alp^0 \, \br g_{i\mu}  \,
(d^i_0 - \Gam\Ga \nu i \, d^\nu) \wed d^\mu$
yield
\beq
- c^2 \, \tau \wed 
\Ome \wed \Ome \wed \Ome =
- c^5 \, (\alp^0)^3 \, \tau_\lam \, 
\br g_{i_1\mu_1} \, \br g_{i_2\mu_2} \, \br g_{i_3\mu_3} \,
d^\lam \wed 
d^{i_1}_0 \wed d^{\mu_1} \wed
d^{i_2}_0 \wed d^{\mu_2} \wed
d^{i_3}_0 \wed d^{\mu_3} \,.
\eeq

 Hence, by taking into account that the antisymmetrisation makes some
terms vanishing, we obtain
\begin{align*}
- c^2 \, \tau \wed \Ome^3
& =
 c^4 \, (\alp^0)^4 \, 
\br g_{0\lam} \, g_{i_1\mu_1} \, g_{i_2\mu_2} \, g_{i_3\mu_3}  \,
d^\lam \wed 
d^{i_1}_0 \wed d^{\mu_1} \wed
d^{i_2}_0 \wed d^{\mu_2} \wed
d^{i_3}_0 \wed d^{\mu_3}
\\[2mm]
&=
 (3!) \, c^4 \, (\alp^0)^4 \, 
g_{0\lam} \, g_{1\mu_1} \, g_{2\mu_2} \, g_{3\mu_3} \,
d^\lam \wed 
d^1_0 \wed d^{\mu_1} \wed
d^2_0 \wed d^{\mu_2} \wed
d^3_0 \wed d^{\mu_3}
\\[2mm]
&=
 (3!) \, c^4 \, (\alp^0)^4 \, 
g_{0\lam} \, g_{1\mu_1} \, g_{2\mu_2} \, g_{3\mu_3} \,
d^\lam \wed 
d^1_0 \wed d^{\mu_1} \wed
d^2_0 \wed d^{\mu_2} \wed
d^3_0 \wed d^{\mu_3}
\\[2mm]
&=
 (3!) \, c^4 \, (\alp^0)^4 \, |g| \,
d^0 \wed 
d^1_0 \wed d^1 \wed
d^2_0 \wed d^2 \wed
d^3_0 \wed d^3 \,. 
\end{align*}
\vglue-1.5\baselineskip
\end{proof}

\bLm\label{phase space contravariant volume form}
 The section
$- \fr1{c^2} \, \gam \wed \Lam \wed \Lam \wed \Lam :
\M J_1\f E \to (\B T^{4} \ten \B L^{-8}) \ten \Lam^7 T\M J_1\f E$
is a scaled volume vector, with coordinate expression
\beq
- \fr1{c^2} \, \gam \wed \Lam \wed \Lam \wed \Lam = 
 - 3! \,\fr1{ (c\,
\alp^0)^4} \, |\ba g| \, 
\der^0_1 \wed \der^0_2 \wed \der^0_3 \wed 
\der_0 \wed \der_1 \wed \der_2 \wed \der_3 \,.
\eeq

 Hence, the pair
$(- \tfr1{c^2} \, \gam, \Lam)$
is a regular contravariant pair.
\eLm

\begin{proof}
 The equalities
$\gam = c \, \alp^0 \, (\br\del^\lam_0 \,
\der_\lam + \gam^i_{00} \, \der^0_i)$
and
$\Lam =
\tfr1{c \, \alp^0} \, \br g^{j\mu}  \,
(\der_\mu + \Gam\Ga \mu i \, \der^0_i) \wed \der^0_j$
yield
\bal
- \tfr1{c^2} \, \gam \wed \Lam^3 
&= 
- \tfr1{c^4\,(\alp^0)^2} \,  
\br g^{i_1\mu_1} \, \br g^{i_2\mu_2} \, \br g^{i_3\mu_3} \, 
(\br\del^\lam_0 \,
\der_\lam + \gam^i_{00} \, \der^0_{i_0})
\wed \der_{\mu_1} \wed \der^0_{i_1}
\wed \der_{\mu_2} \wed \der^0_{i_2}
\wed \der_{\mu_3} \wed \der^0_{i_3}
\\
& = 
- \tfr1{c^4 \, (\alp^0)^2} \,  
\br g^{i_1\mu_1} \, \br g^{i_2\mu_2} \, \br g^{i_3\mu_3} \, 
\br\del^\lam_0 \, \der_\lam 
\wed \der_{\mu_1} \wed \der^0_{i_1}
\wed \der_{\mu_2} \wed \der^0_{i_2}
\wed \der_{\mu_3} \wed \der^0_{i_3}
\,.
\end{align*}

 Hence, by the identity
$\br\del^\lam_0 = 
\br g_{0i} \, \br g^{i\lam} - \tfr1{(\alp^0)^2} \, g^{0\lam}$
and the fact that the antisymmetrisation makes some terms vanishing,
we obtain
\bal
- \tfr1{c^2} \, \gam \wed \Lam^3 
&= 
 - \tfr1{c^4 \, (\alp^0)^2} \,  
\br g^{i_1\mu_1} \, \br g^{i_2\mu_2} \, \br g^{i_3\mu_3} \, 
(\br g_{0i} \, \br g^{i\lam} - \tfr1{(\alp^0)^2} \, g^{0\lam}) \,
\\
&\qquad\qquad\qquad
\der_\lam 
\wed \der_{\mu_1} \wed \der^0_{i_1}
\wed \der_{\mu_2} \wed \der^0_{i_2}
\wed \der_{\mu_3} \wed \der^0_{i_3}
\\
&=
 \tfr1{(c\,\alp^0)^4} \,  
\br g^{i_1\mu_1} \, \br g^{i_2\mu_2} \, \br g^{i_3\mu_3} \,
g^{0\lam} \,
\der_\lam 
\wed \der_{\mu_1} \wed \der^0_{i_1}
\wed \der_{\mu_2} \wed \der^0_{i_2}
\wed \der_{\mu_3} \wed \der^0_{i_3}
\\
&=
 \tfr1{(c\,\alp^0)^4} \,  
g^{i_1\mu_1} \, g^{i_2\mu_2} \, g^{i_3\mu_3} \, g^{0\lam} \,
\der_\lam 
\wed \der_{\mu_1} \wed \der^0_{i_1}
\wed \der_{\mu_2} \wed \der^0_{i_2}
\wed \der_{\mu_3} \wed \der^0_{i_3}
\\
&=
 3! \, \tfr1{(c\,\alp^0)^4} \,  
g^{0\lam} \, g^{1\mu_1} \, g^{2\mu_2} \, g^{3\mu_3} \, 
\der_\lam 
\wed \der_{\mu_1} \wed \der^0_{1}
\wed \der_{\mu_2} \wed \der^0_{2}
\wed \der_{\mu_3} \wed \der^0_{3}
\\
&=
 3! \, \tfr1{(c\,\alp^0)^4} \, |\ba g| \,   
\der_0 
\wed \der_{1} \wed \der^0_1
\wed \der_{2} \wed \der^0_2
\wed \der_{3} \wed \der^0_3
\\
&=
- 3! \, \tfr1{(c\,\alp^0)^4} \, |\ba g| \,   
\der^0_1 
\wed \der^0_2 \wed \der^0_3
\wed \der_0 \wed \der_1
\wed \der_2 \wed \der_3
\,. 
\end{align*}
\vglue-1.6\baselineskip
\end{proof}

\bLm\label{Lemma: Lam Sha Ome}
We have
\beq
(\Lam\Sha\ten\Lam\Sha)(\Ome) = -\Lam\,,\qquad
(\Ome\Fla\ten\Ome\Fla)(\Lam) = - \Ome\,.
\eeq
\eLm

\begin{proof}
We have
\bal
	\Lam\Sha(d^\lam) 
& = 
	\fr1{c\,\alp^0}\br g^{j\lam}\der^0_j
\,,\quad
	\Lam\Sha(d^i_0)
= 
	\fr1{c\,\alp^0}\big(-\br g^{i\mu}\der_\mu
		+(\br g^{j\rho}\Gam\Ga\rho i 
		- \br g^{i\rho}\Gam\Ga\rho j)\,\der^0_j\big)\,	
\end{align*}
and
\bal
	\Ome\Fla(\der_\lam)
& =
	c\,\alp^0\big(-\br g_{i\lam} d^i_0
	+ (\br g_{p\lam}\Gam\Ga\mu p - \br g_{p\mu}\Gam\Ga\lam p)\,d^\mu\big)\,,
\quad
	\Ome\Fla(\der^0_i)
= 
	c\,\alp^0\br g_{i\mu} d^\mu\,.			
\end{align*}
Then
\bal
	(\Lam\Sha\ten\Lam\Sha)(\Ome) 
& = 
	c\,\alp^0\,\big(\Lam\Sha(d^i_0) 
	- \Gam\Ga\nu i\,\Lam\Sha(d^\nu)\big)\wed \Lam\Sha(d^\mu)
\\
& =
	\fr1{c\,\alp^0} \br g_{i\mu}\,\big(- \br g^{i\lam}\,\der_\lam
		+ (\br g^{j\rho}\,\Gam\Ga\rho i 
		- \br g^{i\rho}\,\Gam\Ga\rho j)\,\der^0_j
		- \br g^{j\rho}\,\Gam\Ga\rho i\, \der^0_j\big) \wed
		(\br g^{k\mu}\der^0_k)
\\
& =
	\fr1{c\,\alp^0} \,\big(- \br g^{i\lam}\,\der_\lam		
		- \br g^{i\lam}\,\Gam\Ga\lam j \,\der^0_j\big) \wed
		\der^0_i\,
\end{align*}
and
\bal
	(\Ome\Fla\ten\Ome\Fla)(\Lam) 
& =
	\fr1{c\,\alp^0} \br g^{j\lam}\, \big(\Ome\Fla(\der_\lam)	
		+ \Gam\Ga\lam i \,\Ome\Fla(\der^0_i)\big) \wed
		\Ome\Fla(\der^0_j)
\\
& =
	{c\,\alp^0} \br g^{i\lam}\,\big(-\br g_{p\lam} \, d^p_0
	+ (\br g_{p\lam}\,\Gam\Ga\mu p 
	- \br g_{p\mu}\,\Gam\Ga\lam p )\,d^\mu
	+ \Gam\Ga\lam i \, \br g_{i\mu}\, d^\mu
	\big) \wed (\br g_{j\nu} d^\nu)
\\
& =
	{c\,\alp^0}\, \big(-\br g_{i\lam}\, d^i_0 
	+ \br g_{i\lam}\,\Gam\Ga\mu i \, d^\mu
	\big) \wed d^\lam \,. 
\end{align*}
\vglue-1.6\baselineskip
\end{proof}

\bPr
 The structures
$(- c^2 \, \tau \,,\, \Ome)$ 
and
$(- \tfr1{c^2} \, \gam, \, \Lam)$ 
are mutually dual if and only if 
$\gam = \gam[\Gam] \,.$
\ePr

\begin{proof}
 Indeed, we have 
\beq
i_{-c^2\,\tau}\Lam = -c^2\, i_\tau\Lam = 0
\ssep{and} 
i_{-\tfr1{c^2}\,\gam}(-c^2\, \tau) = i_\gam\tau = 1 \,.
\eeq

 Moreover,
$i_{-\tfr1{c^2}\gam}\Ome = 0 \,,$ 
if and only if 
$i_\gam\Ome = 0 \,,$  
i.e. if and only if 
$\gam = \gam[\Gam] \,.$

 The fact that 
the maps
$\Ome\Fla_{|\im\Lam\Sha} : \im\Lam\Sha \to
\im\Ome\Fla$
and
$\Lam\Sha_{|\im\Ome\Fla} : \im\Ome\Fla \to
\im\Lam\Sha$
are isomorphisms and
\beq
(\Ome\Fla_{|\im\Lam\Sha})^{-1} = 
\Lam\Sha_{|\im\Ome\Fla} \,,
\quad
(\Lam\Sha_{|\im\Ome\Fla})^{-1} = 
\Ome\Fla_{|\im\Lam\Sha} \,
\eeq 
follows, by \cite{JanMod07}, from the fact that dual structures  
are characterized by
\beq
(\Lam\Sha\ten\Lam\Sha)(\Ome) = -\Lam\,,\qquad
(\Ome\Fla\ten\Ome\Fla)(\Lam) = - \Ome\,.
\eeq
Then Proposition follows from
Lemma  \ref{Lemma: Lam Sha Ome}. 
\end{proof}

\bRm
 Let us remark that dual pairs
$(- c^2 \, \tau \,,\, \Ome)$ 
and
$(- \tfr1{c^2} \, \gam, \, \Lam)$
are characterised by the the following identities \cite{JanMod07}
\beq
\ker (-c^2\,\tau) = \im \Lam\Sha\,,\quad
\ker (-\tfr1{c^2}\,\gam )= \im\Ome\Fla\,.
\eeq
\vglue-1.5\baselineskip{\ }\hfill\ENDE
\eRm
\subsection{Splittings of the phase tangent and cotangent
bundles}
\label{Splittings of the phase tangent and cotangent bundles}
 Next, we study the splitting of the scaled tangent bundle of the
phase  space induced by the pair
$(- \tfr1{c^2} \, \gam, \, \Lam)$ 
and the splitting of the scaled cotangent bundle
of the phase space induced by the pair
$(- c^2 \, \tau \,,\, \Ome) \,.$

\bDf
 We define the 
$\gam$--horizontal and the
$\gam$--vertical subbundles to be, respectively, the vector
subbundles
\bal
H_\gam \M J_1\f E & \byd \lang - \fr1{c^2}\,\gam \rang \subseteq
		     \B T\ten \B L^{-2} \ten T\M J_1\f E\,,
\\
V^*_\gam \M J_1\f E & \byd \ker (- \fr1{c^2}\,\gam) \subseteq
		     \B T^*\ten \B L^{2} \ten T^*\M J_1\f E\,.
\end{align*}

 We define the 
$\tau$--horizontal and the
$\tau$--vertical subbundle
to be, respectively, the vector subbundles
\bal
H^*_\tau \M J_1\f E & \byd \lang - {c^2}\,\tau \rang \subseteq
		     \B T^*\ten \B L^{2} \ten T^*\M J_1\f E\,,
\\[2mm] 
V_\tau \M J_1\f E & \byd \ker (- {c^2}\,\tau) \subseteq
		     \B T\ten \B L^{-2} \ten T\M J_1\f E\,.
\end{align*}
\vglue-1.5\baselineskip{\ }\hfill\ENDE
\eDf

\bPr
We have the linear splittings
\beq
\B T\ten \B L^{-2}\ten T\M J_1\f E 
= H_\gam\M J_1\f E\oplus V_\tau \M J_1\f E
= \lang -\tfr1{c^2}\,\gam\rang \oplus \ker(-c^2\,\tau)
= \lang -\tfr1{c^2}\,\gam\rang \oplus \im\Lam\Sha\,,
\eeq
\beq
\B T^*\ten \B L^2\ten T^*\M J_1\f E 
= H^*_\tau\M J_1\f E \oplus V^*_\gam \M J_1\f E
= \lang -{c^2}\,\tau\rang \oplus \ker(-\tfr1{c^2}\,\gam)
= \lang -{c^2}\,\tau\rang \oplus \im\Ome\Fla
	\,.
\eeq

We have the mutually dual local bases of phase vector fields
$(e_0, \, e_i,\, e^0_i)$
and of phase 1--forms
$(\eps^0, \, \eps^i,\,\eps^i_0)$
adapted to the above splittings, where
\bEq\label{Eq: der to e}
e_0 \byd \br\del^\lam\,(\der_\lam + \Gam\Ga\lam i\,\der^0_i)\,,
\qquad 
e_i \byd (\del^\lam_i + (\alp^0)^2\,\br g_{0i}\, \br\del^\lam_0) \,
(\der_\lam + \Gam\Ga\lam i\,\der^0_i) \,,
\qquad 
e^0_i \byd \der^0_i \,,
\eEq
and
\bEq\label{Eq: d to eps}
\eps^0 \byd - (\alp^0)^2\,\br g_{0\lam} \, d^\lam \,,
\qquad
\eps^i \byd d^i - x^i_0\, d^0 \,,
\qquad
\eps^i_0 \byd d^i_0 - \Gam\Ga\lam i\, d^\lam \,.
\eEq

\smallskip

 We have the inverse relations
\bEq\label{Eq: e to der}
\der_\lam = - (\alp^0)^2\,\br g_{0\lam}\,e_0 
	+ \br\del^i_\lam\,e_i - \Gam\Ga\lam i\,e^0_i\,,
	\qquad 
\der^0_i = e^0_i\,.
\eEq
and
\bAl\label{Eq: eps to d}
d^0  
&= 
\eps^0 + (\alp^0)^2\,\br g_{0i} \, \eps^i \,,
\qquad
d^i = \eps^i + x^i_0\,(\eps^0 + (\alp^0)^2\,\br g_{0j}\,\eps^j) \,,
\qquad
\\ \nonumber
d^i_0  
&= 
\eps^i_0 + (\Gam\Ga j i 
	+ (\alp^0)^2\,\br g_{0j}\,\br\del^\rho_0\,\Gam\Ga\rho i) \, \eps^j
	+ \br\del^\rho_0\, \Gam\Ga\rho i\,\eps^0 \,.
\end{align}
\vglue-1.5\baselineskip{\ }\hfill\ENDE
\ePr

\bPr
 The projection 
$\, \B T\ten \B L^{-2}\ten T\M J_1\f E \to H_\gam\M J_1\f E \,$ 
is given by 
\beq
X\mto (- c^2\, \tau)(X)\,(-\fr1{c^2}\,\gam) = \tau(X)\,\gam
\eeq
and the projection
$\, \B T\ten \B L^{-2}\ten T\M J_1\f E  \to V_\tau \M J_1\f E \,$ 
is given by 
\beq
X\mto (X-\tau(X)\,\gam)\,,
\eeq
for each 
$X\in \B T\ten \B L^{-2}\ten T\M J_1\f E \,.$

 The projection
$\B T^*\ten \B L^2\ten T^*\M J_1\f E \to H^*_\tau\M J_1\f E$ 
is given by
\beq
\phi \mto (-\fr1{c^2}\,\gam)(\phi)\,(-c^2\,\tau) = \gam(\phi)\,\tau
\eeq
and the projection
$\B T^*\ten \B L^2\ten T^*\M J_1\f E \to  V^*_\gam \M J_1\f E$
is given by
\beq
\phi \mto \phi - \gam(\phi)\,\tau \,,
\eeq
for each 
$\phi \in \B T^*\ten \B L^{2}\ten T^*\M J_1\f E \,.$
\hfill\ENDE
\ePr

\bCr
 Each 
$X \in \sec(\M J_1\f E, \, \B T \ten \B L^{-2} \ten T\M J_1\f E)$ 
can be uniquely split as
\beq
X = \tau(X) \, \gam + (X - \tau(X)\,\gam) \,.
\eeq

 If the coordinate expression of
$X$
in a spacetime chart is
$X = 
X^{0\lam} \, u_0 \ten \der_\lam + X^{0}{}^i_0 \,u_0 \ten \der^0_i \,,$
with
$X^{0\lam}, X^{0}{}^i_0 
\in \map(\M J_1\f E, \, \B L^{-2} \ten \Rn) \,,$
then the expression of its splitting in the adapted base 
\eqref{Eq: der to e} is
\beq
X = - (\alp^0)^2\,\br g_{0\lam}\, X^{0\lam}\, u_0\ten e_0
	+ \br\del^i_\lam\, X^{0\lam}\, u_0\ten e_i
	+ (X^0{}^i_0 - \Gam\Ga\lam i\, X^{0\lam})\, u_0\ten e^0_i\,.
\eeq

 Each 
$\phi\in\sec(\M J_1\f E, \, \B T^*\ten \B L^{2}\ten T^*\M J_1\f E)$ 
can be uniquely split as
\beq
\phi = \gam(\phi) \, \tau + (\phi - \gam(\phi) \, \tau) \,.
\eeq

 If the coordinate expression of
$\phi$
in a spacetime chart is
$\phi = 
\phi_{0\lam} \, u^0\ten d^\lam + \phi_{0}{}_i^0 \, u^0 \ten d_0^i \,,$
with
$\phi_{0\lam}, \phi_{0}{}_i^0
\in \map(\M J_1\f E, \, \B L^{2} \ten \Rn) \,,$
then the expression of its splitting in the adapted base 
\eqref{Eq: d to eps} is
\bal
\phi 
& = 
\br\del^\rho_0\,(\phi_{0\rho} + \phi_0{}^0_p\,\Gam\Ga\rho p)\,
		u^0\ten \eps^0
\\
&\quad
	+ \big(\phi_{0i} + \phi_0{}^0_p\,\Gam\Ga i p 
		+ (\alp^0)^2\,\br g_{0i}\, \br\del^\rho_0\,(\phi_{0\rho}
		+ \phi_0{}^0_p\,\Gam\Ga\rho p)\big) \, u^0\ten \eps^i
	+ \phi_0{}^0_i\,\, u_0\ten \eps^i_0\,.
\end{align*}  
\vglue-1.6\baselineskip{\ }\hfill\ENDE
\eCr

\bCr
 If the coordinate expression of
$X$
in a spacetime chart and in a adapted base are
\bal
X
&=
X^{0\lam}\,u_0 \ten \der_\lam +  X^{0}{}^i_0\,u_0 \ten \der^0_i \,,
\\
&=
\wti 	X^{00}\,u_0\ten e_0 + \wti 	X^{0i}\,u_0\ten e_i
+ \wti X^{0}{}^i_0\,u_0\ten e^0_i \,,
\end{align*}
with
$X^{0\lam},\, X^{0}{}^i_0,\,\wti X^{00},\,\wti X^{0i},
\, \wti X^{0}{}^i_0\in \map(\M J_1\f E, \, \B L^{-2} \ten \Rn) \,,$
then we have the equalities
\bal
\wti X^{00} & = - (\alp^0)^2\, \br g_{0\lam}\, X^{0\lam},\quad
\wti X^{0i} =  \br\del^i_\lam\, X^{0\lam},\quad
\wti X^{0}{}^i_0  = X^0{}^i_0 - \Gam\Ga\lam i\, X^{0\lam},
\\
X^{00} & = \wti X^{00} + (\alp^0)^2\, \br g_{0p}\, \wti X^{0p},\quad
X^{0i} = \wti X^{0i} + x^i_0\,(\wti x^{00} 
         + (\alp^0)^2\, \br g_{0p}\, \wti X^{0p},
\\
X^{0}{}^i_0 & = \wti X^{0}{}^i_0 + (\Gam\Ga p i 
         + (\alp^0)^2\, \br g_{0p}\, \br\del^\rho_0\,\Gam\Ga\rho i)
         \, \wti X^{0p} 
         + \br\del^\rho_0\,\Gam\Ga\rho i\, \wti X^{00}  \,.
\end{align*}

 If the coordinate expression of
$\phi$
in a spacetime chart and in a adapted base are
\bal
\phi
&=
\phi_{0\lam}\,u^0\ten d^\lam +  \phi_{0}{}_i^0\,u^0\ten d_0^i
\\
&=
\wti 	\phi_{00}\,u^0\ten \eps^0 + \wti 	\phi_{0i}\,u^0\ten \eps^i
+ \wti \phi_{0}{}_i^0\,u^0\ten \eps_0^i\,,
\end{align*}
with
$\phi_{0\lam},\, \phi_{0}{}_i^0,\,\wti \phi_{00},\,
\wti \phi_{0i},
\, \wti \phi_{0}{}_i^0\in \map(\M J_1\f E, \B L^{2} \ten \Rn) \,,$
then we have the equalities
\bal
\wti \phi_{00} 
&=
\br\del^\lam_0\,(\phi_{0\lam} + \phi_0{}_p^0\, \Gam\Ga\lam p) \,,
\qquad
\wti \phi_{0i} 
= \phi_{0i}
+ \phi_0{}_p^0\, \Gam\Ga i p
+ (\alp^0)^2\,\br g_{0i}\, \br\del^\lam_0\,(\phi_{0\lam}
+ \phi_0{}_p^0\, \Gam\Ga\lam p) \,,
\qquad
\\
\wti \phi_{0}{}_i^0 & = \phi_0{}_i^0 ,
\\
\phi_{0\lam} 
&= 
- (\alp^0)^2\, \br g_{0\lam}\, \wti \phi_{00} 
- \br\del^p_\lam\,\wti \phi_{0p}
-  \wti \phi_0{}_p^0\, \Gam\Ga\lam p \,,
\qquad
\phi_{0}{}_i^0 = \wti \phi_{0}{}_i^0  \,.
\end{align*}
\vglue-1.5\baselineskip{\ }\hfill\ENDE
\eCr

\bPr
 The musical morphisms 
$\Lam\Sha : T^*\M J_1\f E \to \B T \ten \B L^{-2} \ten T\M J_1\f E$
and 
$\Ome\Fla : T\M J_1\f E \to \B T^* \ten \B L^{2} \ten T^*\M J_1\f E$
can be naturally identified, respectively, with the morphisms
$\Lam\Sha : \B T^* \ten \B L^{2} \ten T^*\M J_1\f E \to T\M J_1\f E$ 
and 
$\Ome\Fla : \B T\ten\B L^{-2} \ten T\M J_1\f E \to T^*\M J_1\f E \,.$

 Moreover, these morphisms can be naturally extended, respectively, to
the morphisms, denoted by the same symbols,
$\Lam\Sha : \B T^* \ten \B L^{2} \ten T^*\M J_1\f E
	\to \B T \ten \B L^{-2} \ten T\M J_1\f E$
and \linebreak
$\Ome\Fla : \B T \ten \B L^{-2} \ten T\M J_1\f E
	\to \B T^* \ten \B L^{2} \ten T^*\M J_1\f E \,.$

 Furthermore, the restrictions 
$\Lam\Sha : V^*_\gam\M J_1\f E \to V_\tau\M J_1\f E$
and
$\Ome\Fla : V_\tau\M J_1\f E \to V^*_\gam\M J_1\f E$
are mutually inverse isomorphisms.
\hfill\ENDE
\ePr

\bPr
 We have the morphisms
\bat{4}
\gam\Sha 
&: \B T^* \ten \B L^{2} \ten T^*\M J_1\f E
&&	\to \B T \ten \B L^{-2} \ten T\M J_1\f E
&&:
\phi 
&&\mto \phi(- \tfr1{c^2} \, \gam) \, (- \tfr1{c^2} \, \gam) \,,
\\
\tau\Fla
&: \B T \ten \B L^{-2} \ten T\M J_1\f E
&&	\to \B T^* \ten \B L^{2} \ten T^*\M J_1\f E
&&:
X 
&&\mto (- {c^2} \, \tau)(X) \, (- {c^2} \, \tau) \,.
\end{alignat*}

Moreover, the restrictions
$\gam\Sha: H^*_\tau\M J_1\f E\to H_\gam\M J_1\f E$
and 
$\tau\Fla: H_\gam\M J_1\f E\to H^*_\tau\M J_1\f E$
are mutually inverse isomorphisms 
given by
$f(-c^2\,\tau)\mto f(- \tfr1{c^2}\,\gam)$
and
$f(- \tfr1{c^2}\,\gam) \mto f(-c^2\,\tau) \,,$ 
respectively,
with $f\in\map(\M J_1\f E, \Rn) \,.$
\hfill\ENDE
\ePr

\bNt
 By considering the Planck constant
$\h$
and a particle of mass
$m \,,$
we define also the rescaled morphisms, denoted by the same symbol,
\bat{3}
\gam\Sha 
&: T^*\M J_1\f E 
&&	\to \B T \ten \B L^{-2} \ten T\M J_1\f E 
&&:
\vphi \mto \fr{\h}{m \, c^4} \, \gam(\vphi) \, \gam \,,
\\
\tau\Fla 
&: T\M J_1\f E 
&&\to \B T^* \ten \B L^{2} \ten T^*\M J_1\f E 
&&:
Y \mto \fr{m \, c^4}{\h} \, \tau(Y) \, \tau \,.
\end{alignat*}
\vglue-1.5\baselineskip{\ }\hfill\ENDE
\eNt

\bPr
 The phase tangent and cotangent splittings and the isomorphisms 
$\gam\Sha_{|H^*_\tau\M J_1\f E} \,,$
$\Lam\Sha_{|V^*_\gam\M J_1\f E} \,,$
$\tau\Fla_{|H_\gam\M J_1\f E}$ 
and 
$\Ome\Fla_{|V_\tau\M J_1\f E}$ 
define the mutually inverse isomorphisms
\bat{2}
{}\Sha 
&: \B T^* \ten \B L^{2} \ten T^*\M J_1\f E
&&	\to \B T \ten \B L^{-2} \ten T\M J_1\f E \,,
\\
{}\Fla 
&: \B T \ten \B L^{-2} \ten T\M J_1\f E
&&	\to \B T^* \ten \B L^{2} \ten T^*\M J_1\f E \,,
\end{alignat*}
hence, the mutually inverse rescaled isomorphisms
\bat{2}
{}\Sha
&: T^*\M J_1\f E
	\to \B T \ten \B L^{-2} \ten T\M J_1\f E \,,
\\
{}\Fla
&: \B T \ten \B L^{-2} \ten T\M J_1\f E 
	\to T^*\M J_1\f E\,,
\\[3mm]
{}\Sha
&: \B T^* \ten \B L^{2} \ten T^*\M J_1\f E 
	\to T\M J_1\f E\,,
\\
{}\Fla
&: T\M J_1\f E
	\to \B T^*\ten\B L^{2} \ten T^*\M J_1\f E \,.
\end{alignat*} 
\vglue-1.5\baselineskip{\ }\hfill\ENDE
\ePr

\bCr
 In the adapted bases \eqref{Eq: der to e} and 
\eqref{Eq: d to eps} we have
\bgt
{}\Sha(\eps^0) \byd \fr{\h}{m\, c^4} \,\gam(\eps^0) \, \gam =
\fr{\h\,(\alp^0)^2}{m\,c^2} \, e_0 \,,
\qquad
{}\Sha(\eps^i) \byd i_{\eps^i}\Lam =
\fr1{c\,\alp^0}\,\ha g^{ij}\, e^0_j \,,
\\ 
{}\Sha(\eps^i_0) \byd i_{\eps^i_0}\Lam =
- \fr1{c\,\alp^0} \, \ha g^{ij} \, e_j \,,
\qquad
\end{gather*}
and 
\bgt
{}\Fla(e_0) \byd \fr{m\,c^4}{\h} \,\tau (e_0) \, \tau =
\fr{m\,c^2}{\h\,(\alp^0)^2} \, \eps^0 \,,
\qquad 
{}\Fla(e_i) \byd i_{e_i}\Ome = 
- {c\,\alp^0} \, \ha g_{ij} \, \eps^0_j\,,
\\  
{}\Fla(e_i^0) \byd i_{e_i^0}\Ome = 
					{c\,\alp^0}\,\ha g_{ij}\, \eps^j\,.
\end{gather*}

 Thus, if 
$\vphi \in \sec(\M J_1\f E, \, T^*\M J_1 \f E)$ 
and
$Y \in \sec(\M J_1\f E, \, T^*\M J_1 \f E)$ 
have the coordinate expressions
$\vphi = \wti \vphi_\lam \, \eps^\lam + \wti \vphi_i^0 \, \eps^i_0$
and 
$Y = \wti Y^\lam \, e_\lam + \wti Y^i_0 \, e^0_i \,,$
with
$\wti \vphi_\lam, \wti \vphi_i^0, \wti Y^\lam, \wti Y^i_0
\in \map (\M J_1\f E, \Rn) \,,$
then
\bal
{}\Sha(\vphi) 
&= 
\fr{\h\,(\alp^0)^2}{m\,c^2} \, \wti\vphi_0 \, e_0
- \fr1{c\,\alp^0} \, \ha g^{ij} \, \wti\vphi_j^0 \, e_i
+ \fr1{c\,\alp^0} \, \ha g^{ij} \, \wti\vphi_j \, e^0_i
\\
{}\Fla(Y) 
&= 
\fr{m\,c^2}{\h\,(\alp^0)^2} \, \wti Y^0 \, \eps^0
- {c\,\alp^0}\, \ha g_{ij} \, \wti Y^j_0 \, \eps^i
+ {c\,\alp^0}\, \ha g_{ij} \, \wti Y^j \, \eps^i_0 \,.
\end{align*}
\vglue-1.5\baselineskip{\ }\hfill\ENDE
\eCr
\subsection{Almost--cosymplectic--contact structure}
\label{Almost--cosymplectic--contact structure}
Then, we study the conditions by \\ which
$\Ome$
is closed.

\bLm\label{Lm4.1}
 The scaled phase 2--form
$\Ome$
is closed if and only if the following conditions are satisfied
\bGt\label{Eq4.1}
  \der_\lam (\alp^0 \, \br g_{i\mu})
+ \der^0_i (\alp^0 \, \br g_{j\mu} \, \Gam\Ga\lam j)
- \der_\mu (\alp^0 \, \br g_{i\lam})
- \der^0_i (\alp^0 \, \br g_{j\lam} \, \Gam\Ga\mu j)
=
0 \,,
\\
\label{Eq4.2}
\br g_{j\mu} \, R\Gcu\nu\lam j +
\br g_{j\nu} \, R\Gcu\lam\mu j +
\br g_{j\lam} \, R\Gcu\mu\nu j =
0 \,.
\end{gather}
\eLm

\begin{proof}
 From the coordinate expression of
$\Ome$
we obtain
\bAl\label{Eq4.5}
d\Ome
&=
c \, \der^0_i (\alp^0 \, \br g_{j\mu}) \, 
d^i_0 \wed d^j_0 \wed d^\mu 
- c\, \big(\der_\lam (\alp^0 \, \br g_{i\mu})
+ \der^0_i (\alp^0 \, \br g_{j\mu} \, \Gam\Ga\lam j)\big) \,
d^i_0 \wed d^\lam \wed d^\mu
\\ \nonumber
&\quad
-
c \, \der_\nu(\alp^0 \, \br g_{j\mu} \, \Gam\Ga\lam j) \,
d^\nu \wed d^\lam \wed d^\mu\,.
\end{align}

 Hence,
$d\Ome = 0$
if and only if
\bAl\label{Eq4.6}
\der^0_i (\alp^0 \, \br g_{j\mu}) - \der^0_j(\alp^0 \, \br g_{i\mu})
&=
0\,,
\\[2mm] \label{Eq4.7}
  \der_\lam (\alp^0 \, \br g_{i\mu})
+ \der^0_i (\alp^0 \, \br g_{j\mu} \, \Gam\Ga\lam j)
- \der_\mu (\alp^0 \, \br g_{i\lam})
- \der^0_i (\alp^0 \, \br g_{j\lam} \, \Gam\Ga\mu j)
&=
0\,,
\\[2mm] \label{Eq4.8}
  \der_\lam (\alp^0 \, \br g_{j\mu} \, \Gam\Ga\nu j)
- \der_\mu  (\alp^0 \, \br g_{j\lam} \, \Gam\Ga\nu j)
+ \der_\nu  (\alp^0 \, \br g_{j\lam} \, \Gam\Ga\mu j)
- \der_\lam (\alp^0 \, \br g_{j\nu} \, \Gam\Ga\mu j)
+
&
\\ \nonumber
+ \der_\mu(\alp^0 \, \br g_{j\nu} \, \Gam\Ga\lam j)
- \der_\nu(\alp^0 \, \br g_{j\mu} \, \Gam\Ga\lam j)
&=
0 \,.
\end{align}

 We have
\beq
\der^0_i(\alp^0 \br g_{j\mu}) = (\alp^0)^3
(
\br g_{0i} \, \br g_{j\mu} +
\br g_{0j} \, \br g_{i\mu} +
\ha g_{ij} \, \br g_{0\mu}\big) \,.
\eeq
 Hence,
$\der^0_i(\alp^0 \br g_{j\mu})$
is symmetric with respect to
$i,j \,,$
which implies that \eqref{Eq4.6} is satisfied identically.

 On the other hand, we can rewrite \eqref{Eq4.8} as
\bAl\label{star}
  \alp^0 \, \br g_{j\mu} \, 
(\der_\lam \Gam\Ga\nu j - \der_\nu \Gam\Ga \lam j)
+ \alp^0 \, \br g_{j\nu} \, 
(\der_\mu \Gam\Ga\lam j - \der_\lam \Gam\Ga \mu j)
+ \alp^0 \, \br g_{j\lam} \, 
(\der_\nu \Gam\Ga\mu j - \der_\mu \Gam\Ga \nu j)
+ &
\\ \nonumber
+ \big(
 \der_\lam (\alp^0 \, \br g_{j\mu})
- \der_\mu (\alp^0 \, \br g_{j\lam})\big) \, \Gam\Ga\nu j
+ \big(\der_\nu(\alp^0 \, \br g_{j\lam})
- \der_\lam (\alp^0 \, \br g_{j\nu})\big) \, \Gam\Ga\mu j
+ &
\\ \nonumber
+ \big(\der_\mu (\alp^0 \,\br g_{j\nu})
- \der_\nu (\alp^0 \, \br g_{j\mu})\big) \, \Gam\Ga\lam j
& = 0
\end{align}
and \eqref{Eq4.7} as
\bEq\label{star star}
\der_\lam (\alp^0 \br g_{i\mu}) - \der_\mu(\alp^0 \br g_{i\lam}) =
\der^0_i (\alp^0 \, \br g_{j\lam} \, \Gam\Ga\mu j) -
\der^0_i (\alp^0 \, \br g_{j\mu} \, \Gam\Ga\lam j) \,.
\eEq

 Then, by inserting \eqref{star star} and its permutations into
\eqref{star}, we get
\bal
\big(
  \der^0_h (\alp^0 \, \br g_{j\lam} \, \Gam\Ga\mu j)
- \der^0_h (\alp^0 \, \br g_{j\mu} \, \Gam\Ga\lam j)\big) \,
  \Gam\Ga\nu h
+ \big(\der^0_h (\alp^0 \, \br g_{j\nu} \, \Gam\Ga\lam j)
- \der^0_h (\alp^0 \, \br g_{j\lam} \, \Gam\Ga\nu j)\big) \,
  \Gam\Ga\mu h
+
&
\\
+ \big(
  \der^0_h(\alp^0 \, \br g_{j\mu} \, \Gam\Ga\nu j)
- \der^0_h(\alp^0 \, \br g_{j\nu} \, \Gam\Ga\mu j)\big) \,
  \Gam\Ga\lam h
+ \alp^0 \, \br g_{j\mu} \, (\der_\lam \Gam\Ga\nu j
- \der_\nu \Gam\Ga \lam j)
+
&
\\
+ \alp^0 \, \br g_{j\nu} \, 
(\der_\mu \Gam\Ga\lam j - \der_\lam \Gam\Ga \mu j)
+ \alp^0 \, \br g_{j\lam} \, 
(\der_\nu \Gam\Ga\mu j - \der_\mu \Gam\Ga \nu j)
&=
\\[2mm]
  \alp^0 \, \br g_{j\mu} \, 
( \der_\lam \Gam\Ga\nu j
- \der_\nu \Gam\Ga \lam j
+ \Gam\Ga \lam h \der^0_h \Gam\Ga\nu j
- \Gam\Ga \nu h \der^0_h \Gam\Ga\lam j)
+
&
\\
+ \alp^0 \, \br g_{j\nu} \, 
( \der_\mu \Gam\Ga\lam j
- \der_\lam \Gam\Ga \mu j
+ \Gam\Ga \mu h \der^0_h \Gam\Ga\lam j
- \Gam\Ga \lam h \der^0_h \Gam\Ga\mu j)
+
&
\\
+ \alp^0 \, \br g_{j\lam} \, (\der_\nu \Gam\Ga\mu j
- \der_\mu \Gam\Ga\nu j
+ \Gam\Ga\nu h \der^0_h \Gam\Ga\mu j
- \Gam\Ga\mu h \der^0_h \Gam\Ga\nu j)
+
&
\\
+ (\Gam\Ga \lam h \, \Gam\Ga \kap j
- \Gam\Ga \kap h \, \Gam\Ga \lam j ) \,
  \der^0_q (\alp^0 \, \br g_{j\mu})
+ (\Gam\Ga \mu h \, \Gam\Ga \lam j
- \Gam\Ga \lam h \, \Gam\Ga \mu j ) \,
 \der^0_h (\alp^0 \, \br g_{j\nu})
  +
&
\\
+ (\Gam\Ga \nu h \, \Gam\Ga \mu j
- \Gam\Ga \mu h \, \Gam\Ga \nu j ) \,
  \der^0_h (\alp^0 \, \br g_{j\lam})
&=
\\[2mm]
  \alp^0 \, \big( \br g_{j\mu} \, R\Gcu\nu\lam j
+ \br g_{j\nu} \, R\Gcu\lam\mu j
+ \br g_{j\lam} \, R\Gcu\mu\nu j\big)
&=
0 \,.
\end{align*}

 Hence, \eqref{Eq4.7} and \eqref{Eq4.8}
are equivalent to
\eqref{Eq4.1} and \eqref{Eq4.2}. 
\end{proof}

\bNt\label{Nt4.2}
 By using the above computations we obtain the coordinate expression
\bal
d\Ome 
& =
\tfr12 c \, \alp^0 \br g_{j\mu} \, R\Gcu\nu\lam j \,
d^\nu \wed d^\lam \wed d^\mu 
-
c \, \big(\alp^0 \, \br g_{j\mu} \, \der^0_i \Gam\Ga\lam j +
\der_\lam (\alp^0 \, \br g_{i\mu})\big) \,
(d^i_0 - \Gam\Ga\nu i \, d^\nu) \wed d^\lam \wed d^\mu
\\
&\quad
+ c \, \der^0_i(\alp^0 \, \br g_{j\mu}) \,
d^i_0 \wed (d^j_0- \Gam\Ga\lam j d^\lam) \wed d^\mu
\,.
\end{align*}
\vglue-1.5\baselineskip{\ }\hfill\ENDE
\eNt

 Now, we provide a geometric interpretation of identities
\eqref{Eq4.1} and \eqref{Eq4.2}. 
 For this purpose, let us note that we can define the Lie derivatives
of 
$\tau$ 
with respect to 
$\Gam$ 
and 
$R$ 
according to \eqref{definitions of Lie derivatives}.

\bPr\label{Pr4.3}
 The scaled phase 2--form 
$\Ome$ 
is closed if and only if
\bEq\label{conditions for cosymplectic structure; general case}
L_{\nu_\tau(X)} \, L_\Gam \, \tau = 0 \,,
\quad
\Al \, X \in \sec(\f E,T\f E ) \,,
\ssep{and}
L_R \, \tau = 0 \,.
\eEq
\ePr

\begin{proof}
 The sections
\beq
L_\Gam \, \tau : \M J_1\f E \to \B T \ten \Lam^2 T^*\f E 
\ssep{and}
L_R \, \tau : \M J_1 \f E \to \B T \ten \Lam^3 T^*\f E
\eeq
have the coordinate expressions
\bAl\label{Eq4.9}
L_\Gam \, \tau 
&=
\big(\der_\lam\tau_\mu +\Gam\Ga\lam j \, \der^0_j\tau_\mu \big)\, 
d^\lam\wed d^\mu =
- \tfr1c \, \big(\der_\lam (\alp^0 \, \br g_{0\mu}) 
	+ \Gam\Ga\lam j \, \der^0_j (\alp^0 \, \br g_{0\mu}) \big)\, 
d^\lam\wed d^\mu
\\ \nonumber
&=
- \tfr1c \, \big(\der_\lam (\alp^0\, \br g_{0\mu}) 
	+ \alp^0 \, \br g_{j\mu} \, \Gam\Ga\lam j \big) \, 
d^\lam\wed d^\mu
\\ \nonumber
&=
\tfr{\alp^0}{c} \big(
\der_\mu\br g_{0\lam} - g_{j\mu}\, \Gam\Ga\lam j +
(\alp^0)^2 \br g_{0\lam} \,
(\tfr 12 \der_\mu {\ha g_{00}} + \br g_{0p} \, \Gam\Ga\mu p)
\big) \, d^\lam\wed d^\mu
\end{align}
and
\bEq\label{Eq4.11}
L_R \, \tau =  
- \tfr{\alp^0}{c} \, \br g_{i\lam}\, R\Gcu\mu\nu i \,
d^\lam \wed d^\mu \wed d^\nu =
- \tfr{\alp^0}{c} \, \ha g_{ij} \, R\Gcu\mu\nu i \,
\bet^j \wed d^\mu \wed d^\nu \,.
\eEq

 Hence,
$L_R \, \tau = 0$ 
if and only if the identity \eqref{Eq4.2} is satisfied.

 Then, for each 
$X = X^\lam \, \der_\lam \,,$
we have the coordinate expression
\bAl\label{coordinate expression of L X L Gam tau} 
L_{\nu_\tau(X)} \, L_\Gam \, \tau 
&=
\tfr1{c\,\alp^0}\, \big[\der^0_i \der_\lam \tau_\mu
+ \der^0_i(\Gam\Ga\lam j \, \der^0_j \tau_\mu)\big]
		\, \ti X^i \, d^\lam \wed d^\mu
\\ \nonumber
&=   
- \tfr1{c^2\,\alp^0}\, \big[\der_\lam \der^0_i(\alp^0 \, 		\br g_{0\mu})
+ \der^0_i 
\big(\Gam\Ga\lam j \, \der^0_j (\alp^0 \, 		\br g_{0\mu})\big)
\big] \, \ti X^i \, d^\lam \wed d^\mu \,,
\end{align}
where 
$\ti X^i = X^i - x^i_0 \, X^0 \,.$

 Moreover, we have
\beq
\der_\lam\der^0_i (\alp^0 \, \br g_{0\mu})
+ \der^0_i (\Gam\Ga\lam j \, \der^0_p(\alp^0 \, \br g_{0\mu}))
=
\der_\lam (\alp^0 \, \br g_{i\mu}) + 
\der^0_i (\alp^0 \, \br g_{j\mu} \,
\Gam\Ga\lam j) \,,
\eeq
which implies
\bEq\label{Eq: Lie nu Lie Gam tau}
L_{\nu_\tau(X)} \, L_\Gam \, \tau =
- \tfr1{c^2 \, \alp^0} \, \big[\der_\lam(\alp^0 \, \br g_{i\mu})
+ \der^0_i(\alp^0 \, \br g_{j\mu} \, \Gam\Ga\lam j)\big]
\, \ti X^i \, d^\lam \wed d^\mu \,.
\eEq

 Hence 
$L_{\nu_\tau(X)} \, L_\Gam \, \tau = 0$
if and only if \eqref{Eq4.1} is satisfied.

 Thus, Lemma \ref{Lm4.1} implies the Proposition. 
\end{proof}

\bTh\label{Conditions ACC}
 The pair
$(- c^2 \, \tau \,, \, \Ome)$
is a scaled almost--cosymplectic--contact structure if and only if
$g$ 
and
$\Gam$
fulfill the conditions
\eqref{conditions for cosymplectic structure; general case}.
\hfill\ENDE
\eTh
\subsection{Almost--coPoisson--Jacobi structure}
\label{Almost--coPoisson--Jacobi structure: general case}
 First, we compute the Schouten bracket between
$\gam$
and
$\Lam \,.$

\bLm\label{Lm4.6}
 For {\em any} dynamical phase connection
$\gam : \M J_1\f E \to \B T^* \ten T\M J_1\f E \,,$
the scaled 2-vector
\beq
[\gam, \Lam] : 
\M J_1\f E \to \B L^{-2} \ten \Lam^2T\M J_1\f E
\eeq
has the coordinate expression
\bAl\label{Eq: [gam,Lam] general}
	[\gam, \Lam] 
& =
\big[
- (\alp^0)^2 \, \big(
  \tfr 12 \br g^{j\lam} \, \br\del^\rho_0 \, \der_\rho {\ha g_{00}} +
\br g^{j\lam} \, \gam^p_{00} \, \br g_{0p} + \br\del^\lam_0 \,
\gam^j_{00}
\big)
\\
&\qquad\nonumber
- 2 \, g^{0\lam} \, \gam^j_{00}
+ \br\del^\rho_0 \, \der_\rho \br g^{j\lam}
- \br g^{p\lam} \, \der^0_p \gam^j_{00}
\big] \,
(\der_\lam + \Gam\Ga\lam i \, \der^0_i) \wed \der^0_j
\\ \nonumber
&\quad+
\big[
- (\alp^0)^2 \, \big(
  \tfr 12 \, \br g^{j\rho}\, \br\del^\lam_0 \, \der_\rho {\ha g_{00}}
+ \br\del^\lam_0 \, \br g^{j\nu} \, \Gam\Ga\nu p \, \br g_{0p}
- \br\del^\lam_0 \, \br\del^\nu_0 \, \Gam\Ga\nu j
\big)
\\
&\qquad\nonumber
+ g^{\lam\nu} \, \Gam\Ga\nu j
- \del^\lam_p \, \br g^{j\nu} \, \Gam\Ga\nu p
\big] \,
\der_\lam \wed \der^0_j
\\ \nonumber
&\quad +
\big[
- (\alp^0)^2 \, \big(
  \tfr 12 \br g^{j\rho} \, \gam^i_{00} \, \der_\rho {\ha g_{00}}
+ \gam^i_{00} \, \br g^{j\lam} \, \br g_{0p}\, \Gam\Ga\lam p
\big)
\\
&\qquad\nonumber
+ \br g^{j\lam} \, \br\del^\rho_0 \, \der_\rho \Gam\Ga\lam i
+ \br g^{j\lam} \gam^p_{00} \, \der^0_p\,\Gam\Ga\lam i
- \br g^{j\lam} \, \der_\lam\gam^i_{00}
+ \br g^{i\lam} \, \Gam\Ga\lam p \,\der^0_p \gam^j_{00}
\big] \,
\der^0_i\wed \der^0_j \,.
\end{align}
\eLm

\begin{proof}
 We have
\bal
[\gam, \Lam] 
&=
\big[
- (\alp^0)^2 \, \big(
  \tfr 12 (\br g^{j\lam}\, \br\del^\rho_0
+ \br g^{j\rho} \, \br\del^\lam_0) \, \der_\rho {\ha g_{00}}
+ \br g^{j\lam} \, \gam^p_{00} \, \br g_{0p}
\\
&\qquad
+ \br\del^\lam_0 \,\gam^j_{00}
+ \br\del^\lam_0 \, \br g^{j\nu} \, \Gam\Ga\nu p \, \br g_{0p}
- \br\del^\lam_0 \, \br\del^\nu_0 \, \Gam\Ga\nu j 
\big)
\\
&\qquad
- 2 \, g^{0\lam} \, \gam^j_{00}
+ g^{\lam\nu} \, \Gam\Ga\nu j
+ \br\del^\rho_0 \, \der_\rho \br g^{j\lam}
- \del^\lam_p \, \br g^{j\nu} \, \Gam\Ga\nu p
- \br g^{p\lam} \,\der^0_p \gam^j_{00}
\big] \, 
\der_\lam \wed \der^0_j
\\
&\quad+
\big[
- (\alp^0)^2 \, \big(
  \br g^{j\lam} \,\Gam_\lam{}^i_0 \,
  (\tfr 12 \, \br\del^\rho_0 \, \der_\rho {\ha g_{00}}
+ \gam^p_{00} \, \br g_{0p})
+ \tfr 12 \br g^{j\rho} \, \gam^i_{00} \, \der_\rho {\ha g_{00}}
\\
&\qquad
+ \gam^i_{00} \, \br g^{j\lam} \, \Gam\Ga\lam p \, \br g_{0p}
+ \gam^j_{00} \, \br\del^{\lam}_0 \,\Gam\Ga\lam i
\big)
- 2 \, \gam^j_{00} \, g^{0\lam} \,\Gam\Ga\lam i
\\
&\qquad
+ \Gam\Ga\lam i \, \br\del^\rho_0 \, \der_\rho\br g^{j\lam}
+ \br g^{j\lam} \, \br\del^\rho_0 \, \der_\rho \Gam\Ga\lam i
+ \br g^{j\lam} \gam^p_{00} \, \der^0_p\,\Gam\Ga\lam i
- \br g^{j\lam} \, \der_\lam\gam^i_{00}
\\
&\qquad
+ \br g^{i\lam} \, \Gam\Ga\lam p \,\der^0_p \gam^j_{00}
- \br g^{p\lam} \,\Gam\Ga\lam i \, \der^0_p \gam^j_{00}
\big] \, 
\der^0_i\wed \der^0_j
\\[2mm]
&=
\big[
- (\alp^0)^2 \, \big(
  \tfr12 \br g^{j\lam}\, \br\del^\rho_0 \, \der_\rho {\ha g_{00}}
+ \br g^{j\lam} \, \gam^p_{00} \, \br g_{0p}
+ \br\del^\lam_0 \,\gam^j_{00}
\big)
\\
&\quad
- 2 \, g^{0\lam} \, \gam^j_{00}
+ \br\del^\rho_0 \, \der_\rho \br g^{j\lam}
- \br g^{p\lam}\,\der^0_p \gam^j_{00}
\big] \,
 (\der_\lam + \Gam\Ga\lam i \, \der^0_i) \wed \der^0_j
\\
&\quad
+ \big[
- (\alp^0)^2 \, \big(
  \tfr12 \, \br g^{j\rho} \, \br\del^\lam_0 \, \der_\rho {\ha g_{00}}
+ \br\del^\lam_0 \, \br g^{j\nu} \, \Gam\Ga\nu p \, \br g_{0p}
- \br\del^\lam_0 \, \br\del^\nu_0 \, \Gam\Ga\nu j
\big)
\\
&\qquad
+ g^{\lam\nu} \, \Gam\Ga\nu j
- \del^\lam_p \, \br g^{j\nu} \, \Gam\Ga\nu p
\big] \, 
\der_\lam \wed \der^0_j
\\
&\quad
+ \big[
- (\alp^0)^2 \, \big(
  \tfr 12 \br g^{j\rho} \, \gam^i_{00} \, \der_\rho {\ha g_{00}}
+ \gam^i_{00} \, \br g^{j\lam} \, \Gam\Ga\lam p \, \br g_{0p}
\big)
\\
&\quad
+ \br g^{j\lam} \, \br\del^\rho_0 \, \der_\rho \Gam\Ga\lam i
+ \br g^{j\lam} \gam^p_{00} \, \der^0_p\,\Gam\Ga\lam i
- \br g^{j\lam} \, \der_\lam\gam^i_{00}
+ \br g^{i\lam} \, \Gam\Ga\lam p \,\der^0_p \gam^j_{00}
\big] \, 
\der^0_i\wed \der^0_j \,. 
\end{align*}
\vglue-1.5\baselineskip
\end{proof}

\medskip
 Now, let us go back to the particular case when 
$\gam \byd \gam[\Gam] \,.$

\bPr\label{Pr4.15}
 We have 
\beq
i_{[\gam,\Lam]} \, \Ome = - i_{\gam \wed \Lam} \, d\Ome\,.
\eeq
\ePr

\begin{proof}
 The Schouten bracket is characterised by the following identity
\cite{Vai94}, for each 2-form 
$\bet \,,$
\beq
i_{[\gam,\Lam]} \, \bet = i_\gam \, di_\Lam \, \bet
	- i_\Lam \, di_\gam \, \bet - i_{\gam \wed \Lam} \, d\bet \,.
\eeq

 Then, our claim follows from the equality
$i_\gam \, \Ome = 0$ 
and Lemma \ref{Lm3.3}. 
\end{proof}

\bLm\label{condition for Schouten of gam and Lam; general case}
 We have the coordinate expression
\bAl\label{Eq: [gam[Gam],Lam] general}
[\gam, \Lam] 
& =
\big[
- (\alp^0)^2 \, (\br g^{j\lam} \, \br\del^\rho_0
+ \br g^{j\rho} \, \br\del^\lam_0) \, (\tfr12 \der_\rho {\ha g_{00}}
+ \br g_{0p} \, \Gam\Ga\rho p)
- \del^\lam_p \, \br g^{j\nu} \, \Gam\Ga\nu p
\\ \nonumber
&\qquad + \br\del^\rho_0 \, (\der_\rho \br g^{j\lam}
- g^{0\lam} \, \Gam\Ga\rho j
- \br g^{p\lam}\, \der^0_p \Gam\Ga\rho j)
\big] \,
(\der_\lam + \Gam\Ga\lam i \, \der^0_i) \wed \der^0_j
\\ \nonumber
&\quad
+ \br g^{j\lam} \, \br\del^\rho_0 \, R\Ga{\lam\rho} i \,
\der^0_i \wed \der^0_j \,.
\end{align}
\eLm

\begin{proof}
 The equalities
$\gam^i_{00} = \br\del^\rho_0 \, \Gam\Ga\rho i$
and
$\der^0_p \gam^i_{00} = \del^\rho_p \, \Gam\Ga\rho i +
\br\del^\rho_0 \, \der^0_p \Gam\Ga\rho i$
yield
\bal
[\gam, \Lam]
&=
\big[
- (\alp^0)^2 \, \big((\br g^{j\lam} \, \br\del^\rho_0
+ \br g^{j\rho} \, \br\del^\lam_0) \, 
(\tfr12 \,\der_\rho {\ha g_{00}} + \br g_{0p} \, \Gam\Ga\rho p)
\big)
\\
&\qquad
+ \br\del^\rho_0 \, \der_\rho \br g^{j\lam}
- g^{0\lam} \, \br\del^\rho_0 \, \Gam\Ga\rho j
- \br g^{p\lam} \, \br\del^\rho_0 \, \der^0_p \Gam\Ga\rho j
- \del^\lam_p \, \br g^{j\nu} \, \Gam\Ga\nu p
\big] \,
(\der_\lam + \Gam\Ga\lam i \, \der^0_i) \wed \der^0_j
\\
&\quad
+ \big[
  \br g^{j\lam} \, \br\del^\rho_0 \, \big(
  \der_\rho \Gam\Ga\lam i
- \der_\lam \Gam\Ga\rho i
+ \Gam\Ga\rho p \, \der^0_p \, \Gam\Ga\lam i
- \Gam\Ga\lam p \, \der^0_p \, \Gam\Ga\rho i
\big)
\big] \,
\der^0_i \wed \der^0_j \,. 
\end{align*}
\vglue-1.5\baselineskip
\end{proof}

\bLm\label{gam and Lam; general case adapted}
 In the adapted base \eqref{Eq: der to e} 
we have the coordinate expression
\bAl\label{Eq: [gam,Lam] adapted}
[\gam, \Lam] 
& = 
	- (\alp^0)^2 \,\big( \tfr12\, \br g^{j\rho}  
	\, \der_\rho {\ha g_{00}}
	+ \br g_{0\lam}\,\br\del^\rho_0\,\der_\rho\br g^{j\lam}
	- \br\del^\rho_0\,\Gam\Ga\rho j
	\big)\,e_0 \wed e^0_j
\\
&\quad\nonumber
	- \big[(\alp^0)^2 \, \hat g^{ij}\,\br\del^\rho_0
	\,(\tfr12 \der_\rho {\ha g_{00}}
	+ \br g_{0p} \, \Gam\Ga\rho p)
\\
&\qquad\nonumber
	+ \br g^{j\rho}\,\Gam\Ga\rho i 
	-\br\del^\rho_0\,\big(\der_\rho\hat g^{ji}
	-\br g^{i0}\,\Gam\Ga\rho j
	- \ha g^{pi}\,\der^0_p\Gam\Ga\rho j\big)	
	\big]\,e_i \wed e^0_j
\\
&\quad\nonumber
		- \br g^{j\lam} \, \br\del^\rho_0 \, R\Ga{\lam\rho} i
	\, e^0_i \wed e^0_j\,.
\end{align}
\eLm

\begin{proof}
 It follows from \eqref{Eq: [gam[Gam],Lam] general} and 
\eqref{Eq: der to e}. 
\end{proof}

\bLm\label{Lemma: gam Lie tau Sha}
 We have the coordinate expression
\bAl\label{Eq: gam wed LamShaLie}
	\gam\wed \Lam\Sha(L_\gam \tau)
& =
	(\alp^0)^2\, 
	\big( \br\del^\lam_0(\tfr12\,\br g^{j\rho}\der_\rho\hat g_{00} 
	+ \br g_{0\rho}\,\br\del^\sig_0\,\der_\sig\br g^{j\rho} 
	- \br\del^\sig_0\, \Gam\Ga\sig j)
	\der_\lam\wed\der^0_j
\\
&\quad
	+ \nonumber
		\br\del^\sig_0\,\Gam\Ga\sig i\,\br g^{j\rho} 
	(\tfr12 \der_\rho\hat g_{00} 
	- \br\del^\tau_0\der_\tau\br g_{0\rho})
	\der^0_i\wed\der^0_j\big),
\end{align}
which, in the adapted base \eqref{Eq: der to e},
reads as
\bAl\label{Eq: gam wed LamShaLie adapted}
	\gam\wed \Lam\Sha(L_\gam \tau)
& =
	(\alp^0)^2 
	\big( \tfr12\,\br g^{j\rho}\der_\rho\hat g_{00} 
	+ \br g_{0\rho}\,\br\del^\sig_0\,\der_\sig\br g^{j\rho} 
	- \br\del^\sig_0\, \Gam\Ga\sig j\big)\,
	e_0\wed e^0_j\,.
\end{align}
\eLm

\begin{proof}
 We have
\bal
	L_\gam\tau 
& =
	- \alp^0 \,\br\del^\rho_0\,\big[\der_\rho(\alp^0\br g_{0\lam})
	- \der_\lam(\alp^0\br g_{0\rho})
		+ \der^0_p(\alp^0 \br g_{0\lam}) \Gam\Ga\rho p\big] \, d^\lam
\\
& =
	- (\alp^0)^2\,\br\del^\rho_0\,\big[\der_\rho\br g_{0\lam} 
- \der_\lam\br g_{0\rho}
	+ \br g_{p\lam} \Gam\Ga\rho p 
	+ \tfr12 (\alp^0)^2 (\br g_{0\lam}\der_\rho\hat g_{00} 
		-  \br g_{0\rho}\der_\lam\hat g_{00})\big]\, d^\lam \,.
\end{align*}

 Then, we have
\bal
	\Lam\Sha(L_\gam \tau)
& =
	- \tfr{\alp^0}{c} \br g^{i\lam}\,\br\del^\rho_0\, 
\big[\der_\rho\br g_{0\lam} - \der_\lam\br g_{0\rho}
	+ \br g_{p\lam} \Gam\Ga\rho p 
	+ \tfr12 (\alp^0)^2 (\br g_{0\lam}\der_\rho\hat g_{00} 
		-  \br g_{0\rho}\der_\lam\hat g_{00})\big]\,\der^0_i
\\
& =
			\tfr{\alp^0}{c}\big[\tfr12 \br g^{i\lam} \der_\lam \hat g_{00} +
			\br g_{0\lam} \br\del^\rho_0 \der_\rho \br g^{i\lam} 
			- \Gam\Ga\rho i\br\del^\rho_0\big] \, \der^0_i\,
\end{align*}
and
\bal
	\gam\wed \Lam\Sha(L_\gam \tau)
& =
	(\alp^0)^2 \,
	\br\del^\lam_0\,(\der_\lam + \Gam\Ga\lam i\,\der^0_i)
\wed(\tfr12\,\br g^{j\rho}\,\der_\rho\hat g_{00} 
	+ \br g_{0\rho}\,\br\del^\sig_0\,\der_\sig\br g^{j\rho} 
	- \br\del^\sig_0\, \Gam\Ga\sig j)\,
	\der^0_j\,. 
\end{align*}
\vglue-1.5\baselineskip
\end{proof}

\bCr\label{Cr: gam Lam = gam Lie Sha}
In the adapted base  \eqref{Eq: gam wed LamShaLie}
we have the coordinate expression
\bAl\label{Eq: [gam,Lam] - gam wed LamShaLie}
	[-\fr1{c^2}\gam,\Lam] & - \fr1{c^2}\gam\wed
\Lam\Sha(L_\gam\,\tau) =
	\big[\fr{(\alp^0)^2}{c^2} \, \hat g^{ij}\,\br\del^\rho_0
	\,(\tfr12 \der_\rho {\ha g_{00}}
	+ \br g_{0p} \, \Gam\Ga\rho p)
\\
&\quad\nonumber
	+ \fr1{c^2}\,\big(\br g^{j\rho}\,\Gam\Ga\rho i 
	-\br\del^\rho_0\,\big(\der_\rho\hat g^{ji}
	-\br g^{i0}\,\Gam\Ga\rho j
	- \ha g^{pi}\,\der^0_p\Gam\Ga\rho j\big)\big)	
	\big]\,e_i \wed e^0_j
\\
&\quad\nonumber
		- \fr1{c^2} \br g^{j\lam} \, \br\del^\rho_0 \, R\Ga{\lam\rho} i
	\, e^0_i \wed e^0_j\,.
\end{align}
\eCr

\begin{proof}
 It follows from \eqref{Eq: [gam,Lam] adapted} 
and \eqref{Eq: gam wed LamShaLie adapted}. 
\end{proof}

 Then, we compute the Schouten bracket between
$\Lam$
and
$\Lam \,.$

\bLm\label{Lm4.8}
 The scaled 3-vector
\beq
[\Lam, \Lam] : 
\M J_1\f E \to (\B T^2\ten \B L^{-4}) \ten \Lam^3T\M J_1\f E \,,
\eeq
has the coordinate expression
\bAl\label{[Lam, Lam]}
[\Lam, \Lam] 
& =
\fr2{c^2}
\big[
\br\del^\lam_0 \, \br g^{j\mu} \,
\der_\lam \wed \der_\mu \wed \der^0_i +
(\br\del^\lam_0\, \br g^{j\rho} -
\br\del^\rho_{0}\, \br g^{j\lam} ) \, \Gam\Ga\rho i \,
\, \der_\lam \wed \der^0_i \wed \der ^0_j
\\ \nonumber
&\quad
+ \br\del^\sig_0 \, \br g^{k\rho} \, \Gam\Ga\sig i \, \Gam\Ga\rho j \,
\der^0_i \wed \der^0_j \wed \der^0_k
\big]
\\ \nonumber
&
+ \fr2{c^2}
\big[\br g^{j\rho}\, \br g^{k\lam} \,
(\tfr12 \der_\rho {\ha g_{00}} + \br g_{0p}\, \Gam\Ga\rho p)
\\ \nonumber
&\quad
+ \fr1{(\alp^0)^2} \, \br g^{k\rho}\,(\der_\rho \br g^{j\lam}
g^{0\lam} \, \Gam\Ga\rho j 
- \br g^{p\lam} \, \der^0_p \Gam\Ga\rho j)\big] \,
(\der_\lam + \Gam\Ga\lam i \, \der^0_i) \wed \der^0_j \wed \der ^0_k
\\ \nonumber
&
+ \fr1{(c \, \alp^0)^2} \,
\br g^{i\rho}\, \br g^{j\sig} \, R\Ga{\rho\sig} k \,
\der^0_i \wed \der^0_j \wed \der^0_k \,.
\end{align}
\eLm

\begin{proof}
 We have
\bal
[\Lam, \Lam] 
& =
\tfr2{c^2} \br\del^\lam_0 \, \br g^{j\mu} \,
\der_\lam \wed \der_\mu \wed \der^0_i
\\
&\quad
+ \tfr2{c^2} \, \big[
  \br g^{i\rho} \, \br g^{j\lam} \, 
(\tfr12 \der_\rho {\ha g_{00}} + \br g_{0p} \, \Gam\Ga\rho p)
+ \br\del^\lam_0 \, \br g^{j\rho} \, \Gam\Ga\rho i
- \br\del^\rho_{0} \, \br g^{j\lam} \, \Gam\Ga\rho i
\\
&\qquad
+ \tfr1{(\alp^0)^2} \, \br g^{j\rho}\,\big(\der_\rho \br g^{i\lam}
- g^{0\lam} \, \Gam\Ga\rho i
- \br g^{p\lam}\, \der^0_p \Gam\Ga\rho i
\big) \big]\, 
\der_\lam \wed \der^0_i \wed \der ^0_j
\\
&\quad
+ \tfr2{c^2} \, \big[
  \br g^{i\sig} \, \br g^{j\rho} \, \Gam\Ga\rho k \,
  (\tfr12 \der_\sig{\ha g_{00}}
+ \br g_{0p} \, \Gam\Ga\sig p)
+ \br\del^\sig_0 \, \br g^{k\rho} \, \Gam\Ga\sig i \, \Gam\Ga\rho j
\\
&\qquad
+ \tfr1{(\alp^0)^2} \, \big(\br g^{i\sig} \,
  \der_\sig (\br g^{k\rho}\, \Gam\Ga\rho j)
+ g^{0\rho} \, \br g^{i\sig} \, \Gam\Ga\rho j \, \Gam\Ga\sig k
\\
&\qquad
+ (\br g^{i\sig}\, \Gam\Ga\sig p
- \br g^{p\sig} \,\Gam\Ga\sig i) \,
  \der^0_p (\br g^{k\rho} \, \Gam\Ga\rho j)
\big)\big]
\der^0_i \wed \der^0_j \wed \der^0_k
\\[2mm]
&=
  \tfr2{c^2} \br\del^\lam_0 \, \br g^{j\mu} \,
  \der_\lam \wed \der_\mu \wed \der^0_i
\\
&\quad
+ \tfr2{c^2} \, \big[\br g^{i\rho} \, \br g^{j\lam} \,
   (\tfr12 \der_\rho {\ha g_{00}}
+ \br g_{0p}\, \Gam\Ga\rho p)
+ (\br\del^\lam_0\, \br g^{j\rho}
- \br\del^\rho_{0}\, \br g^{j\lam} )\, \Gam\Ga\rho i
\\
&\qquad
+ \tfr1{(\alp^0)^2} \, \br g^{j\rho}\,\big(\der_\rho \br g^{i\lam}
- g^{0\lam} \, \Gam\Ga\rho i
- \br g^{p\lam}\, \der^0_p \Gam\Ga\rho i
\big)\big] \, 
\der_\lam \wed \der^0_i \wed \der ^0_j
\\
&\quad
+ \tfr2{c^2} \, \big[\br g^{i\sig}\, \br g^{j\rho} \, \Gam\Ga\rho k
\,
  (\tfr12 \der_\sig{\ha g_{00}}
+ \br g_{0p} \, \Gam\Ga\sig p)
+ \br\del^\sig_0 \, \br g^{k\rho} \, \Gam\Ga\sig i  \, \Gam\Ga\rho j
\\
&\qquad
+ \tfr1{(\alp^0)^2} \, \big(
- \tfr12 \br g^{i\sig} \, \br g^{k\rho} \, R\Ga{\sig\rho} j
\\
&\qquad
+ \br g^{i\sig} \, \Gam\Ga\rho j \,(\der_\sig \br g^{k\rho}
- g^{0\rho}\, \Gam\Ga\sig k
- \br g^{p\rho}\, \der^0_p \Gam\Ga\sig k)
\big)\big]
  \der^0_i \wed \der^0_j \wed \der^0_k \,. 
\end{align*}
\vglue-1.5\baselineskip
\end{proof}

\bLm\label{Lam and Lam: general case adapted}
 In the adapted base \eqref{Eq: der to e} we have the coordinate
expression
\bAl\label{Eq: adapted coordinate expression of [Lam, Lam]}
	[\Lam, \Lam] 
&=
	\tfr2{c^2}\,\big[
	\ha g^{kj}\, e_0\wed e_j\wed e^0_k
	+ \br g^{k\rho}\,\big(\Gam\Ga\rho j
	- \br g_{0\lam}\,\der_\rho\br g^{j\lam}
	\big)\,
	e_0 \wed e^0_j \wed e^0_k
\\
&\quad \nonumber
	+ \big( \ha g^{ki} \,\br g^{j\rho}\, 
	(\tfr12 \der_\rho {\ha g_{00}} + \br g_{0p}\, \Gam\Ga\rho p)
\\
&\qquad \nonumber
	+ \tfr1{(\alp^0)^2}\,\br g^{k\rho}\,
		(\der_\rho\ha g^{ji} - \br g^{i0}\,\Gam\Ga\rho j 
			- \ha g^{pi}\,\der^0_p\Gam\Ga\rho j)
	\big)\,
	e_i \wed e^0_j \wed e^0_k
\\
&\quad \nonumber
	+  \tfr1{(\alp^0)^2} \,
	\br g^{i\rho}\, \br g^{j\sig} \, R\Ga{\rho\sig} k 
		\,
	e^0_i \wed e^0_j \wed e^0_k
	\big]\,.
\end{align}
\eLm

\begin{proof}
 It follows from \eqref{[Lam, Lam]} and \eqref{Eq: der to e}. 
\end{proof}

\bLm\label{Lemma: gam Lam Sha dtau}
 In the adapted base \eqref{Eq: der to e}
we have the coordinate expression
\bAl\label{Eq: gam wed LamShadtau}
	\gam\wed(\Lam\Sha\ten\Lam\Sha)(d\tau)
& =
	\fr1{c^2}\,\big[
	\ha g^{jk}\, e_0\wed e_j\wed e^0_k
\\
&\quad\nonumber
	+ \br g^{j\rho}\,\big(\br g_{0\sig}\,\der_\rho\br g^{k\sig}
	- \br g^{j\rho}\,\Gam\Ga\rho k\big)\,e_0\wed e^0_j\wed e^0_k 
\big]\,.
\end{align}
\eLm

\begin{proof}
We have
\bal
	(\Lam\Sha\ten\Lam\Sha)(d\tau)
& =
	\tfr1{c^3\alp^0}\,\big[\br g^{j\lam}\,\der_\lam\wed\der^0_j
\\
&\quad
	- \big(\br g^{i\lam}\,\br g^{j\mu}\,\der_\lam\br g_{0\mu}
	+ \br g^{i\rho}\,\Gam\Ga\rho j -\br g^{j\rho}\,\Gam\Ga\rho i 
	\big)\,\der^0_i\wed\der^0_j\big]
\end{align*}
Then
\bal
	\gam\wed(\Lam\Sha\ten\Lam\Sha)(d\tau)
& =
	\tfr1{c^2}\,\big(
	\br\del^\nu_0\,(\der_\nu + \Gam\Ga\nu k\,\der^0_k
	\big)\wed\big[\br g^{j\lam}\,\der_\lam\wed\der^0_j
\\
&\quad
	- \big(\br g^{i\lam}\,\br g^{j\mu}\,\der_\lam\br g_{0\mu}
	+ \br g^{i\rho}\,\Gam\Ga\rho j -\br g^{j\rho}\,\Gam\Ga\rho i 
	\big)\,\der^0_i\wed\der^0_j\big]
\\
& =
	\tfr1{c^2}\,
	\big[\br\del^\lam_0\,\br g^{k\mu}\,\der_\lam\wed\der_\mu\wed\der^0_k
\\
&\quad
	+\big(- \br\del^\lam_0\,(\br g^{j\rho}\,\br g^{k\sig}\,\der_\rho\br 
g_{0\sig}
	+ \br g^{j\rho}\,\Gam\Ga\rho k -\br g^{k\rho}\,\Gam\Ga\rho j) 
	+\br\del^\rho_0\, \br g^{j\lam}\,\Gam\Ga\rho k
	\big)\,\der_\lam\wed\der^0_j\wed\der^0_k
\\
& \quad
		-\br\del^\rho_0\, \Gam\Ga\rho k\,\big(\br g^{i\lam}\,\br 
g^{j\mu}\,\der_\lam\br g_{0\mu}
	+ \br g^{i\rho}\,\Gam\Ga\rho j -\br g^{j\rho}\,\Gam\Ga\rho i 
	\big)\,\der^0_i\wed\der^0_j\wed\der^0_k\big]\,.
\end{align*}

 Then, in the adapted base \eqref{Eq: der to e} we obtain the
Lemma. 
\end{proof}

\bLm\label{Lemma: Lam Lam - gam wed Lam Sha dtau}
 In the adapted base \eqref{Eq: der to e} 
we have the coordinate expression
\bAl\label{Eq: coordinate expression of [Lam, Lam]}
	[\Lam, \Lam] 
& - 2\,\gam\wed (\Lam\Sha\ten\Lam\Sha)(d\tau) =
	\fr2{c^2}\,\big[
	\big( \ha g^{ki} \,\br g^{j\rho}\, 
	(\tfr12 \der_\rho {\ha g_{00}} + \br g_{0p}\, \Gam\Ga\rho p)
\\
&\quad \nonumber
	+ \fr1{(\alp^0)^2}\,\br g^{k\rho}\,
		(\der_\rho\ha g^{ji} - \br g^{i0}\,\Gam\Ga\rho j 
			- \ha g^{pi}\,\der^0_p\Gam\Ga\rho j)
	\big)\,
	e_i \wed e^0_j \wed e^0_k
\\
&\quad \nonumber
	+  \fr1{(\alp^0)^2} \,
	\br g^{i\rho}\, \br g^{j\sig} \, R\Ga{\rho\sig} k 
		\,
	e^0_i \wed e^0_j \wed e^0_k
	\big]
\end{align}
\eLm

\begin{proof}
 It follows from 
\eqref{Eq: adapted coordinate expression of [Lam, Lam]}
and
\eqref{Eq: gam wed LamShadtau}. 
\end{proof}

\bLm\label{coordinate conditions for Jacobi: general case}
 The pair 
$(- \tfr1{c^2}\,\gam, \Lam)$ 
is a scaled almost--coPoisson--Jacobi structure along with
the fundamental 1--form $-c^2\,\tau$ if and only
if  the following identities are satisfied
\bAl\label{Eq4.12}
	(\alp^0)^2 \, \ha g^{ji} \, \br\del^\rho_0\, 
		(\tfr12 \,\der_\rho {\ha g_{00}}
	+ \br g_{0p} \, \Gam_\rho{}^p_0 )
&
\\ \nonumber
	- \br\del^\rho_0 \, ( \der_\rho \ha g^{ji}
	- \br g^{i0} \,  \Gam_\rho{}^j_{0}
	- \ha g^{pi}\,  \der^0_p \Gam_\rho{}^j_0)
	+ \br g^{j\rho} \, \Gam_\rho{}^i_0
&=
	0\,,
\\[2mm] \label{Eq4.13}
	\br g^{j\lam} \, \br\del^\rho_0 \, R\Ga{\lam\rho} i
	- \br g^{i\lam} \, \br\del^\rho_0 \, R\Ga{\lam\rho} j
&=
	0 \,,
\\[2mm] \label{Eq4.16}
	(\alp^0)^2\,(\br g^{j\rho}\, \ha g^{ki}
	- \br g^{k\rho}\, \ha g^{ji}) \, 
	(\tfr12 \der_\rho{\ha g_{00}}
	+ \br g_{0p}\, \Gam\Ga\rho p)
&
\\ \nonumber
	+ 
	\br g^{k\rho} \, (\der_\rho \ha g^{ji} 
	- \br g^{i0} \, \Gam\Ga\rho j
	- \ha g^{pi} \, \der^0_p \Gam\Ga\rho j)
	- \br g^{j\rho} \, 
	(\der_\rho \ha g^{ki}
	- \br g^{i0} \, \Gam\Ga\rho k
	- \ha g^{pi}\, \der^0_p \Gam\Ga\rho k)
&=
	0\,,
\\[2mm] \label{Eq4.17}
	\br g^{i\sig} \, \br g^{k\rho} \, R\Ga{\rho\sig} j
	+ \br g^{j\sig} \, \br g^{i\rho} \, R\Ga{\rho\sig} k
	+ \br g^{k\sig} \, \br g^{j\rho} \, R\Ga{\rho\sig} i
&=
 0 \,.
\end{align}
\eLm

\begin{proof}
$(- \tfr1{c^2} \, \gam, \Lam)$ 
is a scaled almost--coPoisson--Jacobi structure
along with the fundamental 1--form $- c^2\,\tau$ if and only if 
$[-\tfr1{c^2}\,\gam, \Lam] =
-(-\tfr1{c^2}\,\gam) \wed 
\Lam\Sha\big(L_{-\tfr1{c^2}\gam}\,(- c^2\,\tau)\big) =
\tfr1{c^2}\,\gam\wed\Lam\Sha(L_\Gam\,\tau)$ 
and 
$[\Lam, \Lam] = 
- \tfr2{c^2} \gam \wed (\Lam\Sha\ten\Lam\Sha)(- c^2\, d\tau)
 = 2\, \gam \wed (\Lam\Sha\ten\Lam\Sha)(d\tau)
\,.$

 But, in virtue of Corollary
\ref{Cr: gam Lam = gam Lie Sha},
$[-\tfr1{c^2}\,\gam, \Lam] 
= \tfr1{c^2}\,\gam\wed \Lam\Sha(L_\gam\,\tau)$  
if and only if \eqref{Eq4.12} and \eqref{Eq4.13} are satisfied.
 Moreover, in virtue of Lemma 
\ref{Lemma: Lam Lam - gam wed Lam Sha dtau}, 
$[\Lam, \Lam] = 2\, \gam \wed  (\Lam\Sha\ten\Lam\Sha)(d\tau)$ 
if and only if \eqref{Eq4.16}
and \eqref{Eq4.17} are satisfied. 
\end{proof} 

\bTh\label{Theorem: conditions for AcoPJ structure; general case}
 The pair
$(- \tfr1{c^2} \gam, \Lam)$ 
is a scaled almost--coPoisson--Jacobi structure along with
the fundamental 1--form $-c^2\,\tau$ if and only if
the conditions
\eqref{conditions for cosymplectic structure; general case}
are satisfied.
\eTh

\begin{proof}
{\em 1st proof.}
 By Theorem \ref{Th1.1} the structure 
$(-c^2\,\tau,\Ome)$
is a scaled almost--cosymplectic--contact structure if and only if
the dual structure
$(-\tfr1{c^2}\,\gam,\Lam)$ 
is a scaled almost--coPoisson--Jacobi structure along with
the fundamental 1--form $-c^2\,\tau \,.$ 
 On the other hand, according Theorem \ref{Conditions ACC}, 
$(-c^2\,\tau,\Ome)$
is a scaled almost--cosymplectic--contact structure if and only if
the conditions
\eqref{conditions for cosymplectic structure; general case}
are satisfied.

{\em 2nd proof.} Next, we prove Thorem directly in local coordinates.

 From Lemma \ref{coordinate conditions for Jacobi: general case}
it folows that 
$(- \tfr1{c^2}\gam,\Lam)$
is a scaled almost--coPoisson--Jacobi structure along with
the fundamental 1--form $-c^2\,\tau$ 
if and only if (\ref{Eq4.12}) --
(\ref{Eq4.17})
are satisfied.

 We have the coordinate expression \eqref{Eq4.11}
of
$L_R \, \tau$ 
and, in the adapted base
\eqref{Eq: d to eps},
we obtain
\bal
L_R \, \tau 
& 
=  
- \tfr{2\,\alp^0}{c} \ha g_{ip}\,\br\del^\rho_0\, 
	R\Gcu\rho k p \, \eps^i \wed \eps^0 \wed \eps^k
\\
&\quad
- \tfr{\alp^0}{c} \,\ha g_{ip} \, \big( 	R\Gcu j k p 
+ (\alp^0)^2\, \br\del^\rho_0\,
(\br g_{0j} \, 
	R\Gcu\rho k p + \br g_{0k}\, 
	R\Gcu j \rho  p)\big) \, \eps^i \wed \eps^j \wed \eps^k
\end{align*}

 Then 
${}\Sha(L_R\,\tau) \,,$ 
in the adapted base 
\eqref{Eq: der to e},
has the coordinate expression
\bal
{}\Sha(L_R \, \tau) & =  
- \tfr{2\,\h\,\alp^0}{m\,c^5} \ha g^{pk}\,\br\del^\rho_0\, 
	R\Gcu\rho p i \, e^0_i \wed e_0 \wed e^0_k
\\
&\quad
- \tfr{1}{c^4\,(\alp^0)^2} \,\big(\ha g^{pj}\,\ha g^{qk}\, 	R\Gcu p q i
+   \br g^{0j}\,\ha g^{pk}\, \br\del^\rho_0\, 	R\Gcu\rho p i
+ \br g^{0k}\,\ha g^{pj}\, \br\del^\rho_0\, 	R\Gcu p \rho i
\big)\, e^0_i \wed e^0_j \wed e^0_k\,.
\end{align*}

The identity
$
\ha g^{ip} = \br\del^p_\lam\, \br g^{i\lam}
$
implies
\beq
\ha g^{ip}\,\br\del^\rho_0\, 	R\Gcu\rho p j
= \br g^{i\lam}\,\br\del^\rho_0\, 	R\Gcu\rho\lam j
- \br g^{0i}\,\br\del^\rho_0\,\br\del^\sig_0\,R\Gcu\rho\sig i
= \br g^{i\lam}\,\br\del^\rho_0\, 	R\Gcu\rho\lam j
\eeq
and, similarly,
\beq
\ha g^{pj}\,\ha g^{qk}\, 	R\Gcu p q i
= \br g^{j\lam}\,\br g^{k\mu}\, (	R\Gcu\lam\mu i
 - \del^0_\lam\,\br\del^\rho_0\, R\Gcu\rho\mu i 
 - \del^0_\mu\,\br\del^\rho_0\, R\Gcu\lam\rho i )\,.
\eeq

 Then,
\bal
{}\Sha(L_R \, \tau) & =  
- \tfr{2\,\h\,\alp^0}{m\,c^5} \br g^{k\lam}\,\br\del^\rho_0\, 
	R\Gcu\rho\lam i \, e^0_i \wed e_0 \wed e^0_k
\\
&\quad
- \tfr{1}{c^4\,(\alp^0)^2} \,\big(\br g^{j\lam}\,\br g^{k\mu}
     \, (	R\Gcu\lam\mu i 
     -\del^0_\lam\,\br\del^\rho_0\, R\Gcu\rho\mu i
     -\del^0_\mu\,\br\del^\rho_0\, R\Gcu\lam\rho i)
\\
&\qquad 
     + \br g^{0j}\,\br g^{k\mu}\, \br\del^\rho_0\, R\Gcu\rho\mu i 
    + \br g^{0k}\,\br g^{j\lam}\, \br\del^\rho_0\, 	R\Gcu\lam\rho i
\big)\, e^0_i \wed e^0_j \wed e^0_k
\\
& = 
- \tfr{2\,\h\,\alp^0}{m\,c^5} \br g^{k\lam}\,\br\del^\rho_0\, 
	R\Gcu\rho\lam i \, e^0_i \wed e_0 \wed e^0_k
- \tfr{1}{c^4\,(\alp^0)^2} \,\br g^{j\lam}\,\br g^{k\mu}
     \, 	R\Gcu\lam\mu i  \, e^0_i \wed e^0_j \wed e^0_k
\,.
\end{align*}

 Then, 
$L_R\,\tau = 0$ 
if and only if 
${}\Sha(L_R \, \tau) = 0 \,,$
i.e., if and only if
\eqref{Eq4.13} and \eqref{Eq4.17}
are satisfied.

\smallskip

 Further, let us consider a scaled vector field 
$X \in \sec(\f E, \B L^{-2} \ten T\f E) \,.$
 Then, 
$\nu_\tau(X)\in\sec(\M J_1\f E, \B T\ten\B L^{-2}\ten V\M J_1\f E)
\sub \sec (\M J_1\f E, V_\tau\M J_1\f E)$ 
and there exists a unique 1--form
$\ome\in \sec (\M J_1\f E, V^*_\gam\M J_1\f E)$
such that 
$\nu_\tau(X) = \Lam\Sha(\ome) \,.$
 In the adapted base
\eqref{Eq: der to e}
\beq
\nu_\tau(X) = \tfr1{c\,\alp^0}\, \wti X^i\,e_i^0
= \tfr1{c\,\alp^0}\, \ha g^{ip}\,\wti \ome_p\,e_i^0= \Lam\Sha(\ome)\,.
\eeq

 Next, let us refer to the coordinate expression 
\ref{Eq: Lie nu Lie Gam tau} of
$L_{\nu_\tau(X)} \, L_\Gam \, \tau \,.$
 Then, in the adapted base 
\eqref{Eq: d to eps},
we obtain
\bal
L_{\Lam\Sha(\ome)} \, L_\Gam \, \tau & =
- \tfr1{c^2 \, \alp^0} \, \br\del^\rho_0\,
\big[\der_\rho(\alp^0 \, \br
g_{pj}) + \der^0_p(\alp^0 \, \br g_{jq} \, \Gam\Ga\rho q)
\\
& \qquad
- \der_j(\alp^0 \, \br
g_{p\rho}) - \der^0_p(\alp^0 \, \br g_{q\mu} \, \Gam\Ga j q)
\big]
\, \ha g^{ip}\, \wti \ome_i \, \eps^0 \wed \eps^j
\\
& \quad
- \tfr1{c^2 \, \alp^0} \, \big[\der_j(\alp^0 \, \br g_{pk})
+ \der^0_p(\alp^0 \, \br g_{qk} \, \Gam\Ga j q)
+ (\alp^0)^2\,\br g_{0j}\,\br\del^\rho_0\,
( \der_\rho(\alp^0 \, 
\br g_{pk}) 
\\
& \qquad
+ \der^0_p(\alp^0 \, \br g_{kq} \, \Gam\Ga\rho q) )
+ (\alp^0)^2\,\br g_{0k}\,\br\del^\rho_0\,
 \der^0_p(\alp^0 \, \br g_{q\mu} \, \Gam\Ga j q))
\big] \, \ha g^{pi}\,\wti \ome_i \, \eps^j \wed \eps^k\,.
\end{align*}

 Then
${}\Sha(L_{\Lam\Sha(\ome)} \, L_\Gam \, \tau)$  
has the coordinate expression, in the adapted base 
\eqref{Eq: der to e},
\bal
{}\Sha(L_{\Lam\Sha(\ome)} \, L_\Gam \, \tau) & =
- \tfr{\h}{m\,c^5} \, \ha g^{ks}\,\ha g^{pi}\, \br\del^\rho_0\,
\big[\der_\rho(\alp^0 \, 
\br g_{ps}) + \der^0_p(\alp^0 \, \br g_{sq} \, \Gam\Ga\rho q)
\\
& \qquad
- \der_s(\alp^0 \, \br
g_{p\rho}) - \der^0_p(\alp^0 \, \br g_{q\rho} \, \Gam\Ga s q)
\big] \, \wti \ome_i \, e_0 \wed e^0_j
\\
& \quad
- \tfr1{c^4 \, (\alp^0)^3} \,\ha g^{pi}\,\ha g^{jr}\,\ha g^{ks}\,
\big[ 
\der^0_p(\alp^0 \, \br g_{qs} \, \Gam\Ga r q)
+ (\alp^0)^2\,\br g_{0s}\,\br\del^\rho_0\,
 \der^0_p(\alp^0 \, \br g_{q\mu} \, \Gam\Ga r q))
\\
& \qquad
+ \der_r(\alp^0\,\br g_{ps})
+ (\alp^0)^2\,\br g_{0r}\,\br\del^\rho_0\,
( \der_\rho(\alp^0 \, 
\br g_{ps}) + \der^0_p(\alp^0 \, \br g_{sq} \, \Gam\Ga\rho q) )
\big] \, \wti \ome_i \, e^0_j \wed e^0_k
\\
& = 
- \tfr{\h\,\alp^0}{m\,c^5} \, \big[	(\alp^0)^2 \, \ha g^{ji} \,
\br\del^\rho_0\, 
		(\tfr12 \,\der_\rho {\ha g_{00}}
	+ \br g_{0p} \, \Gam_\rho{}^p_0 )
\\
&\qquad
	- \br\del^\rho_0 \, ( \der_\rho \ha g^{ji}
	- \br g^{i0} \,  \Gam_\rho{}^j_{0}
	- \ha g^{pi}\,  \der^0_p \Gam_\rho{}^j_0)
	+ \br g^{j\rho} \, \Gam_\rho{}^i_0\big]
\, \wti \ome_i \, e_0 \wed e^0_j
\\
&\quad 
- \tfr1{c^4 \, (\alp^0)^2} \,
\big[(\alp^0)^2\,\ha g^{ki}\,\br g^{j\rho}
\, (\tfr12\,\der_\rho\ha g_{00} + \br g_{0q}\,\Gam\Ga\rho q)
\\
&\qquad
- \br g^{j\rho}\, (\der_\rho\ha g^{ki} - \br g^{i0}\,\Gam\Ga\rho k
- \ha g^{pi}\,\der^0_p\Gam\Ga\rho k)\big]\,
\wti \ome_i \, e^0_j \wed e^0_k
\end{align*}
Then $L_{\nu_\tau(X)} \, L_\Gam \, \tau=0$
if and only if $L_{\Lam\Sha(\ome)} \, L_\Gam \, \tau=0 \,,$ 
i.e., if and only if the equalities
\eqref{Eq4.12} and \eqref{Eq4.16} are satisfied. 
\end{proof}

 We can summarize the main results of the previous two sections as
follows. 

\bTh
The following assertions are equivalent.

\smallskip

{\rm(1)}
$L_{\nu_\tau(X)} \, L_\Gam \, \tau = 0 \,,$
$\Al \, X \in \sec(\f E,T\f E ) \,,$
and
$L_R \, \tau = 0 \,.$

\smallskip

{\rm (2)} 
$d\Ome = 0 \,,$ 
i.e. 
$(- c^2 \, \tau, \, \Ome)$ 
is a (scaled) almost--cosymplectic--contact structure.

\smallskip

{\rm (3)}
$[- \tfr1{c^2} \, \gam, \, \Lam] =  
\tfr1{c^2} \, \gam \wed \Lam\Sha(L_{\gam} \, \tau))$ 
and 
$[\Lam, \, \Lam] = 
2 \, \gam \wed(\Lam\Sha \ten \Lam\Sha)(d\tau)) \,,$ 
i.e. 
$(-\tfr1{c^2} \, \gam, \, \Lam)$ 
is a (scaled) almost--coPoisson--Jacobi structure along with
the fundamental 1--form $- c^2\,\tau \,.$

 Moreover, the almost--cosymplectic--contact pair 
$( - c^2 \, \tau, \, \Ome)$ 
and the almost--coPoisson--Jacobi 3--plet 
$(- \tfr1{c^2} \, \gam, \, \Lam, \, -c^2 \, \tau)$
are mutually dual.
\hfill\ENDE
\eTh
\subsection{Contact structure}
\label{Contact structure}
 Next, we analyse the conditions by which the pair
$(- c^2 \, \tau \,, \, \Ome)$ is a contact pair, i.e. 
$\Ome = - c^2\, d\tau \,.$
 As a direct consequence of Theorem
\ref{Conditions ACC} we obtain that if
$L_\Gam \,\tau = 0 \,,$ 
then the pair
$(- c^2 \, \tau \,, \, \Ome)$
is a scaled almost--cosymplectic--contact structure.
 Indeed, the condition 
$L_\Gam \, \tau = 0$ 
implies 
$L_{\nu_\tau(X)} \, L_\Gam \, \tau = 0$
and
$L_R \, \tau = 0 \,,$ 
because of 
$L_R \, \tau = - L_{[\Gam,\Gam]} \, \tau
= - 2 \, L_\Gam \, L_\Gam \, \tau \,.$ 
 Hence, by Theorem \ref{Conditions ACC}, if 
$L_\Gam \tau = 0 \,,$ 
then the pair
$(- c^2 \, \tau \,, \, \Ome)$
is a scaled almost--cosymplectic--contact structure.
 But, in this case we can prove more. 

\bLm
 We have
\beq
d\Ome = c^2 \, dL_\Gam \, \tau \,. 
\eeq
\eLm

\begin{proof}
 We have
\bal
dL_\Gam \, \tau 
&= 
	- \tfr1c \, \Big[\der_\nu\big(\der_\lam (\alp^0 \, \br g_{0\mu}) 
	+ \Gam\Ga\lam j \, \der^0_j (\alp^0 \, \br g_{0\mu}) \big) \, 
	d^\nu \wed d^\lam \wed d^\mu 
\\
&\quad
	+ \der^0_i\big(\der_\lam (\alp^0 \, \br g_{0\mu}) 
	+ \Gam\Ga\lam j \, \der^0_j (\alp^0 \, \br g_{0\mu}) \big) 	\, 
	d^i_0 \wed d^\lam \wed d^\mu
	\Big]
\\
& =
	- \tfr1c \, \Big[\der_\nu (\alp^0 \, \br g_{j\mu} \, \Gam\Ga\lam j)
	\, 
	d^\nu \wed d^\lam \wed d^\mu 
\\
&\quad
	+ \big(\der_\lam (\alp^0 \, \br g_{i\mu}) 
	+ \der^0_i (\alp^0 \, \br g_{j\mu} \, \Gam\Ga\lam j) \big) \, 
	d^i_0 \wed d^\lam \wed d^\mu
	\Big] \,.
\end{align*}

 Hence, we obtain the Lemma by comparing the above equality with
\eqref{Eq4.5}. 
\end{proof}

\bCr\label{Cr4.9}
 The difference 
$\Ome - c^2 \, L_\Gam \, \tau$ 
is an exact form.
 More precisely, we have 
\beq
\Ome - c^2 \, L_\Gam \, \tau = - c^2\, d\tau \,.
\eeq
\eCr

\begin{proof}
 We have 
\bal
	\Ome - c^2 \, L_\Gam \, \tau 
&= 
	c \, \alp^0 \, \br g_{i\mu} \, 
(d^i_0 - \Gam\Ga\lam i \, d^\lam) \wed d^\mu
		+ c \, (\der_\lam(\alp^0 \, \br g_{0\mu}) 
		+ \alp^0 \, \br g_{p\mu} \, \Gam\Ga\lam p) \, d^\lam \wed d^\mu
\\
&=
	c \, \big(\alp^0 \, \br g_{i\mu} \, d^i_0 \wed d^\mu
	+ \der_\lam(\alp^0 \,\br g_{0\mu}) \, d^\lam \wed d^\mu\big)
\\
&=
	c \, \big(\der^0_i(\alp^0 \, \br g_{0\mu}) \, d^i_0 \wed d^\mu
	+ \der_\lam(\alp^0 \, \br g_{0\mu}) \, d^\lam \wed d^\mu\big)
=
	- c^2 \, d\tau \,. 
\end{align*}
\vglue-1.5\baselineskip
\end{proof}

\bTh\label{Theorem: LGam tau condition for contact structure}
The pair
$(- c^2 \, \tau, \Ome)$ 
is a scaled contact structure if and only if 
$L_\Gam \,\tau = 0 \,.$
\hfill\ENDE
\eTh
\subsection{Jacobi structure}
\label{Jacobi structure: general case}
 Next, we analyse conditions for the pair 
$(- \tfr1{c^2} \gam, \Lam)$ 
to be a scaled Jacobi structure.

\bLm\label{Lemma: coordinate expression of gam wed Lam}
 We have the coordinate expression
\bAl\label{Eq: gam Lam general}
\gam \wed \Lam & =
\br\del^\lam_0 \, \br g^{i\mu} \, 
\der_\lam \wed \der_\mu \wed \der^0_i 
+ (\br\del^\lam_0\, \br g^{j\rho}
- \br\del^\rho_{0}\, \br g^{j\lam}) \, \Gam\Ga\rho i \,
\der_\lam \wed \der^0_i \wed \der ^0_j
\\ \nonumber
&\quad + \br\del^\rho_0 \, \br g^{k\sig} \, \Gam\Ga\rho i 
\, \Gam\Ga\sig j \, \der^0_i \wed \der^0_j \wed \der^0_k \,
\end{align}
and, in the adapted base 
\eqref{Eq: der to e},
\bEq\label{Eq: gam Lam general adapted}
\gam \wed \Lam =  \ha g^{ij} \, e_0\wed e_i\wed e^0_i\,.
\eEq
\eLm

\begin{proof}
 It follows from the coordinate expressions
\beq
\gam = c\,\alp^0\,\br\del^\lam_0\,(\der_\lam + \Gam\Ga\lam
i\,\der^0_i)
     = c\,\alp^0\, e_0
\eeq
and
\beq
\Lam = \fr1{c\,\alp^0}\,\br g^{j\lam}\,(\der_\lam + \Gam\Ga\lam
i\,\der^0_i) \wed \der^0_j
     = \fr1{c\,\alp^0}\, \ha g^{ij} e_i\wed e^0_j\,. 
\eeq
\vglue-1.3\baselineskip
\end{proof}

\bLm\label{Lam and Lam - 2 gam wed Lam}
 In the adapted base
\eqref{Eq: der to e}
we have the coordinate expression
\bAl\label{Eq: Lam and Lam - 2 gam wed Lam}
	[\Lam, \Lam] 
& - \fr2{c^2}\,\gam \wed \Lam =
	\fr2{c^2}\,\big[ \br g^{k\rho}\,\big(\Gam\Ga\rho j
	- \br g_{0\lam}\,\der_\rho\br g^{j\lam}
	\big)\,
	e_0 \wed e^0_j \wed e^0_k
\\
&\quad \nonumber
	+ \big( \ha g^{ki} \,\br g^{j\rho}\, 
	(\tfr12 \der_\rho {\ha g_{00}} + \br g_{0p}\, \Gam\Ga\rho p)
\\
&\quad\quad \nonumber
	+ \fr1{(\alp^0)^2}\,\br g^{k\rho}\,
		(\der_\rho\ha g^{ji} - \br g^{i0}\,\Gam\Ga\rho j 
			- \ha g^{pi}\,\der^0_p\Gam\Ga\rho j)
	\big)\,
	e_i \wed e^0_j \wed e^0_k
\\
&\quad \nonumber
	+  \fr1{(\alp^0)^2} \,
	\br g^{i\rho}\, \br g^{j\sig} \, R\Ga{\rho\sig} k 
		\,
	e^0_i \wed e^0_j \wed e^0_k
	\big]\,.
\end{align}
\eLm

\begin{proof}
It follows from 
\eqref{Eq: adapted coordinate expression of [Lam, Lam]}  and
\eqref{Eq: gam Lam general adapted}. 
\end{proof}

\bLm\label{L Gam tau 0 coordinate}
$L_\Gam\,\tau =0$ if and only if the identities
\bAl\label{Eq5.26}
\tfr12 \br g^{j\rho}\,\der_\rho \ha g_{00} 
+ \br\del^\rho_0\,(\br g_{0\lam}\,\der_\rho\br g^{j\lam}
    - \Gam\Ga\rho j)  & = 0\,,
\\[2mm] \label{Eq5.27}
\br g^{i\rho}\, (\br g_{0\lam}\,\der_\rho\br g^{j\lam}
    - \Gam\Ga\rho j) 
- \br g^{j\rho}\, (\br g_{0\lam}\,\der_\rho\br g^{i\lam}
    - \Gam\Ga\rho i) & = 0\,.
\end{align}
are satisfied.
\eLm

\begin{proof}
 In the adapted base 
\eqref{Eq: der to e} 
we  have
\bal
{}\Sha(L_\Gam \tau)
& =
\tfr{\h\,(\alp^0)^2}{m\, c^4}\, \big(
\tfr12 \,\br g^{j\rho} \, \der_\rho\ha g_{00}
+ \br \del^\rho_0\,(\br g_{0\lam}\,\der_\rho\br g^{j\lam}
- \Gam\Ga\rho j)
\big)\, e_0 \wed e^0_j
\\
& \quad
+ \tfr1{c^3\,\alp^0}\, \br g^{i\rho} \, \big(\br g_{0\lam}\,
    \der_\rho\br g^{j\lam} - \Gam\Ga\rho j\big)
\, e^0_i\wed e^0_j\,.
\end{align*}
 Then, 
$L_\Gam\,\tau = 0$ 
if and only if 
${}\Sha(L_\Gam \tau) = 0 \,,$
i.e. if and only if the identities \eqref{Eq5.26} and
\eqref{Eq5.27} are satisfied. 
\end{proof}

\bLm\label{Lemma: Jacobi coordinate general}
 The pair
$(- \tfr1{c^2} \gam, \Lam)$ 
is a scaled Jacobi structure if and only if the identities
\eqref{Eq5.26} and
\eqref{Eq5.27} are satisfied.
\eLm

\begin{proof}
 The pair
$(- \tfr1{c^2} \gam, \Lam)$ 
is a scaled Jacobi structure if and only if
$[-\tfr1{c^2}\,\gam,\Lam] = 0$ 
and 
$[\Lam,\Lam] = \tfr2{c^2}\,\gam \wed \Lam \,.$
 According to the coordinate expressions in Lemmas
\ref{gam and Lam; general case adapted}
and
\ref{Lam and Lam - 2 gam wed Lam},
these conditions are satisfied if and only if the identities
\eqref{Eq4.12} -- \eqref{Eq4.17}, \eqref{Eq5.26}
and \eqref{Eq5.27} are satisfied.
 But, we can prove that the identities \eqref{Eq5.26}
and \eqref{Eq5.27}
imply the identities \eqref{Eq4.12} -- \eqref{Eq4.17}.

 First, let us prove that the identities \eqref{Eq5.26}
and \eqref{Eq5.27}
imply the identities \eqref{Eq4.13} and \eqref{Eq4.17}.
But it follows from the fact that the identities \eqref{Eq5.26}
and \eqref{Eq5.27} are satisfied if and only if
$L_\Gam \,\tau = 0 \,.$
 But, in this case 
$L_R\,\tau = 
- L_{[\Gam,\Gam]} \,\tau = - 2\,L_\Gam\,L_\Gam\,\tau =0 \,,$
which is equivalent to the identities
\eqref{Eq4.13} and \eqref{Eq4.17}.

 Next, let us recall the identities
$\ha g^{ip}\,\del^\rho_p 
= \br g^{i\rho} - \br g^{i0}\,\br\del^\rho_0 \,,$ 
$\ha g^{ip}\, g_{p\lam} 
= \br\del^i_\lam - \br g^{i0}\,\br g_{0\lam}$
and 
$g^{0\lam} = 
(\alp)^2 \, (\br g_{0p}\,\br g^{p\lam} - \br\del^\lam_0) \,.$
 Hence, if we apply the operator
$\ha g^{ip}\,\der^0_p$ 
on the left hand side of the identity
\eqref{Eq5.26} and we suppose that 
the identities \eqref{Eq5.26}
and \eqref{Eq5.27} are satisfied, then we obtain
\bal
0 =
& - \tfr12 \,\ha g^{ij}\, g^{0\rho}\,\der_\rho \ha g_{00}
+ \ha g^{ip}\, \br g^{j\rho}\,\der_\rho \br g_{0p}
+ (\ha g^{ip}\,\del^\rho_p\,\br g_{0\lam}
	+ \ha g^{ip}\,\br\del^\rho_0\, g_{p\lam})
	\,\der_\rho \br g^{j\lam}
\\
&
- \ha g^{ip}\,\del^\rho_p\,\Gam\Ga\rho j 
- \ha g^{ij}\,\br g_{0\lam}\,\br\del^\rho_0\,\der_\rho g^{0\lam}
- \ha g^{ip}\,\br\del^\rho_0\, \der^0_p\Gam\Ga\rho j
\\
= &
- \tfr12 \,\ha g^{ij}\, g^{0\rho}\,\der_\rho \ha g_{00} 
- \br g^{j\rho}\, \br g_{0p}\,\br\del^p_\lam \, 
	\der_\rho\br g^{i\lam}
- \br g^{j\rho}\,\br g^{i0}\, \der_\rho \ha g_{00} 
+ \tfr1{(\alp^0)^2}\, \br g^{j\rho}\, \der_\rho \br g^{i0}
- \ha g^{ij}\,\br g_{0\lam}\,\br\del^\rho_0\,\der_\rho g^{0\lam}
\\
&
- \ha g^{ip}\,\br\del^\rho_0\, \der^0_p\Gam\Ga\rho j
+ (\br g^{i\rho}\,\br g_{0\lam}
	- 2\, \br g^{i0}\,\br\del^\rho_0\,\br g_{0\lam}
	+ \br\del^i_\lam\,\br\del^\rho_0)\,
	\,\der_\rho \br g^{j\lam}
- (\br g^{i\rho} - \br g^{i0}\,\br\del^\rho_0)
	\,	\Gam\Ga\rho j 
\\
= &
- \tfr12 \,\ha g^{ij}\, g^{0\rho}\,\der_\rho \ha g_{00} 
- \br g^{j\rho}\,\br g^{i0}\, \der_\rho \ha g_{00} 
- \br g_{0\lam}\,(\br g^{j\rho}\,  \der_\rho\br g^{i\lam}
	- \br g^{i\rho}\,\der_\rho \br g^{j\lam})
- \ha g^{ij}\,\br g_{0\lam}\,\br\del^\rho_0\,\der_\rho g^{0\lam}
\\
&
- 2\, \br g^{i0}\,\br\del^\rho_0\,\br g_{0\lam}
	\,\der_\rho \br g^{j\lam}
+ \br\del^\rho_0\,(
	\,\der_\rho \ha g^{ij}
+ \br g^{i0}\,	\Gam\Ga\rho j 
- \ha g^{ip}\, \der^0_p\Gam\Ga\rho j)
- \br g^{i\rho}\,	\Gam\Ga\rho j  
\\
= &
- (\alp^0)^2\,\ha g^{ij}\, \br\del^\rho_0\, 
(\tfr12\,\der_\rho \ha g_{00}  + \br g_{0p}\, \Gam\Ga\rho p)
+ \br\del^\rho_0\,(
	\,\der_\rho \ha g^{ij}
- \br g^{i0}\,	\Gam\Ga\rho j 
- \ha g^{ip}\, \der^0_p\Gam\Ga\rho j)
- \br g^{j\rho}\,	\Gam\Ga\rho i\,. 
\end{align*}
Hence, the identities \eqref{Eq5.26}
and \eqref{Eq5.27}
imply the identity \eqref{Eq4.12}.

 Finally, we apply the same method on the left hand side of the
identity
\eqref{Eq5.27} and obtain
\bat{2}
	0
&=
&&	-\ha g^{ik}\, \,g^{0\rho}\,
	(\br g_{0\lam}\,\der_\rho\br g^{j\lam} - \Gam\Ga\rho j)
	+ \br g^{k\rho}\, (\ha g^{pi}\, g_{p\lam}\, \der_\rho\br g^{j\lam}
	- \ha g^{ij}\, \br g_{0\lam}\, \der_\rho g^{0\lam}
	- \ha g^{ip}\, \der^0_p \Gam\Ga \rho j)
\\
&&&
	+ \ha g^{ij}\, \,g^{0\rho}\,
	(\br g_{0\lam}\,\der_\rho\br g^{k\lam} - \Gam\Ga\rho k)
	- \br g^{j\rho}\, (\ha g^{pi}\, g_{p\lam}\, \der_\rho\br g^{k\lam}
	- \ha g^{ik}\, \br g_{0\lam}\, \der_\rho g^{0\lam}
	- \ha g^{ip}\, \der^0_p \Gam\Ga \rho k)
\\[2mm]
&= 
	&&\quad\,\ha g^{ik}\, \,g^{0\rho}\,
	(\br g^{j\lam}\,\der_\rho\br g_{0\lam} + \Gam\Ga\rho j)
	+ \br g^{k\rho}\, ( \der_\rho\ha g^{ij}
	+ \ha g^{ij}\, g^{0\lam}\, \der_\rho\br g_{0\lam}
	- \ha g^{ip}\, \der^0_p \Gam\Ga \rho j)
\\
&	&&
	-\ha g^{ij}\, \,g^{0\rho}\,
	(\br g^{k\lam}\,\der_\rho\br g_{0\lam} + \Gam\Ga\rho k)
	- \br g^{j\rho}\, ( \der_\rho\ha g^{ik}
	+ \ha g^{ik}\, g^{0\lam}\, \der_\rho\br g_{0\lam}
	- \ha g^{ip}\, \der^0_p \Gam\Ga \rho k)
\\
&&&
	-  \br g^{i0}\,(\br g^{k\rho}\,\br g_{0\lam}\, \der_\rho\br g^{j\lam}
		- \br g^{j\rho}\,\br g_{0\lam}\, \der_\rho\br g^{k\lam})
\\[2mm]
&=
&&	\quad\,(\ha g^{ik}\, g^{0\rho}\,
	\br g^{j\lam}+ \ha g^{ij}\, g^{0\lam}\, \br g^{k\rho})
	\,\der_\rho\br g_{0\lam} 
	+ \ha g^{ik}\, g^{0\rho}\,\Gam\Ga\rho j
	+ \br g^{k\rho}\, ( \der_\rho\ha g^{ij}
	- \br g^{i0}\,\Gam\Ga\rho j
	- \ha g^{ip}\, \der^0_p \Gam\Ga \rho j)
\\
&&&
	- (\ha g^{ij}\, g^{0\rho}\,
	\br g^{k\lam}+ \ha g^{ik}\, g^{0\lam}\, \br g^{j\rho})
	\,\der_\rho\br g_{0\lam} 
	- \ha g^{ij}\, g^{0\rho}\,\Gam\Ga\rho k
	- \br g^{j\rho}\, ( \der_\rho\ha g^{ik}
	- \br g^{i0}\,\Gam\Ga\rho k
	- \ha g^{ip}\, \der^0_p \Gam\Ga \rho k)
\\[2mm]
&=
&&\quad\,	g^{0\rho}\,(\ha g^{ik}\, \br g^{j\lam}
		- \ha g^{ij}\, \br g^{k\lam})
		\,\der_\rho\br g_{0\lam} 
	- g^{0\lam}\,(\ha g^{ik}\, \br g^{j\rho}
		- \ha g^{ij}\, \br g^{k\rho})
		\,\der_\rho\br g_{0\lam} 
	+ g^{0\rho}\, (\ha g^{ik}\, \Gam\Ga\rho j
		- \ha g^{ij}\, \Gam\Ga\rho k)
\\
&&&
	+ \br g^{k\rho}\, ( \der_\rho\ha g^{ij}
	- \br g^{i0}\,\Gam\Ga\rho j
	- \ha g^{ip}\, \der^0_p \Gam\Ga \rho j)
	- \br g^{j\rho}\, ( \der_\rho\ha g^{ik}
	- \br g^{i0}\,\Gam\Ga\rho k
	- \ha g^{ip}\, \der^0_p \Gam\Ga \rho k)
\\[2mm]
&=
&&\quad\,	 (\alp^0)^2\, (\br\del^\lam_0 
		- \br g_{0p}\, \br g^{p\lam})
		\,(\ha g^{ik}\, \br g^{j\rho}
		- \ha g^{ij}\, \br g^{k\rho})
		\,\der_\rho\br g_{0\lam}
\\
&&&
	- (\alp^0)^2\, (\br\del^\rho_0 
		- \br g_{0p}\, \br g^{p\rho})
		\,\big(\br g_{0\lam}\,(\ha g^{ij}\, \der_\rho\br g^{k\lam}
		- \ha g^{ik}\, \der_\rho\br g^{j\lam})	
	+ \ha g^{ik}\, \Gam\Ga\rho j
		- \ha g^{ij}\, \Gam\Ga\rho k\big)
\\
&&&
	+ \br g^{k\rho}\, ( \der_\rho\ha g^{ij}
	- \br g^{i0}\,\Gam\Ga\rho j
	- \ha g^{ip}\, \der^0_p \Gam\Ga \rho j)
	- \br g^{j\rho}\, ( \der_\rho\ha g^{ik}
	- \br g^{i0}\,\Gam\Ga\rho k
	- \ha g^{ip}\, \der^0_p \Gam\Ga \rho k)
\\[2mm]
&=
&&\quad\,	 (\alp^0)^2\, (\ha g^{ik}\, \br g^{j\rho}
		- \ha g^{ij}\, \br g^{k\rho})
		\,\der_\rho\ha g_{00}		
		- (\alp^0)^2\, \br g_{0p}\, \br g_{0\lam}
		\,(\ha g^{ik}\, \br g^{j\rho}
		- \ha g^{ij}\, \br g^{k\rho})
		\,\der_\rho\br g^{p\lam}
\\
&&& 
	- (\alp^0)^2\, (\br\del^\rho_0 
		- \br g_{0p}\, \br g^{p\rho})
		\,\big(\br g_{0\lam}\,(\ha g^{ij}\, \der_\rho\br g^{k\lam}
		- \ha g^{ik}\, \der_\rho\br g^{j\lam})	
	+ \ha g^{ik}\, \Gam\Ga\rho j
		- \ha g^{ij}\, \Gam\Ga\rho k\big)
\\
&&&
	+ \br g^{k\rho}\, ( \der_\rho\ha g^{ij}
	- \br g^{i0}\,\Gam\Ga\rho j
	- \ha g^{ip}\, \der^0_p \Gam\Ga \rho j)
	- \br g^{j\rho}\, ( \der_\rho\ha g^{ik}
	- \br g^{i0}\,\Gam\Ga\rho k
	- \ha g^{ip}\, \der^0_p \Gam\Ga \rho k)
\\[2mm]
&=
&&\quad\,	 (\alp^0)^2\, (\ha g^{ik}\, \br g^{j\rho}
		- \ha g^{ij}\, \br g^{k\rho})
		\,\der_\rho\ha g_{00}	
\\
&	&&		
	- (\alp^0)^2\, \br\del^\rho_0 \,\big(
		\ha g^{ij}\,(\br g_{0\lam}\,\der_\rho\br g^{k\lam}
			- \Gam\Ga\rho k)
		- \ha g^{ik}\,(\br g_{0\lam}\,\der_\rho\br g^{j\lam}
			- \Gam\Ga\rho j)\big)
\\
&	&&
		+ (\alp^0)^2\, \br g_{0p}\,\big(\ha g^{ik}\,
		(\br g^{j\rho}\, \br g_{0\lam}\,\der_\rho\br g^{p\lam}
		 - \br g^{p\rho}\, \br g_{0\lam}\,\der_\rho\br g^{j\lam})
		- \ha g^{ij}\,
		(\br g^{k\rho}\, \br g_{0\lam}\,\der_\rho\br g^{p\lam}
		 - \br g^{p\rho}\, \br g_{0\lam}\,\der_\rho\br g^{k\lam})
		 \big)
\\
&&&
	+ \br g^{k\rho}\, ( \der_\rho\ha g^{ij}
	- \br g^{i0}\,\Gam\Ga\rho j
	- \ha g^{ip}\, \der^0_p \Gam\Ga \rho j)
	- \br g^{j\rho}\, ( \der_\rho\ha g^{ik}
	- \br g^{i0}\,\Gam\Ga\rho k
	- \ha g^{ip}\, \der^0_p \Gam\Ga \rho k)
\\[2mm]
&=
&&	+ (\alp^0)^2\, (\ha g^{ik}\, \br g^{j\rho}
		- \ha g^{ij}\, \br g^{k\rho})
		\,\der_\rho\ha g_{00}	
	+ \tfr12\,(\alp^0)^2\, (
		\ha g^{ij}\,\br g^{k\rho}
		- \ha g^{ik}\,\br g^{j\rho})\,\der_\rho\ha g_{00}
\\
&	&&
		+ (\alp^0)^2\, \br g_{0p}\,\big(\ha g^{ik}\,
		(\br g^{j\rho}\, \Gam\Ga\rho p
		- \br g^{p\rho}\, \Gam\Ga\rho j)
		- \ha g^{ij}\,
		(\br g^{k\rho}\, \Gam\Ga\rho p
		- \br g^{p\rho}\, \Gam\Ga\rho k)
		 \big)
\\
&&&
	+ \br g^{k\rho}\, ( \der_\rho\ha g^{ij}
	- \br g^{i0}\,\Gam\Ga\rho j
	- \ha g^{ip}\, \der^0_p \Gam\Ga \rho j)
	- \br g^{j\rho}\, ( \der_\rho\ha g^{ik}
	- \br g^{i0}\,\Gam\Ga\rho k
	- \ha g^{ip}\, \der^0_p \Gam\Ga \rho k)
\\[2mm]
&=
&&\quad\,	(\alp^0)^2\, (\ha g^{ik}\, \br g^{j\rho}
		- \ha g^{ij}\, \br g^{k\rho})
		\,(\tfr12 \der_\rho\ha g_{00}	
	+ \br g_{0p}\, \Gam\Ga\rho p )
\\
&&&
	+ \br g^{k\rho}\, ( \der_\rho\ha g^{ij}
	- \br g^{i0}\,\Gam\Ga\rho j
	- \ha g^{ip}\, \der^0_p \Gam\Ga \rho j)
	- \br g^{j\rho}\, ( \der_\rho\ha g^{ik}
	- \br g^{i0}\,\Gam\Ga\rho k
	- \ha g^{ip}\, \der^0_p \Gam\Ga \rho k) \,.
\end{alignat*}

 So, the identities \eqref{Eq5.26}
and \eqref{Eq5.27} imply the identity \eqref{Eq4.16}. 
\end{proof}

\smallskip
\bTh\label{Theorem: conditions for Jacobi structure; general case}
 The pair
$(- \tfr1{c^2} \gam, \Lam)$ 
is a scaled Jacobi structure if and only if
\beq
L_\Gam \, \tau = 0 \,.
\eeq
\eTh

\begin{proof}
{\it 1st proof.} 
 By Theorem \ref{Th1.1} the structure 
$(-c^2\,\tau,\Ome)$
is a scaled contact structure if and only if the dual structure
$(-\tfr1{c^2}\,\gam,\Lam)$ 
is a scaled Jacobi structure. 
 But, according to Theorem 
\ref{Theorem: LGam tau condition for contact structure}, 
$(-c^2\,\tau,\Ome)$
is a scaled contact structure if and only if
$L_\Gam\, \tau =0 \,.$

{\it 2nd proof.}
 By Lemma \ref{Lemma: Jacobi coordinate general}
$(-\tfr1{c^2}\,\gam,\Lam)$ is a scaled Jacobi structute if and
only if the identities \eqref{Eq5.26}
and \eqref{Eq5.27} are satisfied but, by Lemma 
\ref{L Gam tau 0 coordinate}, it is equivalent with
$L_\Gam\, \tau = 0 \,.$ 
\end{proof}

 We can summarize the above results as follows.

\bTh\label{Th4.22}
 The following assertions are equivalent.

\smallskip

{\rm (1)} 
$L_\Gam \, \tau = 0 \,.$

\smallskip

{\rm (2)} 
$\Ome = - c^2\, d\tau \,,$ 
i.e. 
$(-c^2\,\tau,\Ome)$ 
is a (scaled) contact structure.

\smallskip

{\rm (3)} 
$[- \tfr1{c^2} \, \gam, \Lam] = 0$ 
and 
$[\Lam,\Lam] = \tfr2{c^2} \, \gam \wed \Lam \,,$ 
i.e. 
$(-\tfr1{c^2} \, \gam, \Lam)$ 
is a (scaled regular) Jacobi structure.

\smallskip

 Moreover, the contact structure 
$(- c^2 \, \tau, \Ome)$ 
and the Jacobi structure
$(-\tfr1{c^2} \, \gam,\Lam)$ 
are mutually dual.
\hfill\ENDE
\eTh

\section{Contact and Jacobi structures: linear case}
\label{Contact and Jacobi structures: linear case}
\setcounter{equation}{0}
 Next, we specialise the results of the previous section
concerning the contact and Jacobi structures, by considering the
phase connection 
$\Gam = \chi(K)$
induced by a linear spacetime connection 
$K \,.$
 Thus, in this section, we refer to the induced objects\linebreak
$\Ome = \Ome[g,K] = \Ome[g, \chi(K)] \,,\;$
$\Lam = \Lam[g,K] = \Lam[g, \chi(K)] \,,\;$
$\gam = \gam[K] = \gam[\chi(K)] \,.$
\subsection{Covariant derivatives and the adapted base}
\label{Covariant derivatives and the adapted base}
 We start with some technical formulas concerning the covariant
derivatives $\nab$ with respect to the connection $K$ of the metric in
the orthogonal adapted bases $(b_0, b_1)$ and $(\bet^0,\bet^i) \,.$

\bLm
 We have
\bat{2}
\nab_\lam \br g_{0\mu}
&\byd
	(\nab_\lam g) (b_0, \der_\mu) 
&&=
\br\del^\nu_0\,\nab_\lam g_{\nu\mu} =
\der_\lam \br g_{0\mu}
+ g_{\rho \mu} \, K\col\lam\rho\sig \, \br\del^\sig_0
+ \br g_{0\rho} \, K\col\lam\rho\mu \,,
\\
	\nab_\lam\br g_{i\mu} 
&\byd
	(\nab_\lam g)(b_i,\der_\mu) 
&&=
	\nab_\lam g_{i\mu} 
	+ (\alp^0)^2 \, \br g_{0i} \, \nab_\lam \br g_{0\mu}
\\
	\nab_\lam \br g^{0\mu}
&\byd
	(\nab_\lam \ba g) (\bet^0, d^\mu) 
&&=
	- (\alp^0)^2\, \br g_{0\nu}\,\big(\der_\lam g^{\nu\mu}
	- g^{\rho\mu} \, K\col\lam\nu\rho
	- g^{\nu\rho}\, K\col\lam\mu\rho\big) \,,
\\
	\nab_\lam \br g^{i\mu}
&\byd
		(\nab_\lam \ba g) (\bet^i, d^\mu)
&&=
\br\del^i_\nu\,\nab_\lam g^{\nu\mu} =
	\der_\lam \br g^{i\mu}
	- \br\del^i_\nu\, g^{\rho\mu} \, K\col\lam\nu\rho
	- \br g^{i\rho}\, K\col\lam\mu\rho \,.
\end{alignat*}

 Moreover, we have the identities
\bal
	\br g_{0\mu} \, \nab_\lam \br g^{0\mu} 
& =
	- \br g^{0\mu} \, \nab_\lam \br g_{0\mu}\,, \qquad
	\br g_{0\mu} \, \nab_\lam \br g^{i\mu} =
	- \br g^{i\mu} \, \nab_\lam \br g_{0\mu}\,, 
\\
	\br g_{i\mu} \, \nab_\lam \br g^{0\mu} 
& =
	- \br g^{0\mu} \, \nab_\lam \br g_{i\mu}\,, \qquad
	\br g_{i\mu} \, \nab_\lam \br g^{j\mu} =
	- \br g^{j\mu} \, \nab_\lam \br g_{i\mu} \,.
\end{align*}
\eLm

\begin{proof}
 We have
\beq
	\nab_\lam g =
	\nab_\lam g_{\nu\mu}\, d^\nu\ten d^\mu =
	\nab_\lam g_{\nu\mu}\, \big(\br \del^\nu_0\, \bet^0 
	+ (\del^\nu_i+ (\alp^0)^2\,\br g_{0i}\, \br\del^\nu_0)\, \bet^i\big)
	\ten d^\mu \,.
\eeq

 Then, we obtain
\beq
(\nab_\lam g)(b_0,\der_\mu) = \br\del^\nu_0 \, \nab_\lam g_{\nu\mu}
\ssep{and}
(\nab_\lam g)(b_i,\der_\mu) = (\del^\nu_i
	+ (\alp^0)^2 \, \br g_{0i}\, \br\del^\nu_0) \, \nab_\lam g_{\nu\mu}
\,.
\eeq

 Analogously, we have
\beq
	\nab_\lam\ba g =
	\nab_\lam g^{\nu\mu}\, \der_\nu\ten\der_\mu =
	\nab_\lam g^{\nu\mu} \, 
\big(\br\del^i_\nu \, b_i - 	(\alp^0)^2 \, \br g_{0\nu} \, b_0	\big) 
\ten \der_\mu	 \,.
\eeq

 Then, we obtain
\beq
\nab_\lam g(\bet^0, d^\mu) = 
- (\alp^0)^2 \, \br g_{0\nu} \, \nab_\lam g^{\nu\mu}
\ssep{and}
\nab_\lam g(\bet^i, d^\mu) = 
\br\del^i_\nu \, \nab_\lam g^{\nu\mu} \,. 
\eeq
\vglue-1.5\baselineskip
\end{proof}

\bLm\label{Lm5.3}
 We have
\bat{2}
\nab_\lam \ha g_{00}
&\byd
(\nab_\lam g) (b_0, b_0)
&&= 
- \nab_\lam (\alp^0)^{-2} =
\der_\lam \ha g_{00} +
2 \, \br g_{0\rho} \, K\col\lam\rho\sig \, \br\del^\sig_0 \,,
\\ 
\nab_\lam \ha g_{i0}
&\byd
(\nab_\lam  g)(b_i,b_0)
&&= 
	\nab_\lam \br g_{0i} + 
(\alp^0)^2 \, \br g_{0i} \, \nab_\lam \ha g_{00} \,,
\\
\nab_\lam \ha g_{ij}
&\byd
(\nab_\lam g)(b_i,b_j)
&&=
	\nab_\lam g_{ij} + (\alp^0)^2 \, (\br g_{0i} \, \nab_\lam\br g_{0j}
		+ \br g_{0j} \, \nab_\lam\br g_{0i}) 
+ (\alp^0)^4 \, \br g_{0i} \, \br g_{0j} \, 			\nab_\lam 
\ha g_{00} 
\\
\nab_\lam \ha g^{00}
&\byd
(\nab_\lam \ba g) (\bet^0, \bet^0)
&&= 
- \nab_\lam (\alp^0)^{2} =
(\alp^0)^4 \,\br g_{0\nu}\,\br g_{0\mu} \, \nab_\lam g^{\nu\mu}
\\ 
&&&=
(\alp^0)^4 \, (\br g_{0\nu} \, \br g_{0\mu} \, \der_\lam g^{\nu\mu} 
	- 2\,\br g_{0\sig} \, \br\del^\rho_0 \, K\col\lam\sig\rho) \,,
\\
\nab_\lam \ha g^{i0}
&\byd
(\nab_\lam \ba g)(\bet^i,\bet^0)
&&= 
	- (\alp^0)^2 \,  \br g_{0\mu} \, \nab_\lam\br g^{i\mu} 
	= - (\alp^0)^2 \,  \big(\br g_{0\mu} \, \der_\lam\br g^{i\mu}
		- (\br\del^\rho_0 \, \br\del^i_\sig 
		+ \br g^{i\rho} \, \br g_{0\sig} \, K\col\lam\sig\rho\big) \,,
\\
\nab_\lam \ha g^{ij}
&\byd
(\nab_\lam \ba g)(\bet^i,\bet^j)
&&=
	\br \del^i_\nu \, \br\del^j_\mu \, \nab_\lam g^{\nu\mu} =
	\der_\lam \ha g^{ij} - (\br g^{j\rho} \, \br\del^i_\sig 
	+ \br g^{i\rho} \, \br\del^j_\sig) \, K\col\lam\sig\rho \,.
\end{alignat*}
\eLm

\begin{proof} 
 We have
\bal
	\nab_\lam g 
&=
	\nab_\lam g_{\nu\mu} \, d^\nu \ten d^\mu 
\\[2mm]
&=
	\nab_\lam g_{\nu\mu} \, 
\big(\br\del^\nu_0 \, \bet^0 +
	(\del^\nu_i + (\alp^0)^2 \,\br g_{0i} \, \br\del^\nu_0) \, \bet^i
\big) 
\ten
\big(\br\del^\mu_0 \, \bet^0 +
(\del^\mu_j + (\alp^0)^2 \, \br g_{0j} \, \br\del^\mu_0) \,
\bet^j\big)
\\[2mm]
&= 	
	\nab_\lam g_{\nu\mu} \, 
\big[\br\del^\nu_0 \, \br\del^\mu_0 \, \bet^0 \ten \bet^0 +
	\br\del^\nu_0 \, (\del^\mu_j 
		+ (\alp^0)^2 \, \br g_{0j} \, \br\del^\mu_0) \, 
\bet^0 \ten \bet^j 
\\
&\quad +
	(\del^\nu_i 
		+ (\alp^0)^2 \, \br g_{0i} \, \br\del^\nu_0) 
		\, \br\del^\mu_0 \,
\bet^i \ten \bet^0
+
	\big(\del^\nu_i 
		+ (\alp^0)^2 \, \br g_{0i} \, \br\del^\nu_0\big) \,
\big(\del^\mu_j +
		(\alp^0)^2 \, \br g_{0j} \, \br\del^\mu_0)\big) \, \bet^i \ten \bet^j
	\big]
\,.
\end{align*}

 Then, we obtain
\bal
	\nab_\lam g(b_0,b_0)
&=
	\br\del^\nu_0\,\br\del^\mu_0\, \nab_\lam g_{\nu\mu} \,,
\\
	\nab_\lam g(b_i,b_0)
&=
	\big(\del^\nu_i \, \br\del^\mu_0 +
	(\alp^0)^2 \, \br g_{0i} \, \br\del^\nu_0 \, \br\del^\mu_0\big) \,
	\nab_\lam g_{\nu\mu} \,,
\\
	\nab_\lam g(b_i,b_j)
&=
	\big(\del^\nu_i \, \br\del^\mu_j +
	(\alp^0)^2 \, (\br g_{0i} \, \br\del^\nu_0 \, \del^\mu_j +
	\br g_{0j} \, \br\del^\mu_0 \, \del^\nu_i) +
	(\alp^0)^4 \,\br g_{0i} \, \br g_{0j} \, \br\del^\nu_0 \,
\br\del^\mu_0 	
	\big) \,
	\nab_\lam g_{\nu\mu} \,.
\end{align*}

 Analogously, we have
\bal
	\nab_\lam\ba g
&=
	\nab_\lam g^{\nu\mu} \, \der_\nu\ten\der_\mu
\\
&=
	\nab_\lam g^{\nu\mu} \, \big(\br\del^i_\nu \, b_i -
	(\alp^0)^2 \, \br g_{0\nu} \, b_0
	\big)\ten\big(	\br\del^j_\mu \, b_j -
	(\alp^0)^2 \, \br g_{0\mu} \, b_0
	\big)	
\\
&=
	\nab_\lam g^{\nu\mu} \, 
\big[\br\del^i_\nu \, \br\del^j_\mu \, b_i\ten b_j -
	(\alp^0)^2 \, \big( \br g_{0\nu} \, \br\del^j_\mu \, b_0 \ten b_j +
		\br g_{0\mu} \, \br\del^i_\nu \, b_i \ten b_0 
	\big) 
\\
&\quad+ 
	(\alp^0)^4 \, \br g_{0\nu} \, \br g_{0\mu} \, b_0 \ten b_0
	\big]		 \,.
\end{align*}
 
 Then, we obtain
\bal
	\nab_\lam \ba g(\bet^0,\bet^0) 
&= 
	(\alp^0)^4 \, \br g_{0\nu} \, \br g_{0\mu} \, \nab_\lam g^{\nu\mu} \,,
\qquad
	\nab_\lam \ba g(\bet^i,\bet^0 )
= 
	- (\alp^0)^2 \,  \br g_{0\mu} \, \br\del^i_\nu \, \nab_\lam g^{\nu\mu}
\,,
\\
	\nab_\lam \ba g(\bet^i,\bet^j )
&= 
	\br\del^i_\nu \, \br\del^j_\mu \, \nab_\lam g^{\nu\mu} \,. 
\end{align*}
\vglue-1.5\baselineskip
\end{proof}

\bLm
 We have the following coordinate expressions
\bAl\label{dKg}
d_K g
&=
	2\, (\nab_\lam \br g_{0\mu} - \br g_{0\sig}\,K\col\lam\sig\mu)
		\,(d^\lam\wed d^\mu)\ten \bet^0
\\
&\quad\nonumber
	+ 2 \, \big[\nab_\lam g_{i\mu} -  g_{i\sig} \, K\col\lam\sig\mu
		+ (\alp^0)^2 \, \br g_{0i} \,  
		(\nab_\lam \br g_{0\mu} - \br g_{0\sig} \, K\col\lam\sig\mu)
		\big] \, (d^\lam \wed d^\mu) \ten \bet^i \,,
\end{align}
\bAl\label{gdKg}
	g \ten d_K g 
&=
	2 \, \ha g_{00} \, (\nab_\lam \br g_{0\mu} - 
\br g_{0\sig} \, K\col\lam\sig\mu) \,
		(d^\lam \wed d^\mu) \ten (\bet^0 \ten \bet^0 \ten \bet^0)
\\
&\quad\nonumber
	+  
	2 \, \ha g_{ij} \, (\nab_\lam \br g_{0\mu} 
		- \br g_{0\sig} \, K\col\lam\sig\mu) \,
		(d^\lam \wed d^\mu) \ten (\bet^i \ten \bet^j \ten \bet^0)
\\
&\quad\nonumber
	+ 2 \, \ha g_{00} \, \big[\nab_\lam g_{i\mu} 
		-  g_{i\sig} \, K\col\lam\sig\mu
		+ (\alp^0)^2 \, \br g_{0i} \,  
		(\nab_\lam \br g_{0\mu} - \br g_{0\sig}\,K\col\lam\sig\mu)
		\big]
\\\quad \nonumber
&\qquad\qquad\qquad\qquad\qquad\qquad\qquad
(d^\lam \wed d^\mu) 		\ten (\bet^0 \ten \bet^0 \ten \bet^i)
\\
&\quad\nonumber
	+ 2 \, \ha g_{ij} \, \big[\nab_\lam g_{k\mu} 
		-  g_{k\sig} \, K\col\lam\sig\mu
	+ (\alp^0)^2 \, \br g_{0k} \,  
		(\nab_\lam \br g_{0\mu} - \br g_{0\sig} \, K\col\lam\sig\mu)
		\big]
\\ \nonumber
&\qquad\qquad\qquad\qquad\qquad\qquad\qquad
(d^\lam \wed d^\mu) \ten (\bet^i \ten \bet^j \ten \bet^k) \,.
\end{align}
\eLm

\begin{proof}
We have
\bal
d_K g 
&=
	2 \, (\nab_\lam g_{\rho\mu} - g_{\sig\rho} \, K\col\lam\sig\mu) \,
	 (d^\lam \wed d^\mu) \ten d^\rho
\\[2mm]
&=
	2 \, (\nab_\lam \br g_{0\mu} - \br g_{0\sig} \, K\col\lam\sig\mu) \,
		(d^\lam \wed d^\mu) \ten \bet^0
\\
&\quad	+ 
2 \, \big[\nab_\lam g_{i\mu} -  g_{i\sig} \, K\col\lam\sig\mu
		+ (\alp^0)^2 \, \br g_{0i} \,  
		(\nab_\lam \br g_{0\mu} - \br g_{0\sig} \, K\col\lam\sig\mu)
		\big] \, (d^\lam \wed d^\mu) \ten \bet^i \,.
\end{align*}

 We can write
\beq
g = g_{\lam\mu} \, d^\lam \ten d^\mu = 
\ha g_{00} \, \bet^0 \ten \bet^0
	+ \ha g_{ij} \, \bet^i \ten \bet^j \,,
\eeq
hence
\bal
	g \ten d_K g 
&=
	2 \,g_{\rho\sig} \, (\nab_\lam g_{\tau\mu} 
		- g_{\ome\tau} \, K\col\lam\ome\mu) \,
	 (d^\lam \wed d^\mu) \ten (d^\rho \ten d^\sig \ten d^\tau)
\\[2mm]
&=
	2 \, \ha g_{00} \, 
(\nab_\lam \br g_{0\mu} - \br g_{0\sig} \, K\col\lam\sig\mu) \,
		(d^\lam \wed d^\mu) \ten (\bet^0 \ten \bet^0 \ten \bet^0)
\\
&\quad	+  
	2 \, \ha g_{ij} \, (\nab_\lam \br g_{0\mu} 
		- \br g_{0\sig} \, K\col\lam\sig\mu) \,
		(d^\lam \wed d^\mu) \ten (\bet^i \ten \bet^j \ten \bet^0)
\\
&\quad
	+ 2 \, \ha g_{00} \, \big[\nab_\lam g_{i\mu} 
		-  g_{i\sig} \, K\col\lam\sig\mu
		+ (\alp^0)^2 \, \br g_{0i}\,  
		(\nab_\lam \br g_{0\mu} - \br g_{0\sig} \, K\col\lam\sig\mu)
		\big] \, 
\\
&\qquad\qquad\qquad \qquad\qquad\qquad\qquad		
(d^\lam \wed d^\mu) 		\ten (\bet^0 \ten \bet^0 \ten \bet^i)
\\
&\quad
	+ 2 \, \ha g_{ij} \, \big[\nab_\lam g_{k\mu} 
		-  g_{k\sig} \, K\col\lam\sig\mu
	+ (\alp^0)^2 \, \br g_{0k} \,  
		(\nab_\lam \br g_{0\mu} - \br g_{0\sig} \, K\col\lam\sig\mu)
		\big] 
\\
&\qquad\qquad\qquad \qquad\qquad\qquad\qquad
(d^\lam \wed d^\mu) 		\ten (\bet^i \ten \bet^j \ten \bet^k) \,. 
\end{align*}
\vglue-1.5\baselineskip
\end{proof}

\bLm\label{Lemma: dKg(X,Y)(d)}
 For each
$X, Y \in \sec(\f E, T\f E) \,,$
we have
\beq
d_K g(X,Y)(\K d) = 2 \, (\K d^*L_{\ti K} \, \ti g\Fla) (X,Y) \,,
\eeq
where
$\ti g\Fla \byd \id \ten g\Fla: 
\B T^* \ten T\f E \to \B T^* \ten (\B L^2 \ten T^*\f E) \,.$
\eLm

\begin{proof}
 The claim follows from the coordinate expressions
\bal
d_K g(X,Y) (\K d) 
&= 
c \, \alp^0 \, \big(\der_\lam g_{\mu\rho} 
	+ g_{\mu\sig} \, K\col\lam\sig\rho
	- \der_\mu g_{\lam\rho} - g_{\lam\sig} \, K\col\mu\sig\rho	\big) \,
\br\del^\rho_0 \, X^\lam \, Y^\mu \,,
\\
\K d^*L_{\ti K} \, \ti g\Fla 
&= 
c \, \alp^0 \, (\der_\lam g_{\mu\rho} 
	+ g_{\mu\sig} \, K\col\lam\sig\rho) \,
\br\del^\rho_0 \, d^\lam\wed d^\mu 	\,. 
\end{align*}
\vglue-1.5\baselineskip
\end{proof}

\bLm\label{Lemma: d* Ups prl (X,Y)}
 $\K d^* \, \wti\Ups\prl$ 
is a horizontal 2-form and, for each
$X, Y \in \sec(\f E, T\f E) \,,$
we have
\beq
\K d^* \, \wti\Ups\prl(X,Y) = 
\fr1{4\,c^2} \, \big(
	g(\K d, Y)(\nab_X g)(\K d, \K d) 	- g(\K d, X)(\nab_Y g)(\K d, \K d)
\big) \,.
\eeq
\eLm

\begin{proof}
 The coordinate expression of
$\K d^* \, \wti\Ups\prl$ 
from Proposition \ref{Pr4.11}
can be rewritten as
\beq
\K d^* \, \wti\Ups\prl = 
\tfr12 c \, (\alp^0)^3 \,
\br g_{0\mu} \, \nab_\lam \ha g_{00} \, d^\lam \wed d^\mu \,.
\eeq

 Then, we obtain
\bal
	\K d^* \, \wti\Ups\prl(X,Y) 
&= 
\fr14 \, c\,(\alp^0)^3 \, (
	\br g_{0\mu}\nab_\lam \ha g_{00} 
- \br g_{0\lam}\nab_\mu \ha g_{00}) \, X^\lam \, Y^\mu
\\
&= 
	\fr1{4\,c^2} \, \big(
	g(\K d, Y)(\nab_X g)(\K d, \K d) - g(\K d, X)(\nab_Y g)(\K d, \K d)
\big) \,. 
\end{align*}
\vglue-1.5\baselineskip
\end{proof}

\bLm
 We have the coordinate expressions
\bAl\label{Eq5.5}
   L_{\chi(K)} \, \tau
& = 
   \fr{\alp^0}{c} \, \big(\nab_\mu \br g_{0\lam}
   - \br g_{0\rho} \, K\col\mu\rho\lam
   + \tfr12 (\alp^0)^2 \, \br g_{0\lam} \, \nab_\mu {\ha g_{00}}
   \big) \, d^\lam \wed d^\mu\,,
\\[2mm] \label{Eq5.6}
  L_{R[\chi(K)]} \, \tau 
& = 
   - \fr{\alp^0}{c} \, \br g_{\lam\rho} \,
   R[K]\col{\mu\nu}\rho\sig \, \br\del^\sig_0 \,
   d^\lam\wed d^\mu\wed d^\nu\,,
\\[2mm] \label{Eq5.7}
   L_{\nu_{\tau}(X)} \, L_{\chi(K)} \, \tau 
& =
   \fr{1}{c^2} \, \big[\nab_\mu g_{i\lam} - g_{i\rho} \,
   K\col\mu\rho\lam
\\  
& \quad \nonumber
   + (\alp^0)^2\, \big(\br g_{0i} \, (\nab_\mu\br g_{0\lam}
   - \br  g_{0\rho} \, K\col\mu\rho\lam)
   + \br g_{0\lam} \, \nab_\mu \br g_{0i}
   + \tfr12 \, \br g_{i\lam} \, \nab_\mu \ha g_{00}
  \big)
\\ \nonumber
  & \quad 
  + (\alp^0)^4\, \br g_{0i} \,
  \br g_{0\lam} \, \nab_\mu{\ha g_{00}}
  \big] \, \ti X^i\, d^\lam \wed d^\mu\,.
\end{align}
\eLm

\begin{proof}
 The above equalities follow from
\eqref{Eq4.9}, 
\eqref{Eq4.11} 
and
\eqref{coordinate expression of L X L Gam tau},
by taking into account the equality
$\Gam = \chi(K)$ 
and the identities
$g_{p\mu} \, \br \del^p_\nu = 
g_{\nu\mu} - \br g_{0\mu} \, \del^0_\nu \,,$
$\br g_{0 p} \, \br \del^p_\nu = 
\br g_{0 \nu} - {\ha g_{00}}\, \del^0_\nu \,,$
and
$\ha g_{i p} \, \br \del^p_\nu = 
\br g_{i \nu} \,.$ 
\end{proof}

\bLm\label{Lemma: expression of L Gam tau}
 For each
$X, Y \in \sec(\f E, T\f E) \,,$
we have 
\bal
L_{\chi(K)} &\, \tau(X,Y) 
= 
 \fr1{c^2}(\K d^* L_{\ti K}\ti g\Fla) (X,Y) 
	- \fr1{c^2} \K d^* \wti\Ups\prl(X,Y)
\\
& = 
- \fr1{2\,c^2} \, d_K g(X,Y)(\K d) 
	+ \fr1{4\,c^4}\, \big(g(\K d,X)\,(\nab_Y g)(\K d, \K d)
		- g(\K d,Y)\,(\nab_X g)(\K d, \K d)\big) \,.
\end{align*}
\eLm

\begin{proof}
 It follows from the coordinate expressions
\eqref{dKg}, \eqref{Eq5.5} and 
Lemmas \ref{Lemma: dKg(X,Y)(d)} and
\ref{Lemma: d* Ups prl (X,Y)}. 
\end{proof}

\bLm\label{Lm5.5}
$L_{\chi(K)} \, \tau = 0$ 
if and only if, for each
$X, Y \in \sec(\f E, T\f E) \,,$
\bEq\label{000}
g(b_0,b_0) \, d_K g(X,Y)(b_0) + 
\tfr12 g(b_0,X) \, (\nab_Y g)(b_0,b_0) -
	\tfr12 g(b_0,Y) \, (\nab_X g)(b_0,b_0) = 0 \,.
\eEq
\eLm

\begin{proof}
 In virtue of the above 
Lemma \ref{Lemma: expression of L Gam tau}, 
we have
$L_{\chi(K)} \, \tau = 0$ 
if and only if
\beq
- \tfr1{2\,c^2} \, d_K g(X,Y)(\K d)
	+ \tfr1{4\,c^4} \, \big(g(\K d,X) \, (\nab_Y g)(\K d, \K d)
		- g(\K d,Y) \, (\nab_X g)(\K d, \K d)\big) = 0 \,.
\eeq

 Moreover, by multiplying the above equality by 
$2 \, c^4$ 
and by taking into account the fact that 
$g(\K d,\K d) = - c^2 \,,$ 
the above equality is equivalent to
\beq
g(\K d,\K d) \, d_K g(X,Y)(\K d)
	+ \tfr12 \big(g(\K d,X) \, (\nab_Y g)(\K d, \K d)
		- g(\K d,Y) \, (\nab_X g)(\K d, \K d)\big) = 0 \,,
\eeq
and, by recalling the coordinate expression
$\K d = c \, \alp^0 \, b_0 \,,$ 
we obtain \eqref{000}. 
\end{proof}

\bLm\label{Lm5.7}
$L_{\chi(K)} \,\tau = 0$ 
implies 
\bAl\label{00i}
g(b_0,b_0) &\, d_K g(X,Y)(b_i) 
 + 2\, g(b_0,b_i) \, d_K g(X,Y)(b_0) 
\\
& +g(b_0,X) \, (\nab_Y g)(b_0,b_i) - g(b_0,Y)\,(\nab_X g)(b_0,b_i)
\nonumber
\\
& 
	+ \tfr12  g(b_i,X) \, (\nab_Y g)(b_0,b_0) 
	- \tfr12  g(b_i,Y) \, (\nab_X g)(b_0,b_0)
= 0 \,,\nonumber
\\[2mm]
2\,g(b_i,b_j) &\, d_K g(X,Y)(b_0)  \label{0ij}
+ 2\,g(b_0,b_i) \, d_K g(X,Y)(b_j) + 2\,g(b_0,b_j) \, d_K g(X,Y)(b_i)  
\\
& + \nonumber
g(b_i,X) \, (\nab_Y g)(b_0,b_j) + g(b_j,X) \, (\nab_Y g)(b_0,b_i) 
\\ 
& \nonumber
- g(b_i,Y) \, (\nab_X g)(b_0,b_j) -  g(b_j,Y) \, (\nab_X g)(b_0,b_i)
\\ \nonumber
&
	+ g(b_0,X) \, (\nab_Y g)(b_i,b_j) 
	- g(b_0,Y) \, (\nab_X g)(b_i,b_j)
= 0 \,,
\\[2mm]
\label{ijk}
2\,g(b_i,b_j) & \, d_K g(X,Y)(b_k) 
+ 2\,g(b_k,b_i) \, d_K g(X,Y)(b_j) + 2\,g(b_k,b_j) \, d_K g(X,Y)(b_i)  
\\
& \nonumber
+ g(b_i,X) \, (\nab_Y g)(b_k,b_j) + g(b_j,X) \, (\nab_Y g)(b_k,b_i) 
\\
& \nonumber
- g(b_i,Y) \, (\nab_X g)(b_k,b_j) -  g(b_j,Y) \, (\nab_X g)(b_k,b_i)
\\ \nonumber
&
	+ g(b_k,X) \, (\nab_Y g)(b_i,b_j) 
	- g(b_k,Y) \, (\nab_X g)(b_i,b_j)
= 0 \,.
\end{align}
\eLm

\begin{proof}
$L_{\chi(K)} \, \tau = 0$ 
implies
$L_{\nu_\tau(Z)}^r \, L_{\chi(K)} \, \tau = 0 \,,$
for each
$Z \in \sec(\f E, T\f E)$
and for each integer
$r \geq 1 \,.$
 Moreover, 
$L_{\nu_\tau(Z)}^r \, L_{\chi(K)} \, \tau$ 
is
a horizontal 2--form and, for each,
$X,Y \in \sec(\f E,T\f E) \,,\, $
we have 
$(L_{\nu_\tau(Z)}^r \, L_{\chi(K)} \, \tau)(X,Y) =
\nu_\tau(Z)^r.L_{\chi(K)} \, \tau (X,Y) \,.$
 Let us denote the left hand side of
\eqref{000} by 
$K(X,Y) \,.$

 Then, from Lemma \ref{Lemma: expression of L Gam tau},
\bal
	\nu_\tau(Z)^r.L_{\chi(K)}\,\tau (X,Y)
&=
	\fr1{2\, c^2}\,\nu_\tau(Z)^{r}.\big( (c\,\alp^0)^3 K(X,Y)\big)
\\
&= 
		\fr {c}{2}\,\sum_{k=0}^{r} (\nu_\tau(Z)^{r-k}.(\alp^0)^3) \,
			(\nu_\tau(Z)^{k}.K(X,Y)) \,.
\end{align*}

 Now, 
$L_{\chi(K)} \, \tau = 0$ 
and  Lemma \ref{Lemma: expression of L Gam tau}
imply 
$\nu_\tau(Z)^{k}.K(X,Y) = 0 \,,\,$
$k = 1,\dots,r \,.$
 
 For 
$k =1 \,,$ 
we have
\bal
	\nu_\tau(Z).K(X,Y) 
& =
	\fr1{c\,\alp^0}\,\ti Z^i\,\der^0_i. K(X,Y)
\\
& =
	\fr1{c\,\alp^0}\,\ti Z^i\,\big[ g(b_0,b_0) \, d_K g(X,Y)(\der_i) 
		+ 2\, g(b_0,\der_i)  \, d_K g(X,Y)(b_0) 
\\
	&\quad	
		+ g(b_0,X) \, (\nab_Y g)(b_0,\der_i) 
- g(b_0,Y) \, (\nab_X g)(b_0,\der_i)			
\\
	& \quad
	+ \tfr12 g(\der_i,X) \, (\nab_Y g)(b_0,b_0) 
		- \tfr12 g(\der_i,Y) \, (\nab_X g)(b_0,b_0)	
		\big] 
\\
& =
	\fr1{c\,\alp^0}\,\ti Z^i\,\big[ g(b_0,b_0) \, d_K g(X,Y)(b_i) 
		+ 2\, g(b_0,b_i)  \, d_K g(X,Y)(b_0) 
\\
	&\quad	
		+ g(b_0,X) \, (\nab_Y g)(b_0,b_i) 
- g(b_0,Y) \, (\nab_X g)(b_0,b_i)			
\\
	& \quad
	+ \tfr12 g(b_i,X) \, (\nab_Y g)(b_0,b_0) 
		- \tfr12 g(b_i,Y) \, (\nab_X g)(b_0,b_0)	
\\
& \quad
  - 3\, (\alp^0)^2\, \br g_{0i}\, K(X,Y)
  \big] 
\end{align*}
which vanishes if and only if \eqref{00i} is satisfied.

 Similarly, the condition 
$\nu_\tau(Z)^{2}.K(X,Y) = 0$ 
implies
\eqref{0ij} and finally the condition 
$\nu_\tau(Z)^{3}.K(X,Y) = 0$ 
implies
\eqref{ijk}. 
\end{proof}

\bTh\label{Theorem: conditions for L Gam tau = 0}
$L_{\chi(K)} \, \tau  = 0$
if and only if the following condition hold, for each
$X, Y, Z \in \sec(\f E, T\f E) \,,$
\bEq\label{C}
g(Z,Z) \, d_K g(X,Y)(Z) + \tfr12 g(Z,Y) \, (\nab_{X} g)(Z,Z) -
\tfr12 g(Z,X) \, (\nab_{Y} g)(Z,Z) = 0 \,.
\eEq
\eTh

\begin{proof}
 In virtue of 
Lemma \ref{Lemma: expression of L Gam tau},
the condition \eqref{C} implies 
$L_{\chi(K)}\,\tau = 0 \,.$

 On the other hand, in an orthogonal adapted base, we have the
coordinate expression
$Z = \ti Z^0 \, b_0 + \ti Z^i \, b_i \,.$
 If 
$L_{\chi(K)} \, \tau = 0 \,,$ 
then, by linearity and in virtue of the above Lemmas \ref{Lm5.5} and
\ref{Lm5.7}, the condition \eqref{C} is satisfied.
\end{proof}

\bCr\label{Corollary: symmetry conditions for LGam tau = 0}
 If additionally
$K$ 
is a torsion free linear spacetime connection such that 
$\nab g$
and
$g \ten \nab g$
are symmetric, then
$L_{\chi(K)} \, \tau = 0 \,.$
\eCr

\begin{proof}
 For a torsion free connection 
$K$ 
the symmetry of 
$\nab g$ 
is equivalent to 
$d_K \, g = 0 \,.$
 Then, the conditions 
$\nab g$ 
and
$g \ten \nab g$ 
being symmetric imply 
\eqref{C} 
and this is equivalent to
$L_{\chi(K)} \, \tau = 0 \,.$ 
\end{proof}
\subsection{Almost-cosymplectic--contact structure}
\label{Almost-cosymplectic--contact structure: linear case}
 Let us analyze conditions for 
$(- c^2\,\tau, \Ome)$
to be a scaled almost--cosymplectic--contact structure.

\bLm\label{Lm5.12}
 We have the coordinate expression
\bAl\label{Eq5.10}
d\Ome
&=
- c \, \alp^0\big[
(\alp^0)^4 \, \br g_{0i} \, \br g_{0\mu} \,
\nab_\lam {\ha g_{00}}
\\ \nonumber
&\quad
+ (\alp^0)^2 \, \big(
\tfr12 \,\br g_{i\mu} \, \nab_\lam {\ha g_{00}}
+ \br g_{0\mu} \, \nab_\lam \br g_{0i}
+ \br g_{0i} \, (\nab_\lam \br g_{0\mu}
- \br g_{0\rho} \, K\col \lam\rho\mu)\big)
\\ \nonumber
&\quad
+ \nab_\lam g_{i\mu} 
- g_{i\rho} \, K\col \lam\rho\mu
\big] \,
( d^i_0 - \br\del^i_\tau\,K\col\nu\rho\sig \,
\br\del^\sig_0 \, d^\nu) \wed d^\lam \wed d^\mu
\\ \nonumber
&\quad
+ \tfr12 c \, \alp^0 \br g_{\mu\rho} \,
R[K]\cul\bet\lam\rho\sig \, \br\del^\sig_0 \,
d^\bet \wed d^\lam \wed d^\mu \,.
\end{align}
\eLm

\begin{proof}
 The identity
$\br g_{p\mu} \, \br\del^p_\nu = \br g_{\nu\mu}$
and 
\eqref{Eq4.5}
yield
\bal
d\Ome 
&=
- c \, \big(\der_\lam(\alp^0 \br g_{i\mu})
+ \der^0_i(\alp^0 \, \br g_{p\mu} \,
\br\del^p_\nu \, K\col\lam\nu\rho \, \br\del^\rho_0)\big) \,
d^i_0 \wed d^\lam \wed d^\mu
\\
&\quad
-  c \, \der_\bet\big(\alp^0 \, \br g_{p\mu} \,
\br\del^p_\nu \, K\col\lam\nu\rho \, \br\del^\rho_0
\big) \, d^\bet \wed d^\lam \wed d^\mu
\\[2mm]
&=
- c \, \alp^0 \big[
\tfr32 (\alp^0)^4 \, \br g_{0i}\, \br g_{0\mu} \, 
\nab_\lam {\ha g_{00}}
+ \nab_\lam g_{i\mu}
- g_{i\rho} \, K\col\lam\rho\mu
\\
&\quad
+ (\alp^0)^2 \, \big(
\tfr12 g_{i\mu} \, \nab_\lam {\ha g_{00}}
+ \br g_{0\mu} \, \nab_\lam \br g_{0i}
+ \br g_{0i} \, 
(\nab_\lam \br g_{0\mu} - \br g_{0\rho} \, K\col\lam\rho\mu)
\big)
\big] \,
d^i_0 \wed d^\lam \wed d^\mu
\\[2mm]
&=
- c \, \alp^0\big[
\tfr 32 (\alp^0)^4 \, \br g_{0\mu} \, \br g_{0\nu} \,
K\col\lam\nu\rho \, \br\del^\rho_0 \, \nab_\bet {\ha g_{00}}
+ (\alp^0)^2 \, \big(
\tfr12 g_{\mu\nu} \, K\col\lam\nu\rho \, \br\del^\rho_0 \,
\nab_\bet {\ha g_{00}}
\\
&\qquad
+ \br g_{0\nu} \, K\col\lam\nu\rho \, \br\del^\rho_0 \,
(\nab_\bet \br g_{0\mu}
- \br g_{0\tau} \, K\col\bet\tau\mu)
+ \br g_{0\mu} \, K\col\lam\nu\rho \, \br\del^\rho_0 \,
(\nab_\bet \br g_{0\nu}
- g_{\tau\nu} \, K\col\bet\tau\sig \, \br\del^\sig_0)
\\
&\qquad
- \tfr12 \br g_{0\mu} \, \br g_{0\nu} \, \br\del^\rho_0 \,
R[K]\cul\bet\lam\nu\rho
\big)
\\
&\quad
+ K\col\lam\nu\rho \, \br\del^\rho_0 \, (\nab_\bet g_{\mu\nu}
- g_{\tau\nu} \, K\col\bet\tau\mu)
- \tfr12 g_{\mu\nu} \, \br\del^\rho_0 \, R[K]\cul\bet\lam\nu\rho
\big] \,
d^\bet \wed d^\lam \wed d^\mu \,. 
\end{align*}
\vglue-1.5\baselineskip
\end{proof}

\bTh
 The pair 
$(-c^2\,\tau, \Ome)$ 
is a scaled almost--cosymplectic--contact structure if and only if
\bEq\label{Eq: conditions ACC linear}
L_{\nu_\tau(X)} \, L_{\chi(K)} \, \tau = 0 \,, 
\quad 
\Al X \in \sec(\f E, T\f E) \,, 
\ssep{and}
L_{R[\chi(K)]} \, \tau = 0 \,.
\eEq
\eTh

\begin{proof}
 It follows immediately from Theorem \ref{Conditions ACC}
by considering 
$\Gam = \chi(K) \,.$ 

 Eventually, we can prove directly Theorem by comparing the
coordinate expression \eqref{Eq5.10} with
\eqref{Eq5.6} and \eqref{Eq5.7}. 
\end{proof}
\subsection{Almost--coPoisson--Jacobi structure}
\label{Almost--coPoisson--Jacobi structure: linear case}
 Let us analyze conditions for 
$(- \tfr1{c^2}\,\gam, \Lam)$
to be a scaled contact structure.

\bLm\label{Lm5.17}
 We have the coordinate expression
\bAl\label{Eq5.13a}
	[\gam , \Lam] 
& =
	\big[
	- \tfr12 (\alp^0)^2\, \br g_{0k} \, (\br g^{j\lam} \, \br g^{k\rho}
	+ \br g^{j\rho} \, \br g^{k\lam}) \, \nab_\rho {\ha g_{00}}	
	+ (\br g^{j\sig} \, \br\del^\rho_0
	- \br g^{j\rho} \, \br\del^\sig_0) \, K\col\rho\lam\sig
\\
& \qquad \nonumber
	+ \br\del^\rho_0 \, \nab_\rho \br g^{j\lam}
	+ \tfr12 (g^{0\rho} \, \br g^{j\lam}
	+ g^{0\lam} \, \br g^{j\rho}) \, \nab_\rho \ha g_{00}
	\big] \,
	(\der_\lam + \br\del^i_\nu \, \br\del^\kap_0 \,
	K\col\lam\nu\kap \, \der^0_i) \wed \der^0_j 
\\
& \quad \nonumber
	+ \br g^{j\lam} \, \br\del^i_\nu \, \br\del^\rho_0 \, \br\del^\sig_0
	R[K]\cul\lam\rho\nu\sig \, \der^0_i \wed \der^0_j \,
\end{align}
and, in the adapted base 
\eqref{Eq: der to e}, 
\bAl
	[\gam, \Lam] 
& = 
	- (\alp^0)^2 \,\big[ \tfr12\, \br g^{j\rho}  
	\, ( \nab_\rho\ha g_{00}
	- 2\, \br g_{0\nu}\,\br\del^\sig_0\, K_\rho{}^\nu{}_\sig )
\\
& \qquad \nonumber
	+ \br g_{0\lam}\,\br\del^\rho_0\, (\nab_\rho\br g^{j\lam}
	+ \br g^{j\sig}\, K\col\rho\lam\sig)
	\big]\,e_0 \wed e^0_j
\\
&\quad \nonumber
	- \big[\tfr12\, (\alp^0)^2 \, \hat g^{ij}\,\br\del^\rho_0
	\, \nab_\rho {\ha g_{00}}
	+ \br g^{j\rho}\,\br\del^i_\nu\, \br\del^\sig_0\,K\col\rho\nu\sig 
\\
&\qquad \nonumber
	-\br\del^\rho_0\,\big( \nab_\rho \ha g^{ji}
	+ \br\del^i_\nu\, \br g^{j\sig} \,  K\col\rho\nu\sig
	+ \ha g^{pi}\,  \br\del^\sig_0\, K\col\rho{0}\sig)	
	\big]\,e_i \wed e^0_j
\\
&\quad \nonumber
	- \br g^{j\lam} \,\br\del^i_\nu\, \br\del^\sig_0\, \br\del^\rho_0 \,
	R[K]\col{\lam\rho}\nu\sig
	\, e^0_i \wed e^0_j\,.
\end{align}
\eLm

\begin{proof}
 From Lemma \ref{Lm4.6} and the identity
$(\alp^0)^2 \, {\ha g_{00}} = -1$
we get
\bal
[\gam , \Lam]
&=
\big(
- (\alp^0)^2 \, (\br g^{j\lam} \, \br\del^\rho_0
+ \br g^{j\rho} \, \br\del^\lam_0) \, (\tfr12 \,\der_\rho {\ha g_{00}}
+ \br g_{0p} \, \br\del^p_\nu \, \br\del^\sig_0 \, K\col\rho{\nu}\sig
)
\\
&\qquad
+ \br\del^\rho_0 \, \der_\rho \br g^{j\lam}
- g^{0\lam} \, \br\del^\rho_0 \, \br\del^j_\nu\,
  K\col\rho{\nu}\sig \, \br\del^\sig_0
- \del^\lam_p \, \br g^{j\nu} \, \br\del^p_\kap\,
  K\col\nu\kap\sig \, \br\del^\sig_0
+ \br g^{j\lam} \, \br\del^\rho_0 \,
  K\col\rho{0}\sig \, \br\del^\sig_0
\\
&\qquad
- \br g^{p\lam} \, \br\del^\rho_0 \, \br\del^j_\nu
  K\col\rho\nu\sig \, \del^\sig_p
\big)
  (\der_\lam + \br\del^i_\nu \br\del^\bet_0 \,
  K\col\lam\nu\bet \, \der^0_i) \wed \der^0_j
\\
&\quad
+ \br g^{j\lam} \, \br\del^i_\nu \, \br\del^\rho_0 \, \br\del^\sig_0\,
R[K]\cul\lam\rho\nu\sig \, \der^0_i \wed \der^0_j
\\[2mm]
& =
\big[
- \tfr12 (\alp^0)^2 \, (\br g^{j\lam} \, \br\del^\rho_0
+ \br g^{j\rho} \, \br\del^\lam_0) \, \nab_\rho {\ha g_{00}}
\\
&\qquad
+ \br\del^\rho_0 \, \nab_\rho \br g^{j\lam}
+ (\br g^{j\sig} \, \br\del^\rho_0
- \br g^{j\rho} \, \br\del^\sig_0) \, K\col\rho\lam\sig
\big]
(\der_\lam + \br\del^i_\nu \, \br\del^\kap_0 \,
K\col\lam\nu\kap \, \der^0_i) \wed \der^0_j
\\
&\quad
+ \br g^{j\lam} \, \br\del^i_\nu \br\del^\rho_0 \, \br\del^\sig_0 \,
R[K]\cul\lam\rho\nu\sig \, \der^0_i \wed \der^0_j \,.
\end{align*}

 Finally, we obtain \eqref{Eq5.13a} by using the identity
$\br\del^\rho_0 = 
\br g_{0k} \, \br g^{k\rho} - \tfr1{(\alp^0)^2} \, g^{0\rho} \,.$ 
\end{proof}

\bLm\label{Lemma: gam Lam = gam Lie Sha: linear case}
 In the adapted base \eqref{Eq: der to e} we have the coordinate
expression
\bAl \label{Eq6.13}
	[-\fr1{c^2}\, \gam, \Lam] 
& 
- \fr1{c^2}\, \gam\wed \Lam\Sha(L_\gam\,\tau) 
=
	\fr1{c^2}\,\big[\fr{(\alp^0)^2}{2\,c^2} \, 
	\ha g^{ij}\,\br\del^\rho_0\, \nab_\rho {\ha g_{00}}
	+ \br g^{j\rho}\,\br\del^i_\nu\, \br\del^\sig_0\,K\col\rho\nu\sig 
\\
& \qquad \nonumber	
	-\br\del^\rho_0\,\big( \nab_\rho \ha g^{ji}
	+ \br\del^i_\nu\, \br g^{j\sig} \,  K\col\rho\nu\sig
	+ \ha g^{pi}\,  \br\del^\sig_0\, K\col\rho{0}\sig
)	
	\big]\, e_i \wed e^0_j
\\
& \quad\nonumber
		+ \fr1{c^2}\, \br g^{j\lam} \,\br\del^i_\nu\, \br\del^\sig_0\,
	\br\del^\rho_0 \, R[K]\col{\lam\rho}\nu\sig
	\, e^0_i \wed e^0_j \,.
\end{align}
\vglue-1.5\baselineskip{\ }\hfill\ENDE
\eLm

\bLm\label{Lemma: Lam Lam - gam wed Lam Sha dtau: linear case}
 In the adapted base \eqref{Eq: der to e} we have the coordinate
expression
\begin{gather}\label{Eq6.14}
	[\Lam, \Lam] 
- 2\,\gam \wed (\Lam\Sha\ten\Lam\Sha)(d\tau) =
\\
\begin{align*}
=
	\fr2{c^2}\,\big[
	\big(&\tfr12 \, \ha g^{ki} \, \br g^{j\rho}\, 
	\nab_\rho {\ha g_{00}} 
	+ \fr1{(\alp^0)^2} \, \br g^{k\rho}\,
		(\nab_\rho\ha g^{ki} 
+ \br\del^i_\nu \, \br g^{j\sig} \, K\col\rho\nu\sig
+ \ha g^{pi} \, \br\del^\sig_0 \, K\col\rho{0}\sig
)
	\big)\,
	e_i \wed e^0_j \wed e^0_k
\\
&
	+ \fr1{(\alp^0)^2} \,
	\br g^{i\rho} \, \br g^{j\sig} \, \br\del^i_\nu \, \br\del^\sig_0 \,
R[K]\col{\rho\sig}\nu\sig 
		\,
	e^0_i \wed e^0_j \wed e^0_k 	\big] \,.
\end{align*}
\end{gather}
\vglue-1.7\baselineskip{\ }\hfill\ENDE
\eLm

\bTh\label{conditions AcoPJ: linear case}
 The pair
$(-\tfr1{c^2} \gam, \, \Lam)$
is a scaled almost--coPoisson--Jacobi pair
if and only if 
the conditions \eqref{Eq: conditions ACC linear}
are satisfied.
\eTh

\begin{proof}
 It follows immediately from Theorem 
\ref{Theorem: conditions for AcoPJ structure; general case}
by considering $\Gam = \chi(K) \,.$ 

 Eventually, we can prove directly Theorem by comparing the
coordinate expressions \eqref{Eq6.13} and 
\eqref{Eq6.14} with
\bal
{}\Sha(L_{R[\Gam]} \, \tau) 
&= 
- \tfr{2\,\h\,\alp^0}{m\,c^5} \br g^{k\lam}\,\br\del^\rho_0\, 
	\br\del^i_\nu\, \br\del^\sig_0\,R\col{\rho\lam}\nu\sig \, 
e^0_i \wed e_0 \wed e^0_k
- \tfr{1}{c^4\,(\alp^0)^2} \,\br g^{j\lam}\,\br g^{k\mu}
     \, \br\del^i_\nu\, \br\del^\sig_0\,
		R\col{\lam\mu}\nu\sig  \, e^0_i \wed e^0_j \wed e^0_k
\,
\end{align*}
and
\bal
	{}\Sha(L_{\Lam\Sha(\ome)} \, L_{\chi(K)} \, \tau) 
& = 
	- \tfr{\h\,\alp^0}{m\,c^5} \, 
	\big[\tfr12\,	(\alp^0)^2 \, \ha g^{ji} \,
	\br\del^\rho_0\, \nab_\rho {\ha g_{00}}
	+ \br g^{j\rho} \,\br\del^i_0\,\br\del^\sig_0\, 
	K\col\rho\nu\sig
\\
&\qquad
	- \br\del^\rho_0 \, ( \nab_\rho \ha g^{ji}
	+ \br\del^i_\nu\, \br g^{j\sig} \,  K\col\rho\nu\sig
	+ \ha g^{pi}\,  \br\del^\sig_0\, K\col\rho{0}\sig)
	\big]\, \wti \ome_i \, e_0 \wed e^0_j
\\
&\quad 
	- \tfr1{c^4 \, (\alp^0)^2} \,
	\big[\tfr12\, (\alp^0)^2\,\ha g^{ki}\,\br g^{j\rho}
	\, \nab_\rho\ha g_{00} 
\\
& \qquad
		- \br g^{j\rho}\, (\nab_\rho\ha g^{ki} 
	+ \br\del^i_\nu\, \br g^{j\sig}\,K\col\rho\nu\sig
	+ \ha g^{pi}\,\br\del^\sig_0\, K\col\rho{0}\sig)\big]\,
	\wti \ome_i \, e^0_j \wed e^0_k\,.
\end{align*}

 Then, 
$(-\tfr1{c^2}\,\gam,\Lam)$
is a scaled almost--coPoisson--Jacobi structure if and only if
\beq
{}\Sha(L_{\Lam\Sha(\ome)} \, L_\Gam \, \tau) = 0 \,,
\sep{for each}
\ome \in \sec(\M J_1\f E, \, V^*_\gam\M J_1\f E) \,,
\sep{and}
{}\Sha(L_{R[\Gam]} \, \tau) = 0 \,,
\eeq
i.e. if and only if
$L_{\nu_\tau(X)} \, L_\Gam \, \tau = 0 \,,$
for each
$X\in\sec(\f E,T\M J_1\f E) \,,$
and 
$L_{R[\Gam]} \, \tau = 0 \,.$ 
\end{proof}

 We can summarize the above results as follows.

\bTh
 The following assertions are equivalent.

\smallskip

{\rm (1)} 
$L_{\nu_\tau(X)}\, L_{\chi(K)} \,\tau = 0 \,,$
for each 
$X \in \sec(\f E, T\f E) \,,$
and 
$L_{R[\chi(K)]}\,\tau =0 \,.$\

\smallskip

{\rm (2)}
$\Ome = 0 \,,$ 
i.e. the pair
$(-c^2 \, \tau, \, \Ome)$ 
is a scaled almost--cosymplectic--contact pair;

\smallskip

{\rm (3)}
$[-\fr1{c^2}\,\gam,\Lam] = \fr1{c^2}\gam\wed\Lam\Sha(L_\gam\,\tau)
\,,$ 
$[\Lam,\Lam] = 2\,\gam \wed (\Lam\Sha\ten\Lam\Sha)(d\tau) \,,$
i.e. the pair
$(-\tfr1{c^2} \gam, \, \Lam)$
is a scaled almost--coPoisson--Jacobi pair.
\hfill\ENDE
\eTh
\subsection{Contact structure}
\label{Contact structure: linear case}
 Let us analyse conditions for 
$(- c^2\,\tau, \Ome)$
to be a scaled contact structure.

\bTh\label{Th6.19}
$\Ome = - c^2\, d\tau \,,$ 
i.e. the pair
$(-c^2 \, \tau, \, \Ome)$ 
is a scaled contact structure,
if and only if the identity \eqref{C}
is satisfied.
\eTh

\begin{proof}
 It follows immediately from 
Theorems \ref{Theorem: LGam tau condition for contact structure} 
and \ref{Theorem: conditions for L Gam tau = 0}. 
\end{proof}

\bCr\label{Corollary: symmetry conditions for contact structure}
 In the particular case when
$K$
is torsion free and the tensors 
$\nab g$ 
and 
$g \ten \nab g$ 
are symmetric, the pair
$(-c^2 \, \tau, \, \Ome)$ 
is a scaled contact structure. 
\eCr

\begin{proof} 
 This follows from 
Theorem \ref{Theorem: LGam tau condition for contact structure}, 
Corollary \ref{Corollary: symmetry conditions for LGam tau = 0} 
and Theorem \ref{Theorem: conditions for L Gam tau = 0}. 
\end{proof}

 Let us recall that, by Theorem \ref{Th3.12} and Proposition
\ref{Pr4.11},
\bEq\label{Eq5.11}
\Ome[g,K] = 
\K d^* \, \wti\Ups\per =
\K d^* \, \wti\Ups - 
\K d^* \, \wti\Ups\prl \,.
\eEq

 Then, we have the following result.

\bLm\label{Lm6.8}
\beq
\Ome = \K d^* \, \wti\Ups
\eeq
if and only if
\beq
g(\K{d},X) \, \nab_Y g(\K{d},\K{d}) = 
g(\K{d},Y) \, \nab_X g(\K{d},\K{d}) \,,
\qquad
\Al X,Y\in \sec(\f E,T\f E) \,.
\eeq
\eLm

\begin{proof}
 The coordinate expression of
$\K d^* \, \wti \Ups\prl$ 
can be rewritten as
\beq
\K d^* \, \wti \Ups\prl = 
\tfr 12 c \, (\alp^0)^3 \,
\br g_{0\mu} \, \nab_\lam \ha g_{00} \, d^\lam \wed d^\mu \,.
\eeq

 Then,
$\Ome = \K d^* \, \wti \Ups$
if and only if
$\K d^* \, \wti \Ups\prl = 0 \,,$ 
i.e., if and only if
\beq
\br g_{0\mu} \, \nab_\lam \ha g_{00} -
\br g_{0\lam} \, \nab_\mu \ha g_{00} = 0 \,,
\eeq
which is equivalent to 
\beq
g(\K d,X) \, \nab_Y g(\K d,\K d) = 
g(\K d,Y) \, \nab_X g(\K d,\K d)\,. 
\eeq
\vglue-1.5\baselineskip
\end{proof}

\bRm\label{Rm6.9}
 By Lemma \ref{Lemma: expression of L Gam tau}, Theorems
\ref{Theorem: conditions for L Gam tau = 0} and \ref{Th6.19}, the two
conditions
\bal
d_K g  & =0\,,
\\
g(\K d,X) \, \nab_Y g(\K d,\K d) & = 
g(\K d,Y) \, \nab_X g(\K d,\K d)
\end{align*}
imply that 
$(-c^2\,\tau,\Ome)$ 
is a scaled contact pair.
 But this fact can be viewed also as a consequence of Lemma
\ref{Lm6.8} and Theorem
\ref{Th2.7}, by considering the fact that 
$d_K g = 0$ 
is equivalent to
$\ti \Ups = d\ti g\Fla \,,$
where 
$\ti g\Fla = \id \ten g\Fla$ 
is the Liouville 1--form on 
$\ti T\f E \,,$
and 
$\K d^* d\ti g\Fla = - c^2\,\tau \,.$
\hfill\ENDE
\eRm
 
\bCr\label{Cr5.9}
 We have
\beq
\K d^* \, \wti\Ups - \K d^* \, \wti\Ups\prl
= - c^2 \, (d\tau + L_\Gam \, \tau) \,.
\eeq
\eCr

\begin{proof}
 It follows from \eqref{Eq5.11} and Corollary \ref{Cr4.9}. 
\end{proof}
\subsection{Jacobi structure}
\label{Jacobi structure: linear case}
 Let us analyse conditions for 
$(- \tfr1{c^2}\,\gam, \Lam)$
to define a scaled Jacobi structure.

\bLm\label{Lm5.11}
 We have the coordinate expression
\bAl
[\Lam, \Lam] - & \fr2{c^2} \gam \wed \Lam
=
\fr2{c^2} \, \big[\tfr12 \br g^{j\rho} \,
\br g^{k\lam} \, \nab_\rho{\ha g_{00}}
\\
&\quad \nonumber
+ \fr1{(\alp^0)^2} \, \br g^{k\rho} \, (\nab_\rho \br g^{j\lam}
+ \br g^{j\sig} \, K\col\rho\lam\sig)\big]
(\der_\lam + \br\del^i_\nu \, K\col\lam\nu\ome \,
\br\del^\ome_0 \, \der^0_i) \wed \der^0_j \wed \der^0_k
\\ \nonumber
&
+ \fr1{(c\,\alp^0)^2} \, \br g^{i\rho} \, \br g^{j\sig} \,
\br\del^k_\nu \, R[K]\cul\rho\sig\nu\ome \, \br\del^\ome_0 \,
\der^0_i \wed \der^0_j \wed \der^0_k \,
\end{align}
and, in the adapted base
\eqref{Eq: der to e}, we have the coordinate expression
\bAl
[\Lam, \Lam] - & \fr2{c^2} \gam \wed \Lam
= \fr2{c^2} \, \big[\tfr12 \br g^{j\rho} \,
\br g^{k\lam} \, \nab_\rho{\ha g_{00}}
- \br g^{k\rho}\, \br g_{0\lam}\, (\nab_\rho\br g^{j\lam} 
+ \br g^{j\sig}\, K\col\rho\lam\sig)
\\
&\quad\nonumber
+ \fr1{(\alp^0)^2} \, \br g^{k\rho} \, (\nab_\rho \ha g^{ij}
+ \br\del^i_\nu\,\br g^{j\sig}\, K\col\rho\nu\sig
+ \ha g^{ij} \,\br\del^\sig_0\, K\col\rho{0}\sig)\big]
\, e_i \wed e^0_j \wed e^0_k
\\ \nonumber
&
+ \fr2{(c\,\alp^0)^2} \, \br g^{i\rho} \, \br g^{j\sig} \,
\br\del^k_\nu \, \br\del^\mu_0 \, R[K]\cul\rho\sig\nu\mu \,
e^0_i \wed e^0_j \wed e^0_k \,.
\end{align}
\eLm

\begin{proof}
 It follows from
\eqref{[Lam, Lam]},
\eqref{Eq: adapted coordinate expression of [Lam, Lam]},
\eqref{Eq: gam Lam general}
and
\eqref{Eq: Lam and Lam - 2 gam wed Lam}. 
\end{proof}

\bTh\label{Th5.12}
The pair
$(- \tfr1{c^2} \gam, \, \Lam)$
is a scaled Jacobi structure if and only if the condition 
\eqref{C} 
is satisfied.
\eTh

\begin{proof}
 It follows from
Theorems \ref{Theorem: conditions for Jacobi structure; general case}
and 
\ref{Theorem: conditions for L Gam tau = 0}. 
\end{proof}

\bCr\label{Corollary: symmetry condition for Jacobi structure}
 In the particular case when
$K$
is torsion free and the tensors
$\nab g$ 
and 
$g \ten \nab g$ 
are symmetric,
$(-\tfr1{c^2} \, \gam, \, \Lam)$
is a scaled Jacobi structure.
\eCr

\begin{proof}
 It follows from 
Corollary \ref{Corollary: symmetry conditions for LGam tau = 0} 
and Theorem \ref{Th5.12}. 
\end{proof}

 We can summarize the main results of the previous two sections as
follows.

\bTh\label{Th5.6}
 The following assertions are equivalent.

\smallskip

{\rm (1)} 
$g(Z,Z) \, d_K g(X,Y)(Z) + \tfr12 g(Z,Y) \, (\nab_{X} g)(Z,Z) -
\tfr12 g(Z,X) \, (\nab_{Y} g)(Z,Z) = 0 \,;$

\smallskip

{\rm (2)}
$\Ome = - c^2 d\tau \,,$ 
i.e. the pair
$(-c^2 \, \tau, \, \Ome)$ 
is a scaled contact pair;

\smallskip

{\rm (3)}
$[-\fr1{c^2}\,\gam,\Lam] = 0 \,,$ 
$[\Lam,\Lam] = \fr2{c^2}\gam \wed \Lam \,,$
i.e. the pair
$(-\tfr1{c^2} \gam, \, \Lam)$
is a scaled Jacobi pair. 
\eTh

\begin{proof}
 It follows immediately from 
Theorems \ref{Th6.19} 
and \ref{Th5.12}. 
\end{proof}

\section{Metric and non-metric phase objects}
\label{Metric and non-metric phase objects}
\setcounter{equation}{0}
 So far, we have analysed the scaled phase 2--form 
$\Ome$
and the scaled phase 2--vector
$\Lam$
associated with the metric 
$g$
and a generic phase connection 
$\Gam \,.$

 On the other hand, the metric
$g$
yields the Levi--Civita spacetime connection\linebreak
$K \byd K[g] \,,$
hence the distinguished metric phase connection
$\Gam[g] \byd \chi(K[g]) \,.$ 
 Accordingly, we obtain the distinguished scaled phase 2--form
$\Ome[g] \byd \Ome[g,\Gam[g]] \,,$
the scaled phase 2--vector
$\Lam[g] \byd \Lam[g,\Gam[g]]$
and the dynamical connection
$\gam[g] \byd \gam[\Gam[g]] \,.$

 Then, any phase connection 
$\Gam$
and the associated scaled phase 2--form
$\Ome$
and scaled phase 2--vector
$\Lam$
split into a metric component and an additional non--metric
component.

 This splitting provides further information on our phase structures.
 In particular, we shall see the effect of an additional closed
spacetime 2--form, i.e. of an electromagnetic field.
\subsection{Metric contact and Jacobi structures}
\label{Metric contact and Jacobi structures}
 We start by considering the metric scaled phase 2--form
$\Ome[g]$
and scaled phase 2--vector
$\Lam[g] \,.$

\bTh\label{Theorem: metric contact and Jacobi structures}
 The pairs
$(- c^2 \, \tau, \, \Ome[g])$
and
$(- \tfr1{c^2} \gam, \, \Lam[g])$
are a scaled contact structure and a scaled Jacobi structure,
respectively, which are mutually dual.
\eTh

\begin{proof}
	 The hypothesis
$\nab g = 0$
and 
Corollary \ref{Corollary: symmetry conditions for contact structure}
imply that
$(- c^2 \, \tau, \, \Ome[g])$
is a contact structure.
 On the other hand, we can also prove that
$\Ome[g] = - d (c^2 \, \tau) \,,$
by a direct computation in coordinates.

 Analogously, the hypothesis
$\nab g = 0$
and 
Corollary \ref{Corollary: symmetry condition for Jacobi structure} 
prove that
$(- \tfr1{c^2} \gam, \, \Lam[g])$
is a Jacobi structure. 
\end{proof}
\subsection{Non--metric contact and Jacobi structures}
\label{Non--metric contact and Jacobi structures}
 Next, we consider any phase connection
$\Gam$
and the induced scaled 2--form
$\Ome \byd \Ome[g,\Gam] \,,$
scaled 2--vector
$\Lam \byd \Lam[g,\Gam]$
and dynamical connection
$\gam \byd \gam[\Gam] \,.$

 Here, we perform a further analysis of the conditions by which the
pairs
$(-c^2 \, \tau, \, \Ome)$
and
$(\tfr1{c^2} \, \gam, \, \Lam)$
are a scaled contact structure and a scaled Jacobi structure,
respectively.

 In virtue of 
Theorems \ref{Th4.22} and \ref{Th5.6}, 
the conditions for the contact structure and for the Jacobi structure 
coincide, for any phase connection
$\Gam \,.$ 
 For this reason, it suffices to prove explicitly only the
conditions for the contact structure, which give the corresponding
dual Jacobi structure, at the same time.

\bPr\label{Proposition: modified objects}
 Any phase connection 
$\Gam$
can be uniquely written as
\beq
\Gam = \Gam[g] + \Sig \,,
\ssep{with}
\Sig : \M J_1\f E \to T^*\f E \ten V\M J_1\f E \,.
\eeq
i.e., in coordinates, as
\beq
\Gam\Ga\lam i = \Gam[g]\Ga\lam i + \Sig\Ga\lam i \,,
\ssep{with}
	\Sig\Ga\lam i \in \map(\M J_1\f E, \Rn)\,.
\eeq

 Correspondingly, the scaled phase 2--form 
$\Ome$
and the scaled phase 2--vector
$\Lam$
split as
\beq
\Ome \byd \Ome[g,\Gam] = \Ome[g] + \Ome\Ela[g, \Sig]
\ssep{and}
\Lam \byd \Lam[g,\Gam] = \Lam[g] + \Lam\Ela[g, \Sig] \,,
\eeq
where
\bat{2}
\Ome\Ela[g,\Sig] 
&\byd
g\con \big((\nu_\tau \com \Sig) \wed \tht) 
&&:
	\M J_1\f E \to (\B T^* \ten \B L^2) \ten \Lam^2 T^*\f E \,,
\\
\Lam\Ela[g,\Sig]
&\byd
\ba g \con (\Sig \wed \nu_\tau) 
&&:
\M J_1\f E \to (\B T \ten \B L^{-2}) \ten \Lam^2V\M J_1\f E \,,
\end{alignat*}
have the coordinate expressions
\beq
\Ome\Ela[g,\Sig] =
- c \, \alp^0 \, \br g_{i\mu} \, \Sig\Ga\lam i \, d^\lam \wed d^\mu
\ssep{and}
\Lam\Ela[g,\Sig] =
\fr1{c \, \alp^0} \, \br g^{j\lam} \, \Sig\Ga\lam i \, 
\der^0_i \wed \der^0_j \,.
\eeq

 Moreover, we have
\beq
- c^2 \, \tau \wed \Ome[g,\Gam] \wed \Ome[g,\Gam] \wed \Ome[g,\Gam] =
- c^2 \, \tau \wed \Ome[g] \wed \Ome[g] \wed \Ome[g] \,.
\eeq
 \vglue-1.5\baselineskip{\ }\hfill\ENDE
\ePr

\bLm\label{Ome and symmetry of Sig}
 Let us define the scaled (0,2)--tensor
\beq
\ul\Sig \byd (g\Fla \com \nu^{-1}_\tau) (\Sig) : 
\M J_1\f E \to (\B T^* \ten \B L^2) \ten T^*\f E \ten T^*\f E \,,
\eeq
with coordinate expression
\beq
\ul\Sig = 
c \, \alp^0 \, \br g_{i\mu} \, \Sig\Ga\lam i \, d^\lam \ten d^\mu \,.
\eeq

 Indeed, we have
\beq
\Ome\Ela[g,\Sig] = - \Alt \ul\Sig \,.
\eeq

 Hence, the following conditions are equivalent:

\smallskip
1) $\Ome[g,\Gam] = \Ome[g] \,;$
\qquad\qquad
2) $\Ome\Ela[g, \Sig] = 0 \,;$
\qquad\qquad
3) $\ul\Sig$
is symmetric.
\eLm

\begin{proof}
 The equality
$\Ome\Ela[g,\Sig] = - \Alt \ul\Sig$
follows from the coordinate expressions of
$\Ome\Ela[g,\Sig]$
and
$\Alt \ul\Sig \,.$
 Then, this equality implies the equivalence of conditions 1), 2) and
3). 
\end{proof}

\bLm\label{LSig tau}
 The scaled 2--form
$L_\Sig \, \tau : \M J_1\f E \to 	\B T \ten \Lam^2T^*\f E$
fulfills the equality
\beq
L_\Sig \, \tau = \fr1{c^2} \, \Ome\Ela[g, \Sig] =
- \fr1{c^2} \, \Alt \ul\Sig \,.
\eeq
\eLm

\begin{proof}
 The equalities
$d\tau = - \fr1{c^2} \, \Ome[g]$
and
$i_\Sig \tau = 0$
imply
\beq
L_\Sig \, \tau =
i_\Sig d\tau - d i_\Sig \tau =
\fr1{c^2} \, i_\Sig \Ome[g] \,,
\eeq
i.e., in coordinates,
\beq
L_\Sig \, \tau = 
- \fr{\alp^0}c \, \br g_{i\mu} \, \Sig\Ga\lam i \, d^\lam \wed d^\mu =
\fr1{c^2} \Ome\Ela[g, \Sig] \,. 
\eeq
 \vglue-1.7\baselineskip
\end{proof}

\bTh\label{Theorem: condition on Sig for non metric contact structure}
 The pairs
$(- c^2 \, \tau, \, \Ome)$
and
$(- \tfr1{c^2} \gam, \Lam)$
are scaled contact and Jacobi structures if and only if
$\ul\Sig$
is symmetric.
\eTh

\begin{proof}
 \emph{1st Proof.}
 In virtue of 
Theorem \ref{Theorem: metric contact and Jacobi structures},
we have
$\Ome[g] = - c^2 \, d\tau \,.$
 Hence, we obtain
$\Ome = - c^2 \, d\tau$
if and only if
$\Ome\Ela[g,\Sig] = 0 \,,$
i.e., in virtue of Lemma \ref{Ome and symmetry of Sig}, if and only if
$\ul\Sig$
is symmetric.

 \emph{2nd Proof.}
 Theorem \ref{Theorem: LGam tau condition for contact structure}
says that the pair
$(- c^2 \, \tau, \, \Ome[g,\Gam])$
is a scaled contact structure if and only if 
$L_\Gam \, \tau = 0 \,.$
 On the other hand, in virtue of 
Corollary \ref{Corollary: symmetry conditions for LGam tau = 0},
we have
$L_{\Gam[g]} \, \tau = 0 \,.$
 Hence, the result follows from
Lemma \ref{LSig tau}. 
\end{proof}

\bNt
 Let 
$\Gam$ 
and 
$\Gam'$ 
be two phase connections. Then, we have 
$\Ome[g,\Gam] = \Ome[g,\Gam'] \,,$
respectively 
$\Lam[g,\Gam] = \Lam[g,\Gam'] \,,$
if and only if the difference tensor
$\Sig \byd \Gam - \Gam' : 
\M J_1\f E \to T^*\f E \ten V\M J_1\f E$  
is such that 
$\bau\Sig \byd g\Fla(\bau \Phi)$
is symmetric.

 Thus, the relations 
$\Ome[g,\Gam] = \Ome[g,\Gam'] \,,$
respectively 
$\Lam[g,\Gam] = \Lam[g,\Gam'] \,,$
define an equivalence relation on the space of phase connections.

 We are mainly interested in the phase connections which are
equivalent to
$\Gam[g] \,,$
because they yield an exact phase 2--form
$\Ome \,.$
\hfill\ENDE
\eNt

\bEx\label{Example: Sig induced by nu tau}
 Let us consider the vertical valued 1--form
\beq
\Sig \byd \fr{c^2\,m}{\h} \, \nu_\tau : 
\M J_1\f E \to T^*\f E \ten V\M J_1\f E \,,
\eeq 
with coordinate expression
\beq
\Sig = \fr{c\,m}{\h\,\alp^0} \, \br\del^i_\mu \, d^\mu \ten 	\der^0_i
\,.
\eeq

 Then, we obtain the symmetric scaled (0,2)--tensor
\beq
\ul\Sig = \fr{c^2\,m}{\h} \, g\per :
\M J_1\f E \to (\B T^* \ten \B L^2) \ten T^*\f E \ten T^*\f E \,,
\eeq
with coordinate expression
\beq
\ul\Sig = 
\fr{c^2 \, m}{\h} \, \br g_{i\mu} \, \br\del^i_\lam \, 
d^\lam \ten d^\mu =
\fr{c^2 \, m}{\h} \, 
(g_{\mu\lam} + (\alp^0)^2 \, \br g_{0\mu} \, \br g_{0\lam}) \, 
d^\lam \ten d^\mu \,,
\eeq
and the vanishing scaled phase 2--form and 2--vector
\bal
\Ome\Ela[g,\Sig] 
&:
	\M J_1\f E \to (\B T^* \ten \B L^2) \ten \Lam^2 T^*\f E \,,
\\
\Lam\Ela[g,\Sig] 
&:
	\M J_1\f E \to (\B T \ten \B L^{-2}) \ten \Lam^2 T\f E \,,
\end{align*}
with coordinate expressions
\bat{3}
\Ome\Ela[g,\Sig] 
&=
- \fr{c^2\,m}\h \, \br g_{i\mu} \, \br\del^i_\lam \, 
d^\lam \wed d^\mu 
&&=
- \fr{c^2 \, m}{\h} \, 
(g_{\mu\lam} + (\alp^0)^2 \, \br g_{0\mu} \, \br g_{0\lam}) \, 
d^\lam \wed d^\mu 
&&= 0 \,,
\\
\Lam\Ela[g,\Sig] 
&=
\fr{m}{\h \, (\alp^0)^2} \, \br g^{j\lam} \, \br\del^i_\lam \, 
\der^0_i \wed \der^0_j
&&=
\fr{m}{\h \, (\alp^0)^2} \, \ha g^{ij} \, \der^0_i \wed \der^0_j
&&=
0 \,.
\end{alignat*}

 Thus, in virtue of 
Theorem \ref{Theorem: condition on Sig for non metric contact
structure}, the induced pairs
$(-c^2 \, \tau, \, \Ome[g,\Gam])$
and
$(- \tfr1{c^2} \gam, \Lam[g,\Gam])$
turn out to be scaled contact and Jacobi structures,
respectively.
\hfill\ENDE
\eEx

\bLm
 Let us consider a scaled (0,3)--tensor
\beq
\phi : \f E \to \B L^2 \ten (T^*\f E \ten T^*\f E \ten T^*\f E)
\eeq
and the induced linear spacetime vertical valued 1--form
\beq
\Phi \byd \Phi[g,\phi] \byd (\ups \com g\Sha^2) (\phi) :
T\f E \to T^*\f E \ten VT\f E \,,
\eeq
with coordinate expression
\beq
\Phi = 
g^{\nu\rho} \, \phi_{\lam\rho\mu} \, \dt x^\mu \,
d^\lam \ten \dt\der_\nu \,.
\eeq

 Then, we obtain the linear spacetime connection
$K \byd K[g,\phi] \byd K[g] + \Phi \,,$
hence the phase connection
$\Gam \byd \Gam[g,\phi] \byd \chi(K) \,,$
which splits as
$\Gam = \Gam[g] + \Sig[g,\phi] \,,$
where the vertical valued 1--form
\beq
\Sig \byd \Sig[g,\phi] : \M J_1\f E \to T^*\f E \ten V\M J_1\f E
\eeq
has the coordinate expression
\beq
\Sig = (\br g^{i\rho}\, \phi_{\lam\rho\sig} \, \br\del^\sig_0) \,
d^\lam \ten \der^0_i \,.
\eeq

 Thus, we obtain the associated scaled phase 2--form and 2--vector
\beq
\Ome \byd \Ome[g, \Gam] = \Ome[g] + \Ome\Ela[g,\Sig]
\ssep{and}
\Lam \byd \Lam[g, \Gam] = \Lam[g] + \Lam\Ela[g,\Sig] \,.
\eeq

Indeed, the pairs
$(- c^2 \, \tau, \, \Ome)$
and
$(- \fr1{c^2} \, \gam, \, \Lam)$
are scaled contact and Jacobi structures if and only if 
$L_\Gam \, \tau = L_\Sig \, \tau =0 \,,$
i.e., by 
Theorem \ref{Theorem: conditions for L Gam tau = 0}, 
if and only if the condition \eqref{C} is satisfied, i.e. if and only if,
for each
$X, Y, Z \in \sec(\f E, T\f E) \,,$
\beq
g(Z, Z ) \, \phi(X,Y,Z) - 
g(Z,Z) \, \phi(Y,X,Z) + 
g(Z,X) \, \phi(Y,Z,Z) -
g(Z,Y) \, \phi(X,Z,Z) = 0.\hfill\ENDE
\eeq
\eLm
\subsection{Further examples of non--metric phase structures}
\label{Further examples of non--metric phase structures}
 Let us discuss further distinguished examples of non--metric phase
structures generated by a horizontal scaled phase (0,2)--tensor.
 In the cases when this tensor is symmetric or antisymmetric we
obtain different results.

Let us start with the general case.

\bLm\label{Lemma: structures generated by sig}
 Given a scaled (0,2)--tensor
\beq
\sig : \M J_1\f E \to
	(\B T^*\ten \B L^2) \ten T^*\f E \ten T^*\f E \,,
\eeq
we obtain the vertical valued 1--form
\beq
\Sig \byd 
\Sig[g,\sig] \byd (\nu_\tau \com g\Sha^2)(\sig) :
\M J_1\f E \to T^*\f E \ten V\M J_1\f E \,,
\eeq
with coordinate expression
\beq
\Sig = 
\fr1{c\, \alp^0} \, \br g^{i\rho} \, \sig_{\lam\rho} \, 
	d^\lam \ten \der^0_i \,.
\eeq

 Hence, according to 
Proposition \ref{Proposition: modified objects}, 
we obtain the phase connection
\beq
\Gam \byd \Gam[g,\sig] \byd \Gam[g] + \Sig[g,\sig]
\eeq
and the associated scaled phase 2--form and 2--vector
\bat{2}
\Ome 
&\byd \Ome[g,\sig] \byd \Ome[g, \Gam] 
&&= 
\Ome[g] + \Ome\Ela[g,\sig] \,,
\\
\Lam 
&\byd \Lam[g,\sig] \byd \Lam[g, \Gam] 
&&= 
\Lam[g] + \Lam\Ela[g,\sig] \,.
\end{alignat*}

 We have
\bat{2}
\Ome\Ela[g,\sig] 
&= - \Alt (\pi^2\per (\sig)) 
&&: 
\M J_1\f E \to 	(\B T^*\ten \B L^2) \ten \Lam^2T^*\f E \,,
\\
\Lam\Ela[g,\sig] 
&= - 
\Alt \big((\nu_\tau \com g\Sha) \ten (\nu_\tau \com g\Sha)\big) (\sig)
&&: 
\M J_1\f E \to 	(\B T^*\ten \B L^2) \ten \Lam^2T^*\f E
\end{alignat*}
(where
$\pi^2\per$
denotes the orthogonal projection of the 2nd factor), i.e., in
coordinates,
\beq
\Ome\Ela[g,\sig] =
- \br g_{i\mu} \, \br g^{i\rho} \, \sig_{\lam\rho}\, d^\lam \wed d^\mu
\ssep{and}
\Lam\Ela[g,\sig] = 
- \fr1{(c \, \alp^0)^2}
\br g^{i\lam} \, \br g^{j\mu} \, \sig_{\lam\mu}
\, \der^0_i \wed \der^0_j \,.
\eeq

 Moreover, we define the scaled (0,2)--tensor
\beq
[\sig] \byd 
\sig - (\K d \con^2 \sig) \ten \tau : 
\M J_1\f E \to 	(\B T^*\ten \B L^2) \ten T^*\f E \ten T^*\f E
\eeq
(where
$\con^2$ 
denotes the insertion on the 2nd factor), with coordinate expression
\beq
[\sig] = 
(\sig_{\lam\mu} + 
(\alp^0)^2 \, 	\sig_{\lam\rho} \, \br\del^\rho_0 \, \br g_{0\mu}) \,
d^\lam \ten d^\mu \,.
\eeq

 Indeed, we have
$L_\Gam \, \tau = 0$
if and only if 
$[\sig]$
is symmetric.
\eLm

\begin{proof}
 We have
$L_\Gam \, \tau =
L_{\Gam[g]} \, \tau + L_\Sig \, \tau =
L_\Sig \, \tau = 
(\sig_{\lam\mu} + 
(\alp^0)^2 \, \br g_{0\mu} \, 	\sig_{\lam\rho} \, \br\del^\rho_0) \, 
d^\lam \wed d^\mu \,.$

 Hence, 
$L_\Gam \, \tau = 0$
if and only if 
$[\sig]$
is symmetric. 
\end{proof}

\bTh\label{Theorem: condition on sig for non metric contact structure}
 The pairs
$(- c^2 \, \tau, \, \Ome[g,\sig])$
and
$(- \fr1{c^2} \, \gam, \, \Lam[g,\sig])$
are scaled contact and Jacobi structures if and only if
$[\sig]$
is symmetric.
\hfill\ENDE
\eTh

\bEx
  Let us consider the scaled (0,2)--tensor
\beq
\sig \byd \fr{c^2\, m}{\h} \, (g + k \, c^2 \, \tau \ten \tau) \,,
\eeq
with
$k \in \map(\M J_1\f E, \Rn) \,,$
and the induced vertical valued form and the scaled 2--form
\beq
\Sig \byd
(\nu_\tau \com g\Sha^2)(\sig) =
\fr{c^2\, m}{\h} \, \nu_\tau \,.
\eeq

 Indeed, we obtain the symmetric scaled (0,2)--tensor (which does not
depend on $k$)
\beq
[\sig] = \fr{c^2\,m}\h \, (g + c^2 \, \tau \ten \tau)
\eeq
and the scaled phase 2--form and 2--vector
\bal
\Ome\Ela[g,\sig] 
&=
- \fr{c^2\,m}\h \, \br g_{i\mu} \, \br g^{i\rho} \, g_{\lam\rho} \,
d^\lam \wed d^\mu 
\\
&=
- \fr{c^2\,m}\h \, 
(\del^\rho_\mu + (\alp^0)^2 \, \br g_{0\mu} \, \br\del^\rho_0) \, 
g_{\lam\rho} \, d^\lam \wed d^\mu
\\
&=
- \fr{c^2\,m}\h \, 
(g_{\lam\mu} + (\alp^0)^2 \, \br g_{0\lam} \, \br g_{0\mu}) \,
d^\lam \wed d^\mu = 
0 \,,
\\
\Lam\Ela[g,\sig]
&=
- \fr1{(c \, \alp^0)^2} \, 
\br g^{i\lam} \, \br g^{j\mu} \, g_{\lam\mu} \, \der^0_i \wed \der^0_j
= 0 \,.
\end{align*}

 Thus, in virtue of 
Theorem \ref{Theorem: condition on sig for non metric contact
structure}
and according to Example \ref{Example: Sig induced by nu tau},
the pairs
$(- c^2 \, \tau, \, \Ome[g,\sig])$
and
$(- \fr1{c^2} \, \gam, \, \Lam[g,\sig])$
are scaled contact and Jacobi structures.
\hfill\ENDE
\eEx

\bLm\label{Lemma: case generated by psi}
 Let us consider a scaled symmetric (0,2)--tensor 
\beq
\psi : \M J_1\f E \to 	(\B T^*\ten \B L^2) \ten S^2T^*\f E
\eeq
and the induced scaled symmetric (0,2)--tensor
\beq
\sig \byd 
- \tfr12 (\psi - 2 \, (\K d \con \psi) \odot \tau) : \M J_1\f E \to
	(\B T^*\ten \B L^2) \ten S^2T^*\f E \,,
\eeq
with coordinate expression
\beq
\sig = 
- \tfr12 \big(\psi_{\lam\mu} + 2 \,
(\alp^0)^2 \, \br g_{0\lam} \, (\psi_{0\mu} + x^i_0 \, \psi_{i\mu})
\big) \, d^\lam \odot d^\mu \,.
\eeq

 Then, the induced vertical valued 1--form turns out to be the section
\beq
\Sig = 
- \tfr12 (\nu_\tau \com g\Sha^2) (\psi - \tau \ten (\K d \con \psi))
\,,
\eeq
with coordinate expression
\beq
\Sig = \fr1{2 \, c \, \alp^0} \, \br g^{i\rho} \, 
\big(
\psi_{\lam\rho} + (\alp^0)^2 \, \br g_{0\lam} \,
(\psi_{0\rho} + \psi_{i\rho} \, x^i_0) 
\big) \, d^\lam \ten \der^0_i \,.
\eeq

 Indeed, the scaled (0,2)--tensor
\beq
[\sig] = - \tfr12 (\psi - 2 \, \tau \odot (\K d \con \psi) 
	+ (\K d \con \K d \con \psi) \, \tau \ten \tau)
\eeq
turns out to be symmetric.
 Hence, in virtue of 
Theorem \ref{Theorem: condition on sig for non metric contact
structure}, we have
\beq
\Ome\Ela[g,\sig] = 0 = \Lam\Ela[g,\sig]
\eeq
and the pairs
$(- c^2 \, \tau, \, \Ome[g,\sig])$
and
$(- \fr1{c^2} \, \gam, \, \Lam[g,\sig])$
are scaled contact and Jacobi structures, respectively.
\hfill\ENDE
\eLm

\bNt
 With reference to the above Lemma, a direct computation gives
\bal
\Ome\Ela[g,\sig] 
&=
- c \, \alp^0 \, \br g_{i\mu} \, \Sig\Ga\lam i \, d^\lam \wed d^\mu
\\
&=
- \tfr12 \br g_{i\mu} \, \br g^{i\rho} \, 
\big(\psi_{\lam\rho}
+ (\alp^0)^2 \, \br g_{0\lam} \, 
(\psi_{0\rho} + x^i_0 \, \psi_{i\rho})
+ (\alp^0)^2 \, \br g_{0\rho} \, 
(\psi_{0\lam} + x^i_0 \, \psi_{i\lam})
\big) \,d^\lam \wed d^\mu
\\
&=
- \tfr12 
(\del^\rho_\mu + (\alp^0)^2 \,\br g_{0\mu} \, \br \del^{\rho}_0) \, 
\big(\psi_{\lam\rho}
+ (\alp^0)^2 \, (\br g_{0\lam} \, 
\psi_{\sig\rho} + \br g_{0\rho} \, 
\psi_{\sig\lam}) \,\br\del^\sig_0
\big) \, d^\lam \wed d^\mu
\\
&=
- \tfr12 
\big(\psi_{\lam\mu}
+ (\alp^0)^2 \, (\br g_{0\lam} \, 
\psi_{\sig\mu} + \br g_{0\mu} \, 
\psi_{\sig\lam}) \, \br\del^\sig_0
+ (\alp^0)^2 \, \br g_{0\mu} \, \psi_{\lam\rho} \, \br \del^\rho_0
\\
&\qquad\quad
+ (\alp^0)^4 \, (\br g_{0\lam} \, \br g_{0\mu} \, \psi_{\sig\rho}
	\,\br\del^\rho_0 \, \br\del^\sig_0
+ \br g_{0\rho} \, \br g_{0\mu} \, \psi_{\sig\lam}
	\,\br\del^\rho_0 \, \br\del^\sig_0)
\big) \, d^\lam \wed d^\mu
\\
&=
- \tfr12 
\big(\psi_{\lam\mu}
+ (\alp^0)^2 \, (\br g_{0\lam} \, 
\psi_{\sig\mu} + \br g_{0\mu} \, 
\psi_{\sig\lam}) \, \br\del^\sig_0
+ (\alp^0)^2 \, \br g_{0\mu} \, \psi_{\lam\rho} \, \br \del^\rho_0
\\
&\qquad\quad
+ (\alp^0)^4 \, \br g_{0\lam} \, \br g_{0\mu} \, \psi_{\sig\rho}
	\,\br\del^\rho_0 \, \br\del^\sig_0
- (\alp^0)^2 \, \br g_{0\mu} \, \psi_{\sig\lam} \, \br\del^\sig_0)
\big) \,d^\lam \wed d^\mu
\\
&= 0 \,.
\end{align*}
\vglue-1.5\baselineskip\hfill\ENDE
\eNt

\bLm\label{Lemma: case generated by phi}
 Let us consider a scaled antisymmetric (0,2)--tensor
\beq
\vphi : \M J_1\f E \to (\B T^* \ten \B L^2) \ten \Lam^2 T^*\f E \,,
\eeq
and the induced scaled antisymmetric (0,2)--tensor
\beq
\sig \byd 
- \tfr12 (\vphi - 2 \, (\K d \con \vphi) \wed \tau) :
\M J_1\f E \to (\B T^* \ten \B L^2) \ten \Lam^2 T^*\f E \,,
\eeq
with coordinate expression
\beq
\sig =
- \tfr12 \big(\vphi_{\lam\mu} - 2 \,
(\alp^0)^2 \, \br g_{0\lam} \, \vphi_{\sig\mu} \,\br\del^\sig_0
\big) \, d^\lam \wed d^\mu \,.
\eeq

 Then, the induced vertical valued 1--form turns out to be the section
\beq
\Sig = 
- \tfr12 (\nu_\tau \com g\Sha^2) (\vphi + \tau \ten (\K d \con \vphi))
\,,
\eeq
with coordinate expression
\beq
\Sig = - \tfr12 \tfr1{c \, \alp^0} \, \br g^{i\rho} \, 
\big(
\vphi_{\lam\rho} - (\alp^0)^2 \, \br g_{0\lam} \,
\vphi_{\sig\rho} \,\br\del^\sig_0 
\big) \, d^\lam \ten \der^0_i \,.
\eeq

 Hence, the induced scaled phase 2--form and (0,2)--vector turn out to
be
\beq
\Ome\Ela[g,\Sig] = \tfr12 \vphi
\ssep{and}
\Lam\Ela[g,\Sig] = 
\tfr12 \, 
\Alt \big((\nu_\tau\com g\Sha) \ten (\nu_\tau \com g\Sha)\big) (\vphi)
\,.
\eeq
\eLm

\begin{proof}
 \emph{1st proof.}
 We have
\bal
	\Ome\Ela[g,\Sig]
&=
- c \, \alp^0 \, \br g_{i\mu} \, \Sig\Ga\lam i \, d^\lam \wed d^\mu
\\
&=
\tfr12 	\br g_{i\mu} \, \br g^{i\rho} \,
\big(\vphi_{\lam\rho}
	- (\alp^0)^2 \, 
(\br g_{0\lam} \, \vphi_{\sig\rho} - 
\br g_{0\rho} \, 		\vphi_{\sig\lam}) \,
\br\del^\sig_0\big) 	\, d^\lam \wed d^\mu
\\
&=
\tfr12 	(\del^\rho_\mu + (\alp^0)^2 \, \br g_{0\mu} \,
\br\del^\rho_0)
\, 	\big(\vphi_{\lam\rho} - 
(\alp^0)^2 \, 
(\br g_{0\lam} \, \vphi_{\sig\rho} - 
\br g_{0\rho} \, \vphi_{\sig\lam}) \,
\br\del^\sig_0\big) \, d^\lam \wed d^\mu 
\\
&=
\tfr12 	\Big(\vphi_{\lam\mu} - 
(\alp^0)^2 \, \br g_{0\lam} \, 	\vphi_{\sig\mu} \,
\br\del^\sig_0
	+ (\alp^0)^2 \, \br g_{0\mu} \, \vphi_{\sig\lam} \,
\br\del^\sig_0
	+ (\alp^0)^2 \,\br g_{0\mu} \, \vphi_{\lam\sig}] \,
\br\del^\sig_0
\\
&\qquad
	- (\alp^0)^4 \, \br g_{0\lam} \, \br g_{0\mu} 
\vphi_{\sig\rho} \, \br\del^\sig_0 \, \br\del^\rho_0
	+ (\alp^0)^4 \, \br g_{0\mu} \, \ha g_{00} 
\vphi_{\sig\lam} \, \br\del^\sig_0\Big) \, d^\lam \wed d^\mu
\\
&=
\tfr12 \vphi_{\lam\mu} \, d^\lam \wed d^\mu 
= 
\tfr12 \vphi \,,
\\[3mm]
\Lam\Ela[g,\Sig] 
& =
\tfr1{c\,\alp^0} \,
\br g^{j\lam} \, \Sig\Ga\lam i \, \der^0_i\wed \der^0_j
\\
& =
- \tfr1{2\,(c\,\alp^0)^2} \, 
\br g^{j\lam} \, \br g^{i\rho} \, (\vphi_{\lam\rho}
	- (\alp^0)^2 \br g_{0\lam} \, \vphi_{\sig\rho} \, \br\del^\sig_0) \,
\der^0_i \wed \der^0_j
\\
&=
- \tfr1{2\,(c\,\alp^0)^2} \, 
\br g^{j\lam} \, \br g^{i\rho} \, \vphi_{\lam\rho} \,
	\der^0_i \wed \der^0_j \,.
\end{align*}

\emph{2nd proof.}
 By applying 
Lemma \ref{Lemma: structures generated by sig} 
and recalling the identity
$\K d \con \tau = 1 \,,$
we obtain
\bal
	\Ome\Ela[g,\Sig] 
&= 
- \Alt (\pi^2\per (\sig))
\\
&=
\tfr12 \Alt \big(\pi^2\per (\vphi - 2 \, (\K d \con \vphi) \wed
\tau)\big)
\\
&=
\tfr12 \Alt \big(\pi^2\per (\vphi + \tau \ten (\K d \con \vphi)
- (\K d\con \vphi)\ten\tau)\big)
\\
&=
\tfr12 \Alt
\big(\vphi - (\K d \con \vphi) \ten \tau + 
\tau \ten (\K d \con \vphi) + (\K d \con \vphi) \ten \tau 
\\
&\quad
-  \tau \ten (\K d \con \vphi) 
\tau - (\K d \con \K d \con \vphi)\tau \ten \tau\big)
=
\tfr12 \vphi \,,
\\[3mm]
\Lam\Ela[g,\Sig]
&=
- \Alt\big((\nu_\tau\com g\Sha) \ten (\nu_\tau\com g\Sha)\big) (\sig)
\\
& = 
\tfr12 
\Alt\big((\nu_\tau\com g\Sha) \ten (\nu_\tau\com g\Sha)\big) (\vphi)
\,. 
\end{align*}
\vglue-1.5\baselineskip
\end{proof}

 In the above Lemmas, we have defined
$\sig$
by using a normalising factor
$\tfr12 \,;$
indeed, this factor might be chosen in an arbitrary way.
 Our choice is selected in such a way that, in the case when the
source of the additional term
$\phi$
is the electromagnetic field, the resulting 
$\Ome\Ela$
and
$\Lam\Ela$
be normalised as it is usually done in the standard literature.

\bTh\label{Th6.10}
 Let us consider a scaled (0,2)--tensor
\beq
\ome = \psi + \vphi : \M J_1\f E \to
	(\B T^*\ten \B L^2) \ten T^*\f E \ten T^*\f E \,,
\eeq
where
$\psi$
and
$\phi$
are the symmetric and antisymmetric components of
$\ome \,,$ respectively,
and the induced scaled (0,2)--tensor
\beq
\sig \byd 
- \tfr12 \big(
\ome - (\K d \con \ome) \ten \tau - \tau \ten (\K d \con^2 \ome)
\big)
: \M J_1\f E \to
	(\B T^* \ten \B L^2) \ten T^*\f E \ten T^*\f E \,,
\eeq
with coordinate expression
\beq
\sig =
- \tfr12 \big(\ome_{\lam\mu} +
(\alp^0)^2 \, \br g_{0\mu} \, (\ome_{0\lam} + \ome_{i\lam} \, x^i_0) +
(\alp^0)^2 \, \br g_{0\lam} \, (\ome_{\mu 0} + \ome_{\mu i} \, x^i_0)
\big) \, d^\lam \ten d^\mu \,.
\eeq

 Then, the induced vertical valued 1--form turns out to be the section
\beq
\Sig[g, \sig] = 
- \tfr12 (\nu_\tau \com g\Sha^2) 
\big(\ome - \tau \ten (\K d \con^2 \ome)\big)
\,,
\eeq
with coordinate expression
\beq
\Sig[g, \sig] = - \fr1{2 \, c \, \alp^0} \, \br g^{i\rho} \, 
\big(
\ome_{\lam\rho} - (\alp^0)^2 \, \br g_{0\lam} \,
\ome_{\sig\rho} \, \br\del^\sig_0) 
\big) \, d^\lam \ten \der^0_i \,.
\eeq

 Hence, the induced scaled phase 2--form and 2--vector turn out to
be 
\beq
\Ome[g,\Sig] = - c^2 \, d\tau + \tfr12 \vphi
\ssep{and}
\Lam[g,\Sig] = 
\Lam[g] + \tfr12 
\Alt\big((\nu_\tau \com g\Sha) \ten (\nu_\tau \com g\Sha)\big) (\vphi)
\,.
\eeq
\eTh

\begin{proof}
 It follows from
Lemma \ref{Lemma: case generated by psi}
and
Lemma \ref{Lemma: case generated by phi}. 
\end{proof}

\bLm
 Let
$A : \M J_1\f E \to (\B T^* \ten \B L^2) \ten T^*\f E$
be a scaled 1--form and 
$\Gam = \Gam[g] + \Sig$
any phase connection.
 Then, we have
\beq
(- c^2\, \tau + A) \wed \Ome[g,\Gam] \wed \Ome[g,\Gam] 
	\wed \Ome[g,\Gam] =
(- c^2\, \tau + A) \wed \Ome[g] \wed \Ome[g] \wed \Ome[g] \,.
\eeq
\eLm

\begin{proof}
 We can easily see, in coordinates, that the terms including
$\Ome\Ela[g,\Sig]$
disappear in the wedge product because they are horizontal 2--forms,
hence they generate horizontal forms of degree greater than 4. 
\end{proof}

 We stress that the above $7$--form 
$(- c^2 \, \tau + A) \wed \Ome[g] \wed \Ome[g] \wed \Ome[g]$
needs not to be a volume form.

\bLm\label{Lm6.7}
 Let us consider an exact scaled 2--form
\beq
\vphi \byd 2 \, dA : 
\M J_1\f E \to (\B T^* \ten \B L^2) \ten \Lam^2T^*\f E
\eeq
and the induced phase 2--form
$\Ome[g,\vphi] \byd \Ome[g] + \Ome\Ela[g,\vphi] \,.$

 Then, we have
\beq
\Ome[g,\vphi] = d (- c^2 \, \tau + A)
\eeq	
if and only if
$A$ 
is the pullback of a scaled spacetime 1-form
$A : \f E \to (\B T^* \ten \B L^2) \ten T^*\f E \,.$
\eLm

\begin{proof}
 In fact, we have 
$\Ome[g,\vphi] = - c^2 \, d\tau$
and
\beq
\Ome\Ela[g,\phi] =
- c \, \alp^0 \, \br g_{i\mu} \, \Sig\Ga\lam i \,
d^\lam \wed d^\mu = 
\der_\lam A_\mu \, d^\lam \wed d^\mu 	+ 
\der^0_i A_\lam \, d^i_0 \wed d^\lam 	\,.
\eeq

 Hence,
$\Ome\Ela[g,\vphi] = dA$
if and only if
$\der^0_i A_\lam = 0 \,.$ 
\end{proof}

 We can summarize the above results as follows.

\bCr
 Let us consider a closed scaled spacetime 2-form
\beq
\vphi : \f E \to (\B T^* \ten \B L^2) \ten \Lam^2T^*\f E
\eeq
and the induced phase 2--form
$\Ome[g,\vphi] \byd \Ome[g] + \Ome\Ela[g,\vphi] =
\Ome[g] + \tfr12\,\vphi \,.$

 Then, the pair
$(- c^2 \, \tau, \, \Ome[g, \vphi])$
turns out to be an almost--cosymplectic--contact structure, i.e.
$\Ome[g,\vphi]$ 
is closed and 
$- c^2\,\tau\wed \Ome \wed \Ome \wed \Ome$
is a volume form. 

 Moreover, if 
$\vphi = 2dA \,,$
where
\beq
A : \f E \to (\B T^* \ten \B L^2) \ten T^*\f E
\eeq
is a scaled spacetime 1-form, then the induced phase 2--form is given
by 
$\Ome[g, 2dA] \byd \Ome[g] + \Ome\Ela[g,2\,dA] 
= d(- c^2\,\tau + A) \,.$

 Hence, the pair
$(- c^2 \, \tau + A \,, \; \Ome[g, 2\,dA])$
turns out to be a contact structure if and only if
$(- c^2 \, \tau + A) \wed \Ome[g] \wed \Ome[g] \wed \Ome[g]$
is everywhere non vanishing.
\hfill\ENDE
\eCr

\bRm
 Let us consider a closed 2--form 
$\vphi: \f E \to (\B T^* \ten \B L^2) \ten \Lam^2T^*\f E \,.$
 Then, the dual almost-coPoisson--Jacobi pair 
$(- \tfr1{c^2}\,\gam[\vphi], \, \Lam[g,\vphi])$ 
is given by
\beq
\gam = \gam[g] + \gam[\vphi] \byd \K d \con (\Gam[g] + \Sig[\vphi])
\eeq
and 
$\Lam[g,\vphi] \byd \Lam[g] + \Lam\Ela[g,\vphi] \,,$ 
where
\beq
\Sig = 
- \tfr12 (\nu_\tau \com g\Sha^2) (\vphi + \tau \ten (\K d \con \vphi))
\eeq
and
\beq
\Lam\Ela[g,\Sig] =
\tfr12 
\Alt\big((\nu_\tau\com g\Sha) \ten (\nu_\tau\com g\Sha)\big) (\vphi)
\,.
\eeq
\vglue-1.5\baselineskip\hfill\ENDE
\eRm

 We can apply the above results to the phase structures associated
with an electromagnetic field represented by a scaled closed 2--form
$F : \f E \to (\B L^{1/2} \ten \B M^{1/2}) \ten \Lam^2T^*\f E \,.$

 For this purpose, it suffices to consider a mass
$m \in \B M$
and a charge
$q \in \B T^{-1} \ten \B L^{3/2} \ten \B M^{1/2}$
and set
$\vphi \byd \tfr qm F \,.$


\end{document}